\documentclass[12pt]{article}
 
\usepackage{amsfonts,amsmath,amssymb,theorem,graphicx,natbib}
\usepackage{mathrsfs} 
\usepackage{multirow}
\usepackage{enumerate}
\usepackage{caption}
\usepackage{subcaption}
\usepackage{latexsym}
\usepackage{nicefrac}
\usepackage{verbatim}
\usepackage{comment}
\usepackage{float}

\newcommand{\balpha}{\boldsymbol{\alpha}}
\newcommand{\btheta}{\boldsymbol{\theta}}
\newcommand{\btau}{\boldsymbol{\tau}}

\newcommand{\bp}{\mathbf{p}}

\newcommand\BE{\mathbb{E}}

\newcommand\bx{\mathbf{x}}

\newcommand\varm{{\mathrm{var}}}
\DeclareMathOperator*{\argmin}{arg\,min}

\newenvironment{proof}{\paragraph{Proof:}}{\hfill$\square$\\}
  
\newtheorem{exemp}{Example}[section]

\newtheorem{theorem}{Theorem}
\newtheorem{corollary}{Corollary}
\newtheorem{lemma}{Lemma}
\newenvironment{example}
{\begin{exemp}\begin{em}}
{\end{em}\end{exemp}}

\begin{document}

\title{\bf Weakly informative reparameterisations for location-scale mixtures}
\author{{\sc Kaniav Kamary}\thanks{
Kaniav Kamary and Christian Robert, CEREMADE, Universit{\' e} Paris-Dauphine, 75775 Paris cedex 16, France, {\sf
kamary,xian@ceremade.dauphine.fr}, Jeong Eun Lee, Auckland University of Technology, New Zealand, {\sf jeong.lee@aut.ac.nz}.  The authors are grateful to Robert Kohn for his helpful comments and to all reviewers for improving the presentation of the paper.}\\
Universit{\'e} Paris-Dauphine, CEREMADE, and INRIA, Saclay\\
{\sc Jeong Eun Lee}\\
Auckland University of Technology, New Zealand\\
and {\sc Christian P.~Robert}\\
Universit{\'e} Paris-Dauphine, and University of Warwick}
\maketitle

\begin{abstract} 
While mixtures of Gaussian distributions have been studied for more than a
century, the construction of a reference Bayesian analysis of
those models remains unsolved, with a general prohibition of 
improper priors \citep{fruhwirth:2006} due to the ill-posed nature of such
statistical objects. This difficulty is usually bypassed by an empirical Bayes
resolution \citep{richardson:green:1997}. By creating a new parameterisation
centred on the mean and possibly the variance of the mixture distribution
itself, we manage to develop here a weakly informative prior for a wide
class of mixtures with an arbitrary number of components. We demonstrate that
some posterior distributions associated with this prior and a minimal sample size are proper. We provide MCMC implementations that exhibit the expected exchangeability.  We only study here the univariate case, the extension to multivariate
location-scale mixtures being currently under study. An R package called Ultimixt is associated with this paper.
\end{abstract}


\section{Introduction}\label{intro}

A mixture density is traditionally represented as a weighted average of densities from standard families, i.e.,
\begin{equation}\label{eq:mix}
f(x|\btheta,\bp) = \sum_{i=1}^k p_if_i(x|\theta_i)\qquad \sum_{i=1}^k p_i=1\,.
\end{equation}
Each component of the mixture is characterised by a component-wise parameter
$\theta_i$ and the weights $p_i$ of those components translate the importance
of each of those components in the model. A more general if rarely considered
mixture model involves different families for the different components. 

This particular representation \ref{eq:mix} gives a separate meaning to each
component through its parameter $\theta_i$, even though there is a well-known
lack of identifiability in such models, due to the invariance of the sum by
permutation of the indices. This issue relates to the equally well-known
``label switching" phenomenon in the Bayesian approach to the model, which
pertains both to Bayesian inference and to simulation of the corresponding posterior
\citep{celeux:hurn:robert:2000,stephens:2000b,fruhwirth:2001,jasra:holmes:stephens:2005}.
From this Bayesian viewpoint, the choice of the prior distribution on the
component parameters is quite open, the only constraint being obviously that the
corresponding posterior is proper. \cite{diebolt:robert:1994} and \cite{wasserman:1999} discussed the alternative
approach of {\em imposing} proper posteriors on improper priors by banning
almost empty components from the likelihood function. While consistent, this
approach induces dependence between the observations, requires a large enough number of observations, higher
computational costs, and does not handle over-fitting very well. 

The prior distribution on the weights $p_i$ is equally open for choice, but a
standard version is a Dirichlet distribution with common hyperparameter $\alpha_0$,
$\text{Dir}(\alpha_0,\ldots,,\alpha_0)$. \ Recently, \cite{rousseau:mengersen:2011}
demonstrated that the choice of this hyperparameter $\alpha_0$ relates to the
inference on the total number of components, namely that a small enough value
of $\alpha_0$ manages to handle over-fitted mixtures in a convergent manner.
In Bayesian non-parametric modelling, \citet{Griffin2010} showed that the prior
on the weights may have a higher impact when inferring about the number of
components, relative to the prior on the component-specific parameters. As
indicated above, the prior distribution on the $\theta_i$'s has received less
attention and conjugate choices are most standard, since they facilitate
simulation via Gibbs samplers
\citep{diebolt:robert:1990a,escobar:west:1995,richardson:green:1997} if not
estimation, since posterior moments remain unavailable in closed form. In
addition, \cite{richardson:green:1997} among others proposed data-based priors
that derive some hyperparameters as functions of the data, towards an automatic
scaling of such priors, as illustrated by the R package, {\ttfamily bayesm} \citep{PTRSI2010}. 

In an objective Bayes perspective \citep{berger:2004,berger:bernardo:sun:2009},
we seek a prior that is minimally informative with respect to the information
brought by the data. This has been formalised in different ways, including the
Jeffreys prior \citep{jeffreys:1939}, reference priors
\citep{berger:bernardo:sun:2009}, maximum entropy priors \citep{rissanen:2012},
matching priors \citep{ghosh:carlin:srivastiva:1995}, which often include Jeffreys priors
\citep{welch:peers:1963}, and other proposals
\citep{kass:wasserman:1996}. In the case of mixture models, very little has
been done, apart from \cite{bernardo:giron:1988}, who derived the Jeffreys
priors for mixtures where components have disjoint supports,
\cite{figueiredo:jain:2002} who used independent Jeffreys prior on components,
and \cite{rubio:steel:2014} which achieve a closed-form expression for the
Jeffreys prior of a location-scale mixture with two disjoint components.
Recently, \citet{grazian:robert:2015} undertook an analytical and numerical
study of Jeffreys priors for Gaussian mixtures, which showed that Jeffreys
priors are almost invariably associated with improper posteriors, whatever the
sample size, and advocated the use of pseudo-priors expressed as conditional
Jeffreys priors for each type of parameters. In this paper, we instead start
from the traditional Jeffrey prior for a location-scale parameter to derive a
joint prior distribution on all parameters by taking advantage of compact
reparameterisation, which allow for uniform distributions and similarly weakly
informative distributions.

In the case when $\theta_i=(\mu_i,\sigma_i)$ is a location-scale parameter,
\cite{mengersen:robert:1996} have already proposed a reparameterisation of \eqref{eq:mix}
that express each component as a local perturbation of the previous one, namely
$\mu_i=\mu_{i-1}+\sigma_{i-1}\delta_i$,
$\sigma_i=\tau_i\sigma_{i-1}$, $\tau_i<1$ $(i>1)$, with $\mu_1$ and $\sigma_1$
being the reference values. Based on this reparameterisation,
\cite{robert:titterington:1998} established that a specific improper prior on
$(\mu_1,\sigma_1)$ leads to a proper posterior in the Gaussian case. We propose here to modify
this reparameterisation by using the mean and variance of the
mixture distribution as reference location and scale, respectively. This
modification has foundational consequences in terms of identifiability and hence of exploiting improper and
non-informative priors for mixture models, in sharp contrast with the existing literature \citep[see,
e.g.][]{diebolt:robert:1994,wasserman:1999}.

Computational approaches to Bayesian inference on mixtures are quite diverse,
starting with the introduction of the Gibbs sampler
\citep{diebolt:robert:1990a,gelman:king:1990,escobar:west:1995}, some concerned
with approximations \citep{roeder:1990,wasserman:1999} and MCMC features
\citep{richardson:green:1997,celeux:hurn:robert:2000}, and others with
asymptotic justifications, in particular when over-fitting mixtures
\citep{rousseau:mengersen:2011}, but most attempting to overcome the
methodological hurdles in estimating mixture models
\citep{chib:1995,neal:1999,berkhof:mechelen:gelman:2003,marin:mengersen:robert:2004,
fruhwirth:2006,lee:marin:mengersen:robert:2008}.
While we do not propose here a novel computational methodology attached with our new priors, we
study the performances of several MCMC algorithms on such targets.

In this paper, we introduce and study a principle of mean-variance or simply mean reparameterisation (Section
\ref{sec:miX}), which main consequence is to constrain all parameters but mean
and variance of the overall mixture model within a compact space. We study
several possible parameterisations of that kind and demonstrate that an
improper Jeffreys-like prior associated with them is proper for a wide variety
of mixture and compound mixture distributions. Taking advantage of constraints
on component-wise parameters, a domain based prior is used. Section
\ref{sec:weak} discusses some properties of the resulting priors in terms of
the modelled densities. In Section \ref{sec:mcmc}, we propose some MCMC
implementations to estimate the parameters of the mixture, discussing label
switching (Section \ref{sec:witch}). Note that a public R package called 
Ultimixt is associated with this approach.  Section \ref{sim_study} describes
several case studies when implementing the reparameterisation principle, and
Section \ref{sec:con} briefly concludes the paper. Proofs of the main results
are available in the Supplementary Material.

\section{Mixture reparameterisation}\label{sec:miX}

\subsection{Mean and variance of a mixture}

Let us first recall how both mean and variance of a mixture distribution with
finite first two moments can be represented in terms of the mean and variance
parameters of the components of the mixture. 

\begin{lemma}\label{lem:mings}
If $\mu_i$ and $\sigma_i^2$ are well-defined as mean and variance of the distribution with density $f_i(\cdot|\theta_i)$,
respectively, the mean of the mixture distribution \eqref{eq:mix} is given by
$$\BE_{\btheta,\bp}[X]=\sum_{i=1}^k p_i\mu_i$$
and its variance by
$$\varm_{\btheta,\bp}(X)=\sum_{i=1}^k p_i\sigma_i^2 + \sum_{i=1}^k p_i(\mu_i^2-\BE_{\btheta,\bp}[X]^2)$$
\end{lemma}

For any location-scale mixture, we propose a reparameterisation of the mixture model that starts by scaling all parameters in terms of its global mean $\mu$ and global variance \footnote{Strictly speaking, the term {\em global} is
superfluous, but we add it nonetheless to stress that those moments are defined in terms
of the mixture distribution, rather than for its components.} 
$\sigma^2$.  For instance, we can switch from the parameterisation in $(\mu_i,\sigma_i)$ to a new
parameterisation in 
$
(\mu,\sigma,\alpha_1,\ldots,\alpha_k,\tau_1,\ldots,\tau_k,p_1,\ldots,p_k)\,,
$
expressing those component-wise parameters as
\begin{equation}\label{eq:locavore}
\mu_i=\mu+\sigma\alpha_i \quad \text{ and } \quad \sigma_i=\sigma\tau_i \qquad 1\le i\le k
\end{equation}
where $\tau_i>0$ and $\alpha_i \in
\mathbb{R}$. This bijective reparameterisation is similar to the one in
\cite{mengersen:robert:1996}, except that these authors put no special meaning
on their location and scale parameters, which are then non-identifiable. Once $\mu$ and $\sigma$ are defined as
(global) mean and variance of the mixture distribution, eqn. \eqref{eq:locavore} imposes compact
constraints on the other parameters of the model. For instance, since the
mixture variance is equal to $\sigma^2$, this implies that
$(\mu_1,\ldots,\mu_k,\sigma_1,\ldots,\sigma_k)$ belongs to an ellipse
conditional on the weights, $\mu$, and $\sigma$, by virtue of Lemma \ref{lem:mings}. 

Considering the $\alpha_i$'s and the $\tau_i$'s in \eqref{eq:locavore} as the
new and local parameters of the mixture components, the following result (Lemma \ref{lem:const}) states
that the global mean and variance parameters are the sole freely varying
parameters. In other words, once both the global mean and variance are defined
as such, there exists a parameterisation such that all remaining parameters of
a mixture distribution are restricted to belong to a compact set, a feature that is most helpful in selecting a non-informative prior distribution.

\begin{lemma}\label{lem:const}
The parameters $\alpha_i$ and $\tau_i$ in \eqref{eq:locavore} are constrained by
$$\sum_{i=1}^k p_i\alpha_i=0 \quad {\mbox and } \quad \sum_{i=1}^k p_i\tau_i^2+\sum_{i=1}^k p_i\alpha_i^2=1\,.$$
\end{lemma}

The same concept applies for other families, namely that one or
several moments of the mixture distribution can be used as a pivot to constrain
the component parameters. For instance, a mixture of exponential distributions
$\mathcal{E}(\lambda_i^{-1})$ or a mixture of Poisson distributions
$\mathcal{P}(\lambda_i)$ can be reparameterised in terms of its mean,
$\mathbb{E}[X]$, through the constraint
$$
\mathbb{E}[X]=\lambda=\sum_{i=1}^k p_i\lambda_i,
$$
by introducing the parameterisation $\lambda_i=\lambda\gamma_i/p_i$,
$\gamma_i>0$, which implies $\sum_{i=1}^k \gamma_i=1$.  As detailed below, this
notion immediately extends to mixtures of compound distributions, which are
scale perturbations of the original distributions, with a fixed distribution on
the scales.

\subsection{Proper posteriors of improper priors}
The constraints in Lemma \ref{lem:const} define a set of values of
$(p_1,\ldots,p_k,\alpha_1,\ldots,\alpha_k,\tau_1,\ldots,\tau_k)$ that is
obviously compact. One sensible modelling approach exploiting this feature
is to resort to uniform or other weakly
informative proper priors for those component-wise parameters, conditional on
$(\mu,\sigma)$. Furthermore, since $(\mu,\sigma)$ is a location-scale
parameter, we invoke \cite{jeffreys:1939} to choose a Jeffreys-like prior
$\pi(\mu,\sigma)=1/\sigma$ on this parameter, even though we stress this is not the
genuine (if ineffective) Jeffreys prior for the mixture model \citep{grazian:robert:2015}. In
the same spirit as \cite{robert:titterington:1998},
we now establish that this choice of prior produces a proper posterior distribution for a minimal sample size of two.

\begin{theorem}\label{th:proper}
The posterior distribution associated with the prior $\pi(\mu,\sigma)=1/\sigma$
and with the likelihood derived from \eqref{eq:mix} is proper when the
components $f_1(\cdot|\mu,\sigma), \dots, f_k(\cdot|\mu,\sigma)$ are Gaussian
densities, provided (a) prior distributions on the other parameters are proper
and independent of $(\mu,\sigma)$, and (b) there are at least two observations
in the sample. 
\end{theorem}

While only handling the Gaussian case is a limitation, 
the above result extends to mixtures of compound Gaussian distributions, which are defined as
scale mixtures, namely $X=\mu+\sigma \xi Z$, $Z\sim\mathrm{N}(0,1)$ and $\xi\sim h(\xi)$,
when $h$ is a probability distribution on $\mathbb{R}^+$ with second moment
equal to $1$. (The moment constraint ensures that the mean and variance of this
compound Gaussian distribution are $\mu$ and $\sigma^2$, respectively.) As
shown by \cite{andrews:mallows:1974}, by virtue of Bernstein's theorem, such
compound Gaussian distributions are identified as completely monotonous
functions and include a wide range of probability distributions like the $t$,
the double exponential, the logistic, and the $\alpha$-stable distributions \citep{feller:1971}.

\begin{corollary}
The posterior distribution associated with the prior $\pi(\mu,\sigma)=1/\sigma$
and with the likelihood derived from \eqref{eq:mix}
is proper when the component densities $f_i(\cdot|\mu,\sigma)$ are all compound Gaussian densities, provided (a) prior distributions on the other parameters are proper and independent of $(\mu,\sigma)$ and (b) there are at least two observations in the sample.
\end{corollary}

The proof of this result is a straightforward generalisation of the one of
Theorem \ref{th:proper}, which involves integrating out the compounding
variables $\xi_1$ and $\xi_2$ over their respective distributions. Note that
the mixture distribution
\eqref{eq:mix} allows for different classes of location-scale distributions to be used in the different components.

If we now consider the case of a Poisson mixture, 
\begin{equation}\label{eq:poimix}
f(x|\lambda_1, \ldots, \lambda_k)=\frac{1}{x!}\sum_{i=1}^k p_i \lambda_i^x e^{-\lambda_i}
\end{equation}
with a reparameterisation as $\lambda=\mathbb{E}[X]$ and $\lambda_i=\lambda
\gamma_i/p_i$, we can use the equivalent to the Jeffreys prior for the Poisson
distribution, namely, $\pi(\lambda)=1/\lambda$, since it leads to a
well-defined posterior with a single positive observation. 

\begin{theorem}\label{thm:poimix}
The posterior distribution associated with the prior $\pi(\lambda)=1/\lambda$
and with the Poisson mixture \eqref{eq:poimix} is
proper provided (a) prior distributions on the other parameters are proper and independent of $\lambda$ and, 
(b) there is at least one strictly positive observation in the sample.
\end{theorem}

Once again, this result straightforwardly extends to mixtures of compound
Poisson distributions, namely distributions where the parameter is random with
mean $\lambda$:
$$\mathbb{P}(X=x|\lambda)=\int \frac{1}{x!} (\lambda\xi)^x\exp\{-\lambda\xi\}\, \text{d}\nu(\xi)\,,$$
with the distribution $\nu$ possibly discrete. In the special case when $\nu$ is on the integers, this representation covers all infinitely exchangeable distributions \citep{feller:1971}.

\begin{corollary}\label{cor:poimix}
The posterior distribution associated with the prior $\pi(\lambda)=1/\lambda$ 
and with the likelihood derived from a mixture of compound Poisson distributions is 
proper provided (a) prior distributions on the other parameters are proper and independent to $\lambda$ and, (b) there is at least one strictly positive observation in the sample.
\end{corollary}

Another instance of a proper posterior distribution is provided by exponential mixtures,
\begin{equation}\label{eq:expmix}
f(x|\lambda_1, \ldots, \lambda_k)= \sum_{i=1}^k \frac{p_i}{\lambda_i} e^{-\nicefrac{x}{\lambda_i}}\,,
\end{equation}
since a reparameterisation via $\lambda=\mathbb{E}[X]$ and $\lambda_i=\lambda\gamma_i/p_i$ leads to
the posterior being well-defined for a single observation.

\begin{theorem}\label{thm:expmix}
The posterior distribution associated with the prior $\pi(\lambda)=1/\lambda$
and with the likelihood derived from the exponential mixture \eqref{eq:expmix} is
proper provided proper distributions are used on the other parameters. 
\end{theorem}

Once again, this result directly extends to mixtures of compound
exponential distributions, namely exponential distributions where the parameter is random with
mean $\lambda$:
$$f(x|\lambda)=\int  \frac{1}{\lambda\xi}\exp\{-\nicefrac{x}{\lambda\xi}\}\, \text{d}\nu(\xi)\,,\quad x>0\,.$$
In particular, this representation contains all Gamma distributions with shape less than one \citep{gleser:1989}
and Pareto distributions \citep{klugman:etal:2004}.

\begin{corollary}\label{cor:expmix}
The posterior distribution associated with the prior $\pi(\lambda)=1/\lambda$
and with the likelihood derived from a mixture of compound exponential distributions is
proper provided proper distributions are used on the other parameters.
\end{corollary}

The parameterisation (\ref{eq:locavore}) is one of many of a Gaussian mixture.
In practice, one could design a normal and a inverse-gamma distribution as
priors for $\alpha_i$ and $\tau_i$, respectively, and make those priors vague
via the choice of hyperparameters derived from the standardised data (borrowing
the idea by \citet{PTRSI2010}). This would make the prior on local parameters a
data-based prior and consequently the marginal prior for $\mu_i$ and $\sigma_i$
would also be a data-based prior. A fundamental difficulty with this scheme is
that the constraints found in Lemma \ref{lem:const} are incompatible with this
independent modelling to each other.  We are thus seeking another
parameterisation of the mixture that allows for a natural and weakly informative
prior (hence not based on the data) while incorporating the constraints. This is
the purpose of the following sections.

\subsection{Further reparameterisations in location-scale models}\label{sec:tion}

We are now building a reparameterisation that will handle the constraints of
Lemma \ref{lem:const} in such a way as to allow for manageable uniform priors
on the resulting parameter set. Exploiting the form of the constrains, we can
rewrite the component
location and scale parameters in (\ref{eq:locavore}) as $\alpha_i=\gamma_i/\sqrt{p_i}$ and $\tau_i=\eta_i/\sqrt{p_i}\,,$
leading to the mixture representation
\begin{equation} \label{eq:rep_mix}
f(x|\btheta,\bp) = \sum_{i=1}^k p_i f_i(x|\mu+\sigma\gamma_i/\sqrt{p_i},\sigma\eta_i/\sqrt{p_i})\,,
\quad\eta_i>0\,,
\end{equation}
Given the weight vector $(p_1,\cdots,p_k)$, these new parameters are constrained by  
\begin{equation} \label{eq:constraint}
\sum_{i=1}^k \sqrt{p_i}\gamma_i=0\quad \text{and } \quad \sum_{i=1}^k (\eta_i^2+\gamma_i^2 ) =1\,,
\end{equation}
which means that $(\gamma_1,\ldots, \gamma_k,\eta_1,\ldots,\eta_k)$ belongs
both to a hyper-sphere of $\mathbb{R}^{2k}$ and to a hyperplane of this space,
hence to the intersection between both. Since the supports of $\eta_i$ and $\gamma_i$ are
bounded (i.e., $0\leq\eta_i\leq 1$ and $-1\leq\gamma_i\leq 1$), any proper
distribution with the appropriate support can be used as prior and extracting
knowledge from the data is not necessary to build vague priors as in \cite{richardson:green:1997}. 

Given these new constraints, the parameter set remains complex and the
ensuing construction of a prior still is delicate. We can however proceed towards
Dirichlet style distributions on the parameter. First, mean and variance
parameters in \eqref{eq:rep_mix} can be separated by introducing a
supplementary parameter, namely a radius $\varphi$ such that
\begin{equation} 
\sum_{i=1}^k \gamma_i^2=\varphi^2\quad \text{ and } \quad \sum_{i=1}^k \eta_i^2=1-\varphi^2\,.
\end{equation}
This decomposition of the spherical constraint then naturally leads to a
hierarchical prior where, for instance, $\varphi^2$ and $(p_1, \ldots, p_k)$ are
distributed as $\mathcal{B}e(a_1,a_2)$ and $\mathcal{D}ir(\alpha_0,
\ldots, \alpha_0)$ variates, respectively, while the vectors
$(\gamma_1,\ldots,\gamma_k)$ and $(\eta_1,\ldots,\eta_k)$ are uniformly
distributed on the spheres of radius $\varphi$ and $\sqrt{1-\varphi^2}$,
respectively, under the additional linear constraint $\sum_{i=1}^k \sqrt{p_i}\gamma_i=0$.  

Since this constraint is an hindrance for easily simulating the $\gamma_i$'s,
we now complete the move to a new parameterisation, based on spherical coordinates, 
which can be handled most straightforwardly.

\subsubsection{Spherical coordinate representation of the $\gamma$'s}\label{sub:single}

In equations (\ref{eq:constraint}), the vector $(\gamma_1,\ldots,\gamma_k)$
belongs both to the hyper-sphere of radius $\varphi$ and to the hyperplane
orthogonal to $(\sqrt{p_1},\ldots,\sqrt{p_k})$. Therefore,
$(\gamma_1,\ldots,\gamma_k)$ can be expressed in terms of 
its spherical coordinates within that hyperplane. Namely, if
$(\digamma_1,\ldots,\digamma_{k-1})$ denotes an orthonormal basis of 
the hyperplane, $(\gamma_1,\ldots,\gamma_k)$ can be written as
$$
(\gamma_1,\ldots,\gamma_k)=\varphi\cos(\varpi_1)\digamma_1+\varphi\sin(\varpi_1)\cos(\varpi_2)\digamma_2+
\ldots+\varphi\sin(\varpi_1)\cdots\sin(\varpi_{k-2})\digamma_{k-1}
$$
with the angles $\varpi_1,\ldots,\varpi_{k-3}$ in $[0,\pi]$ and $\varpi_{k-2}$ in $[0,2\pi]$. The $s$-th orthonormal
base $\digamma_s$ can be derived from the $k$-dimensional orthogonal vectors $\widetilde{\digamma}_{s}$ where 
\[ 
\widetilde{\digamma}_{1,j}=\Bigg \{ \begin{array}{c c l} -\sqrt{p_2}, &  & j=1 \\ \sqrt{p_1}, && j=2 \\ 0, && j>2
\end{array} \]
and the $s$-th vector is given by $(s>1)$
\[ \widetilde{\digamma}_{s,j}=\left \{ \begin{array}{c c l} -(p_j p_{s+1})^{1/2} \Big/
\left(\displaystyle\sum\nolimits^s_{l=1} p_l \right)^{1/2}, && j\leq s \\ \left(\displaystyle\sum\nolimits^s_{l=1} p_l
\right)^{1/2}, && j=s+1 \\
0, && j>s+1
\end{array} \right.\]

Note the special case when $k=2$ when the angle $\varpi_1$ is missing. In
this case, the mixture location parameter is then defined by
$(\gamma_1,\gamma_2)=\varphi\digamma_1$ and $\varphi$ takes both positive and
negative values. In the general case, the parameter vector
$(\gamma_1\cdots,\gamma_k)$ is a bijective transform of
$(\varphi^2,p_1,\cdots,p_k,\varpi_1,\cdots,\varpi_{k-2})$. 

This reparameterisation achieves the intended goal, since a natural reference prior for $\varpi$ is made of uniforms,
$(\varpi_1,\cdots,\varpi_{k-3}) \sim \mathcal{U}([0,\pi]^{k-3})$, and $\varpi_{k-2}\sim
\mathcal{U}[0,2\pi]$, although other choices are obviously possible and should
be explored to test the sensitivity to the prior. Prior distributions on
the other parameterisations we encountered can then be derived by a change of
variables.

\subsubsection{Dual spherical representation of the $\eta_i$'s}\label{sub:duale}

The vector of the component variance parameters $(\eta_1,\cdots,\eta_k)$ belongs to the $k$-dimension sphere of radius
$\sqrt{1-\varphi^2}$. A natural prior is a Dirichlet distribution with common hyperparameter $a$,
\[ \pi(\eta_1^2,\cdots,\eta^2_k,\varphi^2) = \mathcal{D}ir(\alpha,\cdots,\alpha) \]
For $k$ small enough, $(\eta_1,\cdots,\eta_k)$ can easily be simulated from the corresponding posterior. 
However, as $k$ increases, sampling may become more delicate by a phenomenon of concentration of the posterior 
distribution and it benefits from a similar spherical reparameterisation.
In this approach, the vector $(\eta_1,\cdots,\eta_k)$ is also rewritten through spherical coordinates with angle components 
$(\xi_1,\cdots, \xi_{k-1})$,
\[ \eta_i = \left\{ \begin{array}{l c}  \sqrt{1-\varphi^2}\cos(\xi_i)\, , & i=1 \\  
\sqrt{1-\varphi^2}\displaystyle\prod^{i-1}_{j=1} \sin(\xi_j)\cos(\xi_i)\, , & 1<i < k  \\ 
\sqrt{1-\varphi^2}\displaystyle\prod^{i-1}_{j=1} \sin(\xi_j)\, , & i=k
\end{array}  \right . \]
Unlike $\varpi$, the support for all angles $\xi_1,\cdots,\xi_{k-1}$ is limited to $[0,\pi/2]$, due to the positivity
requirement on the $\eta_i$'s. In this case, a reference prior on the angles is
$
(\xi_1,\cdots,\xi_{k-1})\sim\mathcal{U}([0,\pi/2]^{k-1})\,,
$
while obviously alternative choices are possible.

\subsection{Weakly informative priors for Gaussian mixture models}\label{sec:weak}

The above introduced two new parameterisations of a Gaussian mixture model,
based on equation (\ref{eq:rep_mix}) and its spherical representations in
Sections \ref{sub:single} and \ref{sub:duale}. Under the constraints imposed by the first two moment
-parameters, two weakly informative priors, called the (i) {\it single} and (ii)
{\it double uniform} priors, are considered below and their impact on the resulting marginal
prior distributions is studied. 

For the parameterisation constructed in Section \ref{sub:single}, 
$\gamma$ is expressed in terms of its spherical coordinates
over the $k-1$ subset, uniformly distributed over $[0,\pi]^{k-3}\times[0,2\pi]$,
and $\eta$ is associated with a uniform distribution over the $\mathbb{R}^k$
simplex, conditional on $\varphi$. This prior modelling is called the {\it single uniform} prior.
Another reparametrisation is based on the spherical representations of both $\gamma$ and $\eta$
leading to the {\it double uniform} prior, a uniform distribution on all angle
parameters, $\varpi_i$'s and $\eta_i$'s. Both priors can be argued to be weakly informative
priors, relying on uniforms for a given parameterisation. 

\begin{figure}[!h] 
\centering\setlength{\unitlength}{1cm}
\begin{picture}(10,7.6)
\put(-1,4){\includegraphics[width=6cm,height=4cm]{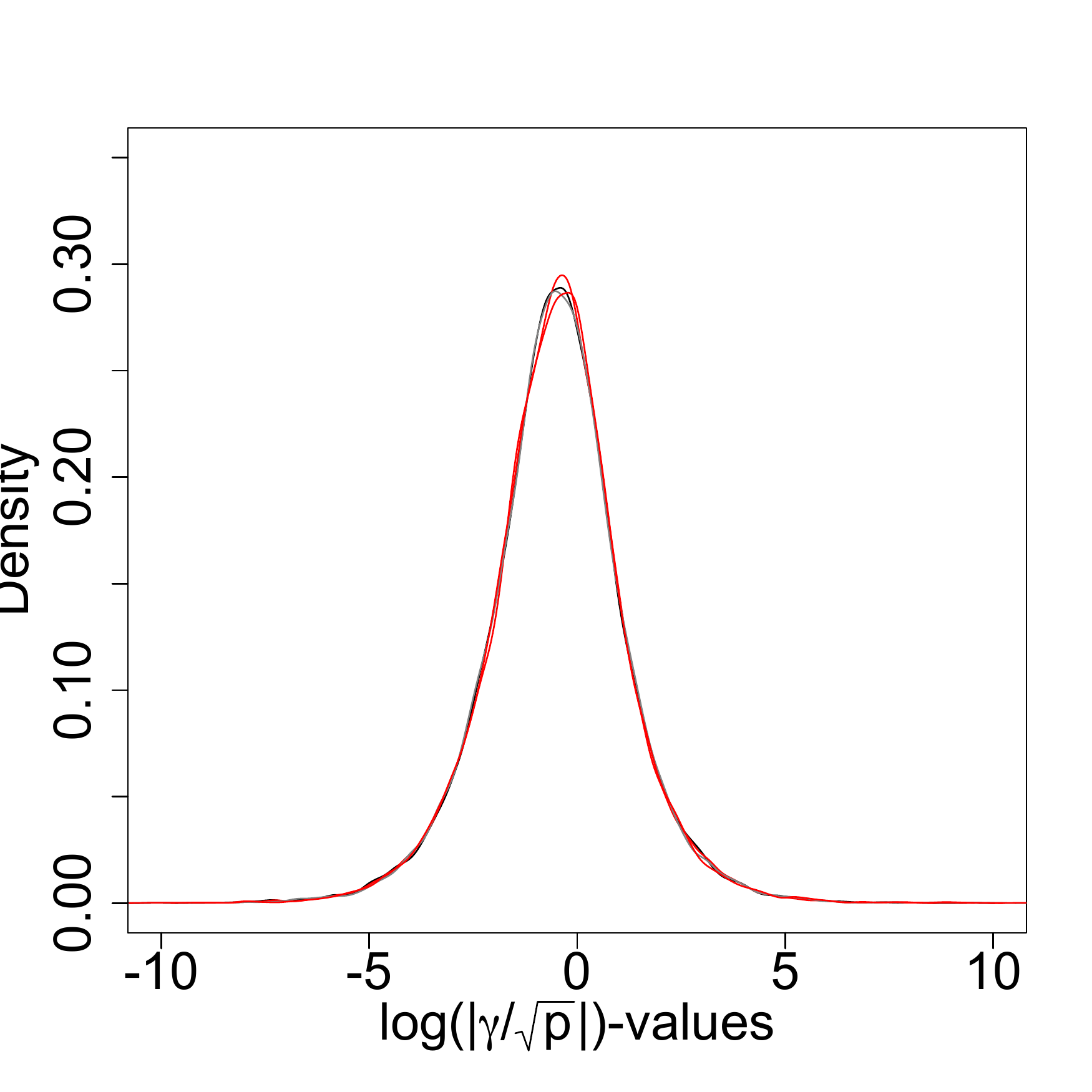}}
\put(5,4){\includegraphics[width=6cm,height=4cm]{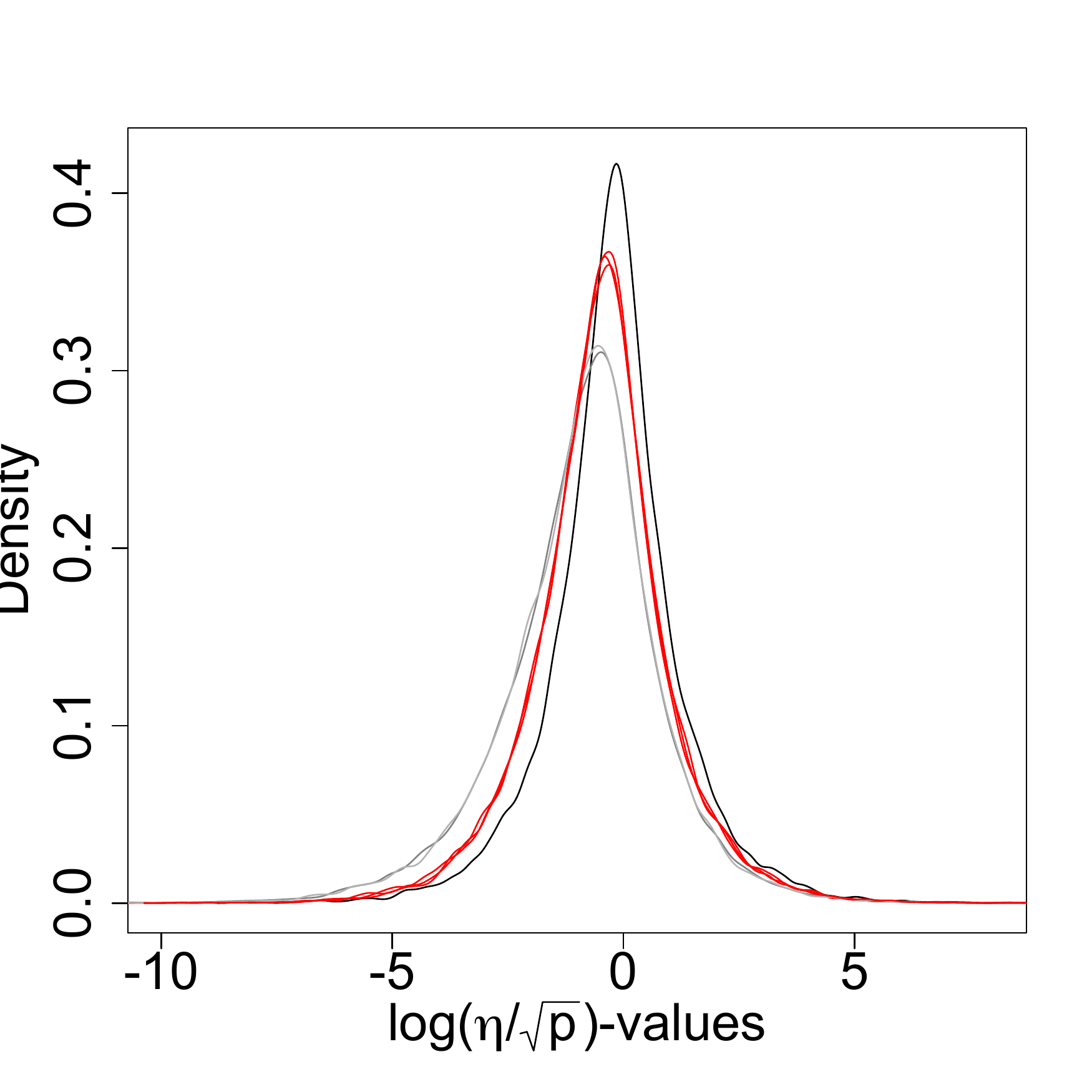}} 
\put(-1,0){\includegraphics[width=6cm,height=4cm]{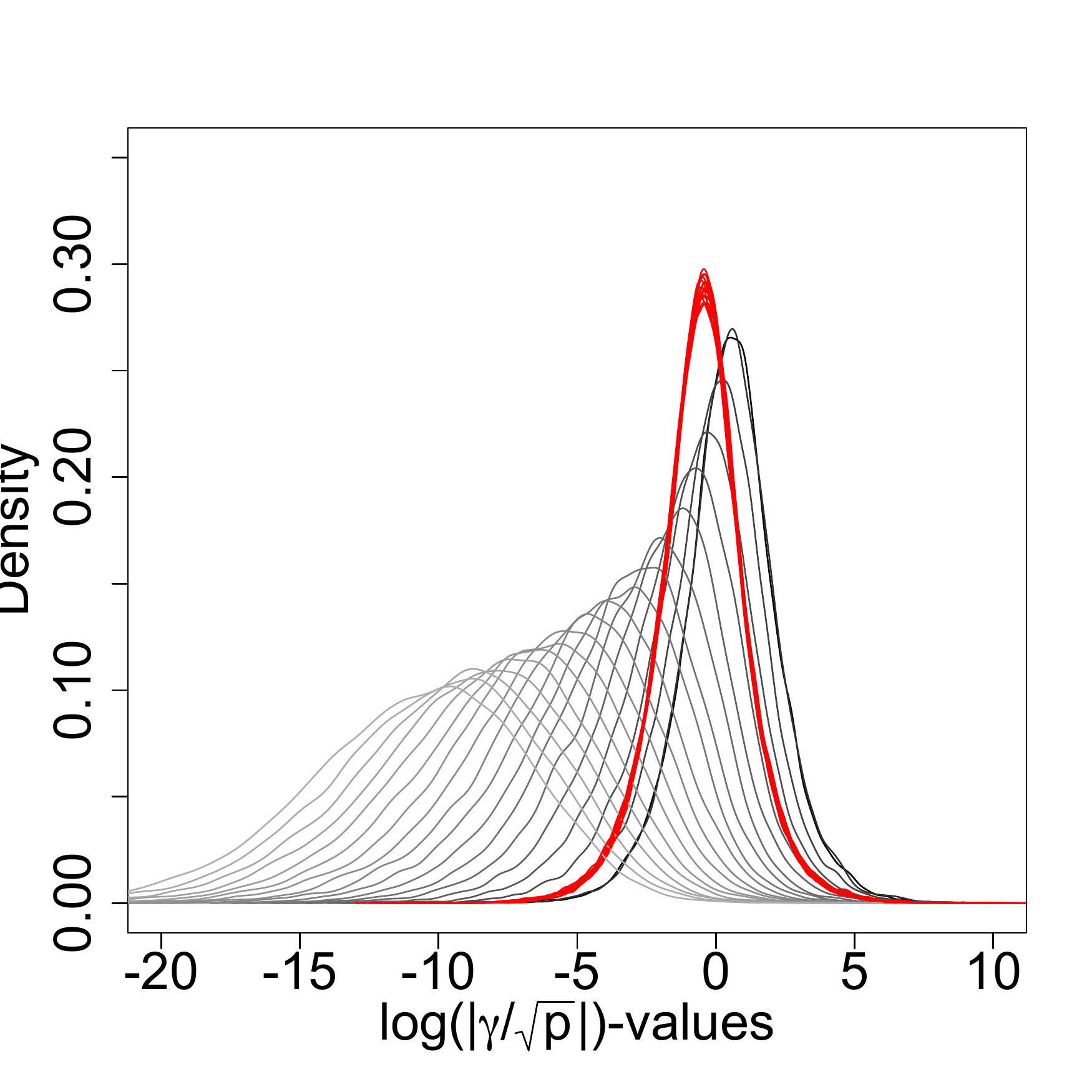}}
\put(5,0){\includegraphics[width=6cm,height=4cm]{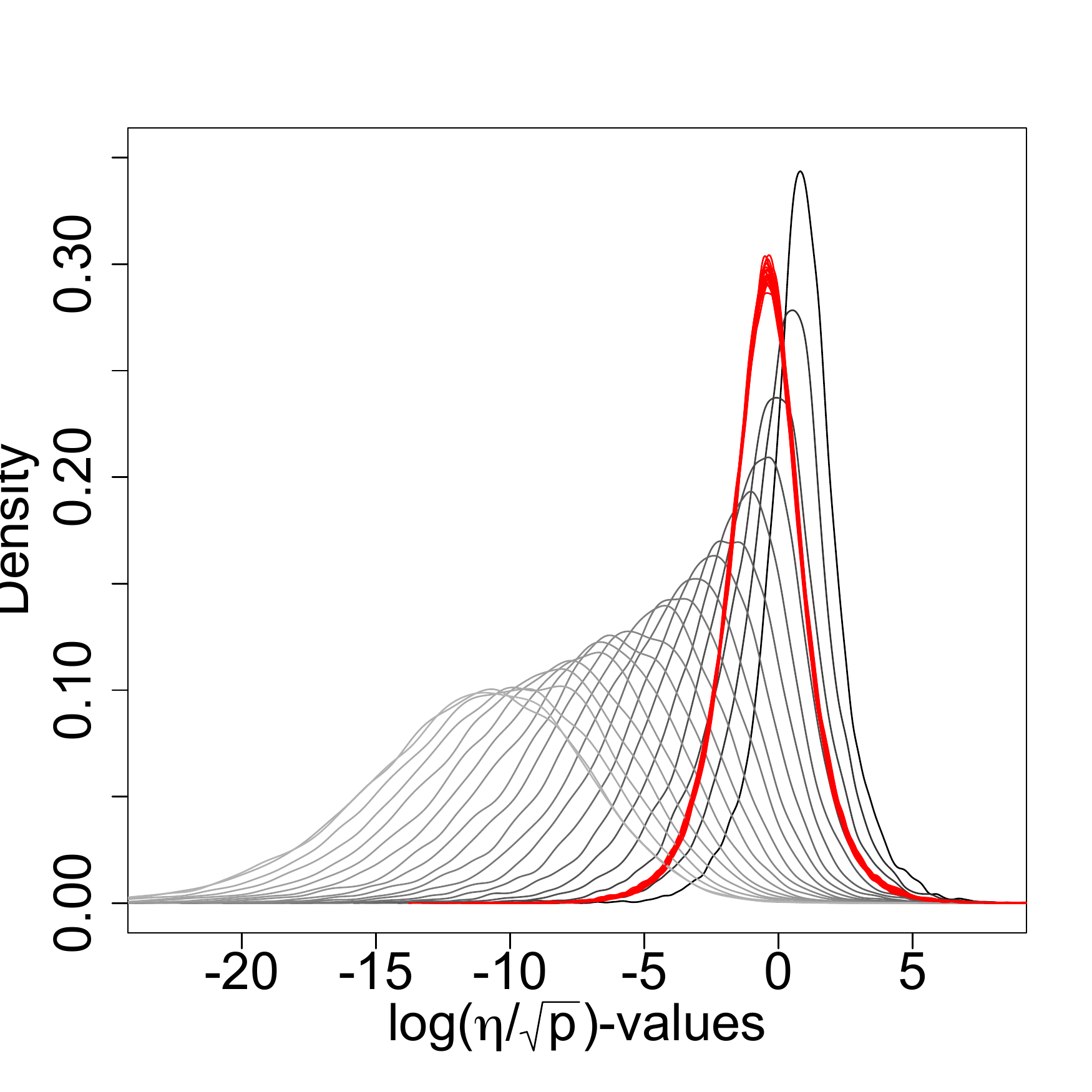}} 
\end{picture} \caption{Density estimate of 20,000 draws of $ \log(|\gamma_{i}/\sqrt{p_i}|)$ and
$\log(\eta_{i}/\sqrt{p_i})$ from the {\it single uniform prior} (red lines) and the {\it double uniform prior} (grey lines)
when $k=3$ {\em (top)} and $k=20$ {\em (bottom)}. Different grey lines indicate
the density estimates for $i=1,\ldots,k$. }\label{prior}
\end{figure}

\begin{figure}[H]
\centering\setlength{\unitlength}{1cm}
\begin{picture}(14,7.6)
\put(-1,4){\includegraphics[width=5cm,height=4cm]{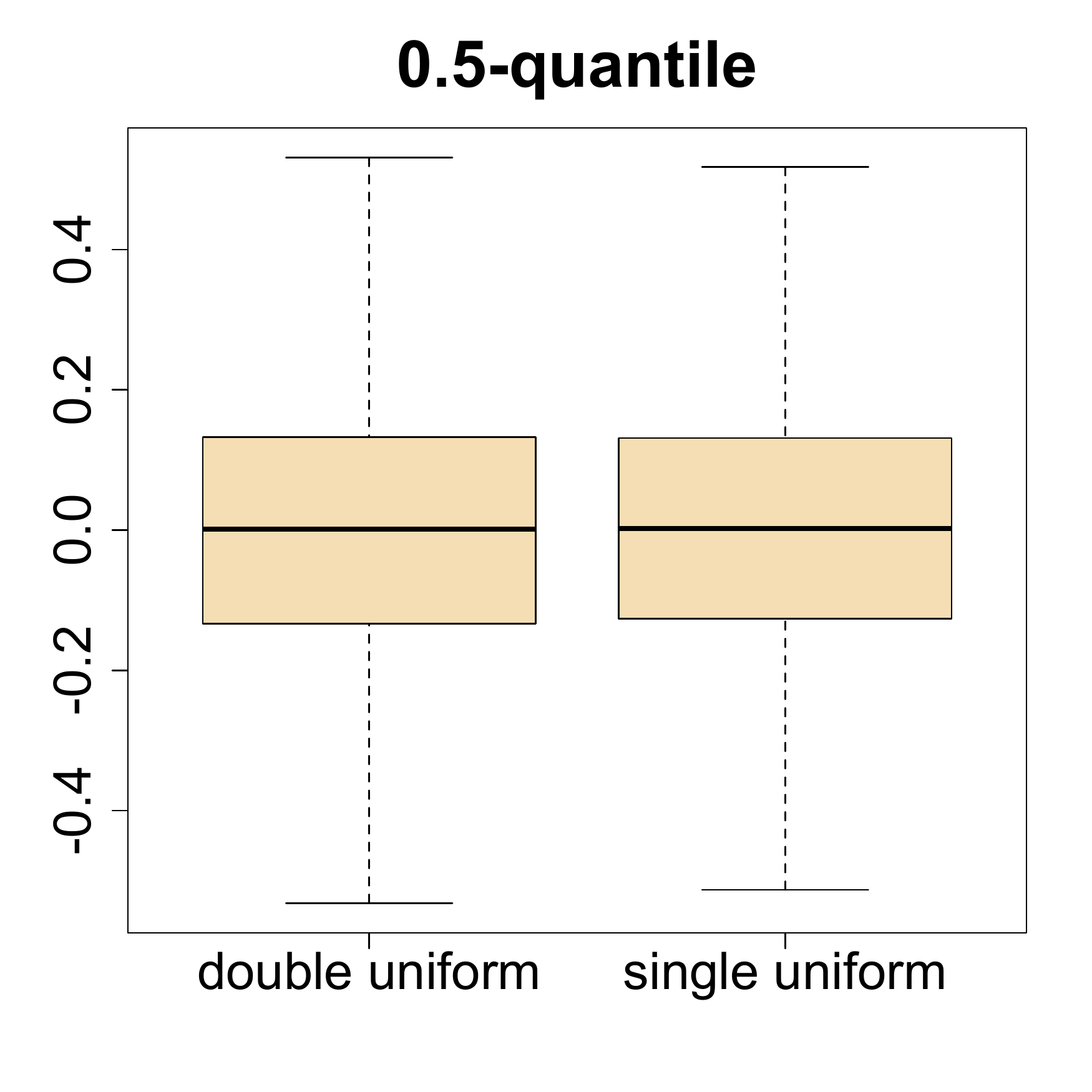}}
\put(4,4){\includegraphics[width=5cm,height=4cm]{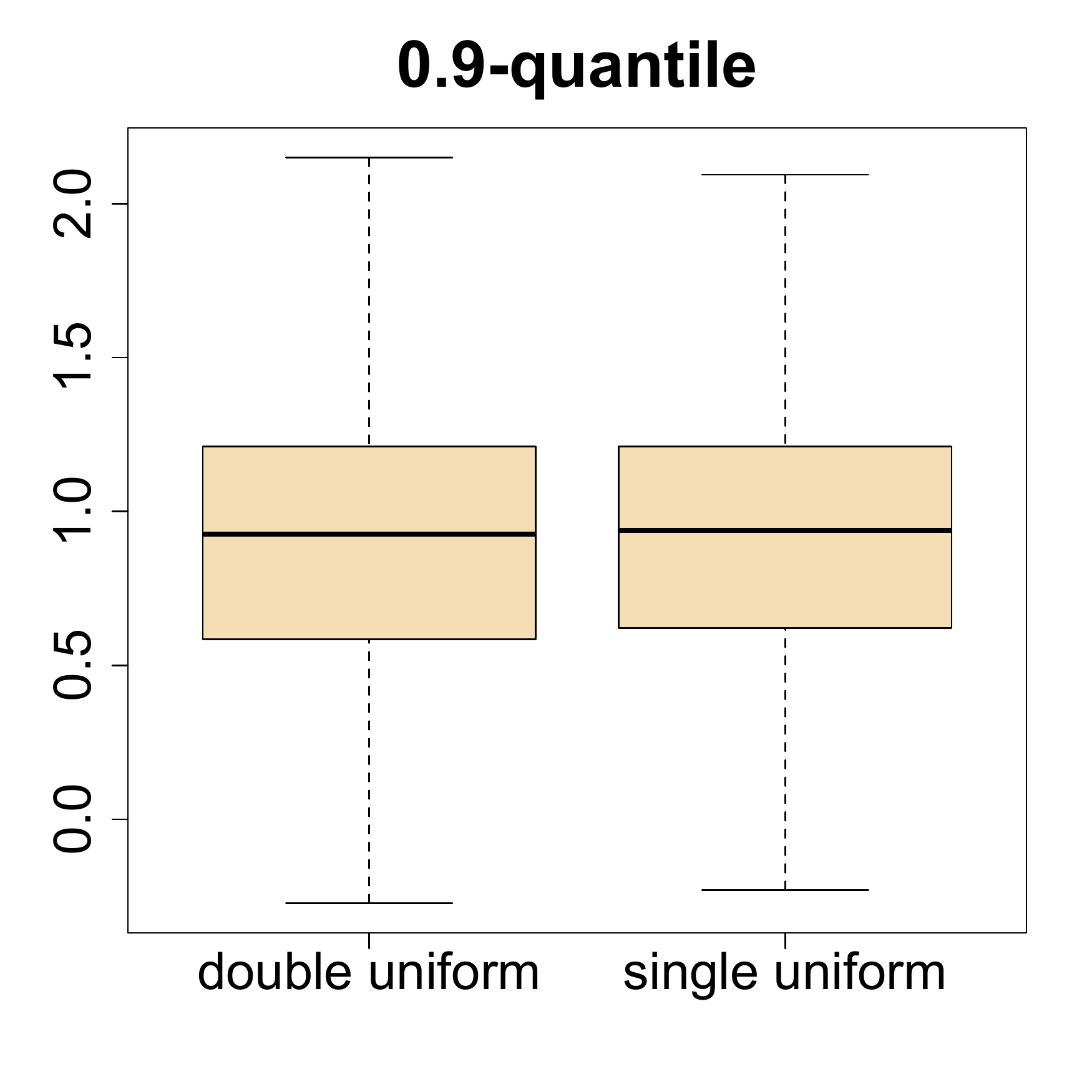}}
\put(9,4){\includegraphics[width=5cm,height=4cm]{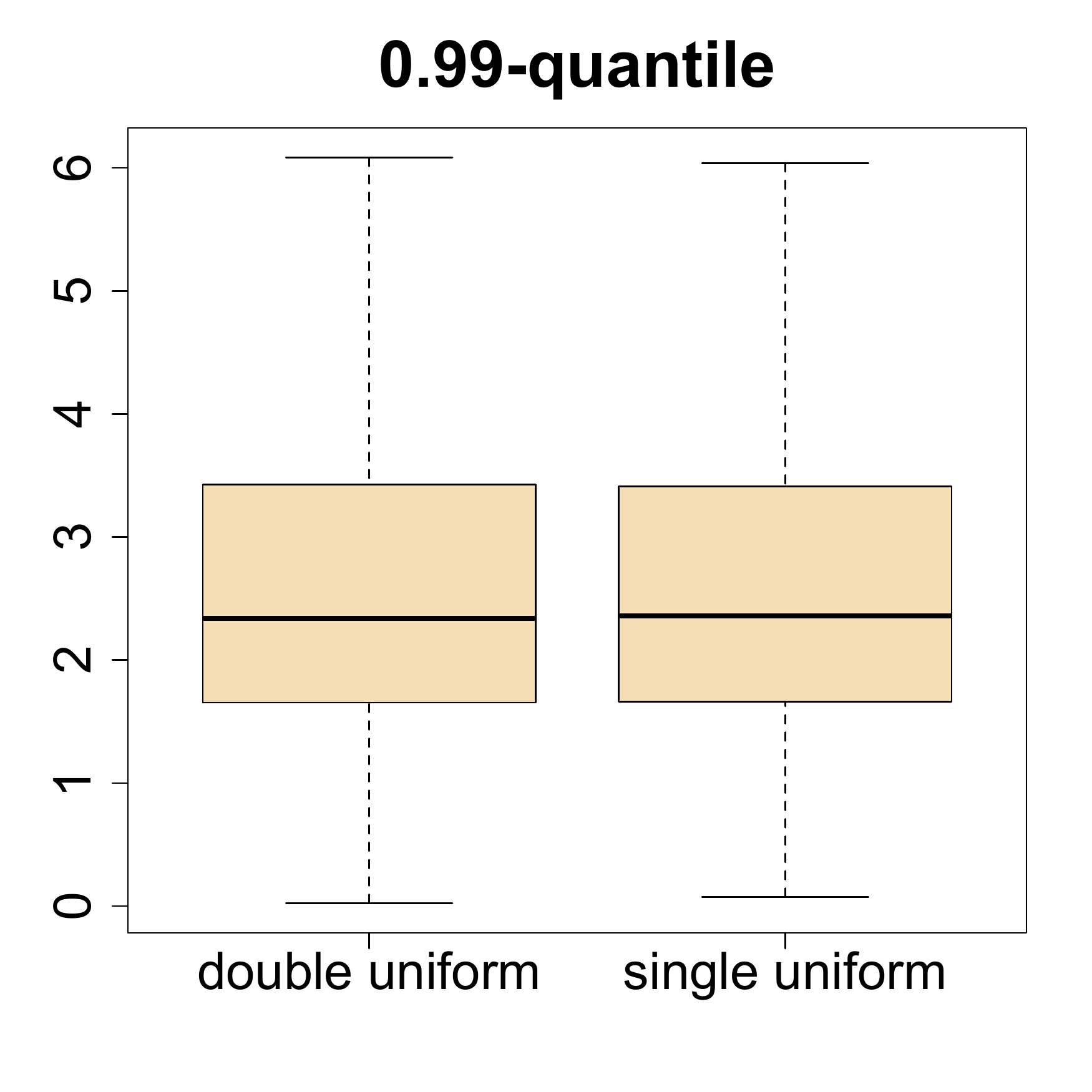}}
\put(-1,0){\includegraphics[width=5cm,height=4cm]{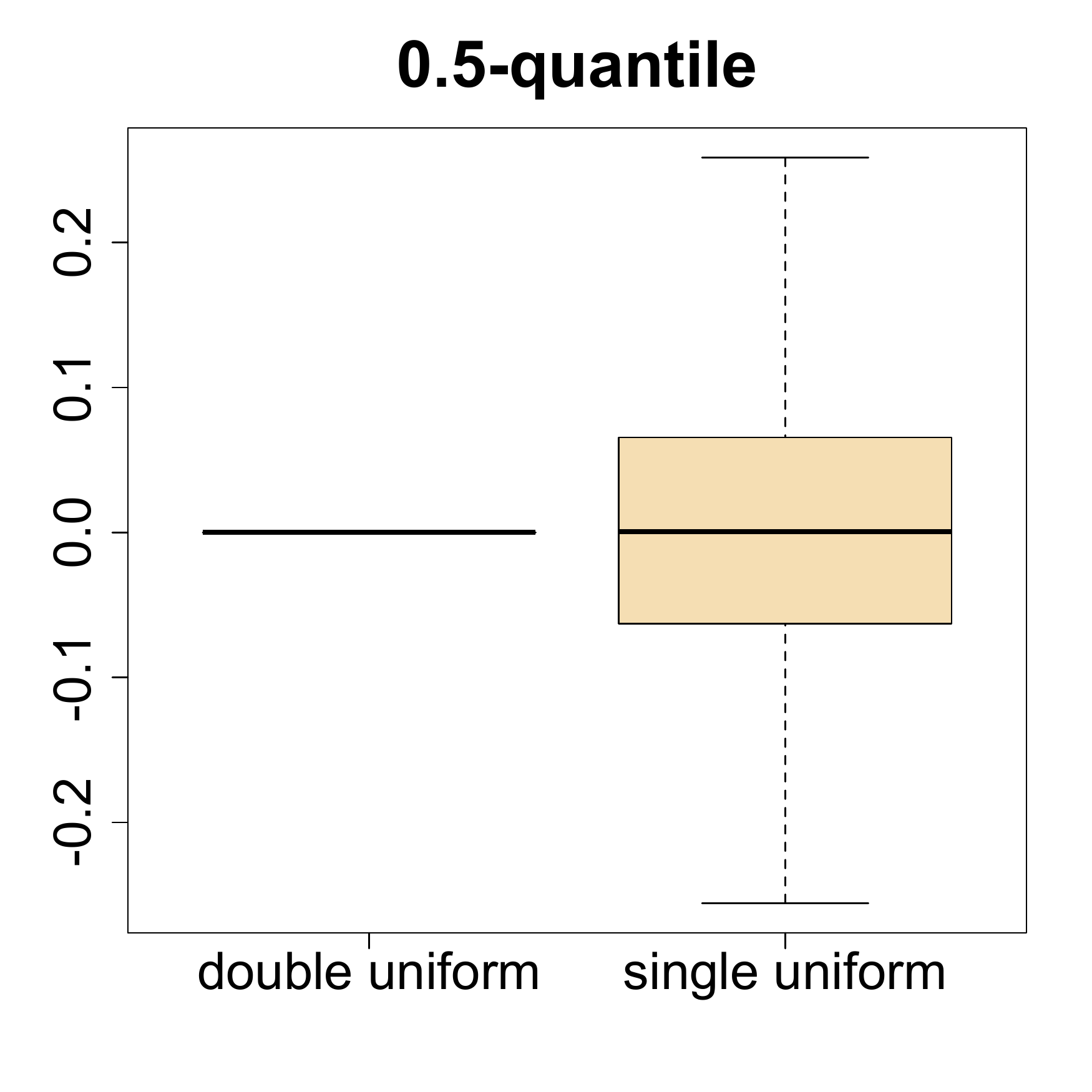}}
\put(4,0){\includegraphics[width=5cm,height=4cm]{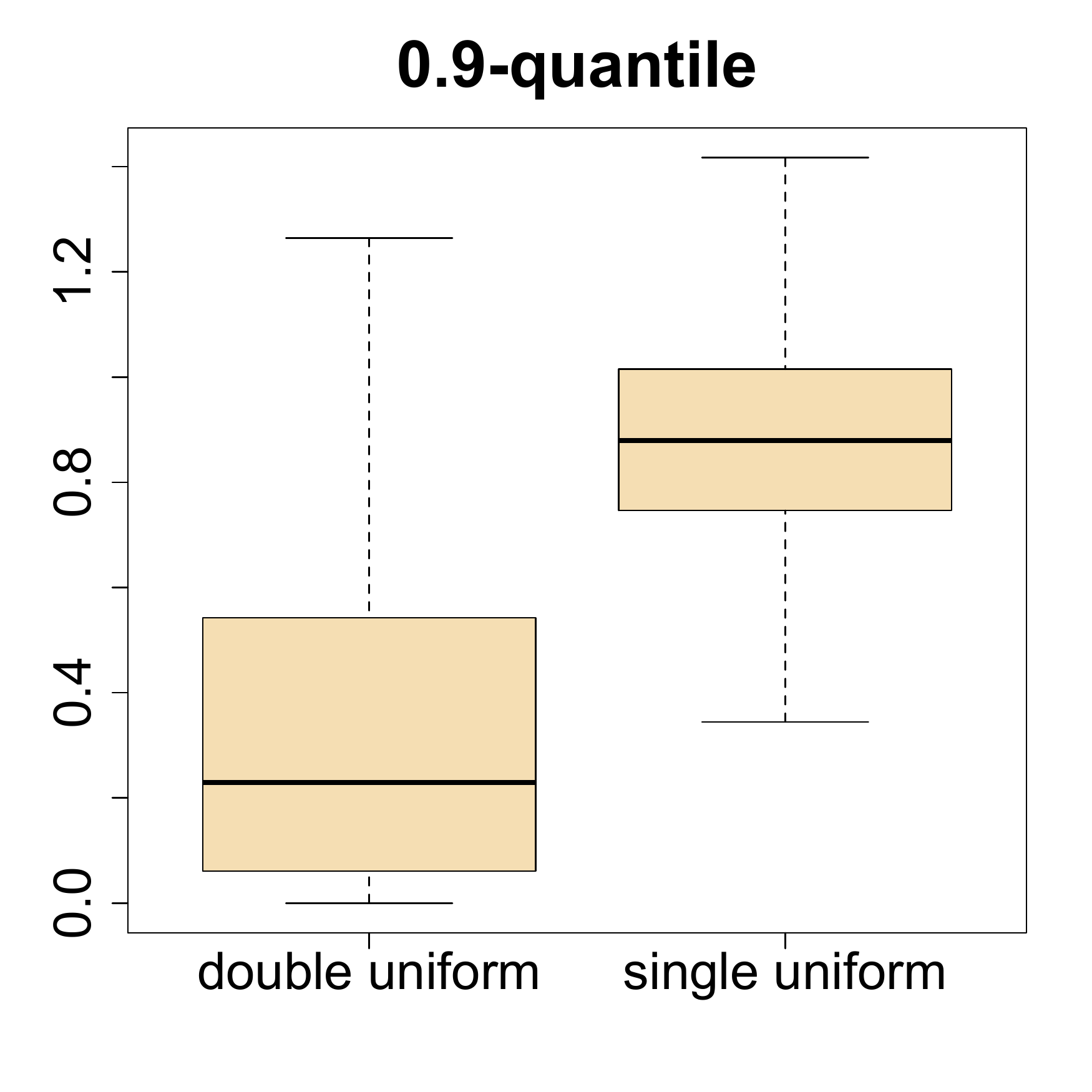}}
\put(9,0){\includegraphics[width=5cm,height=4cm]{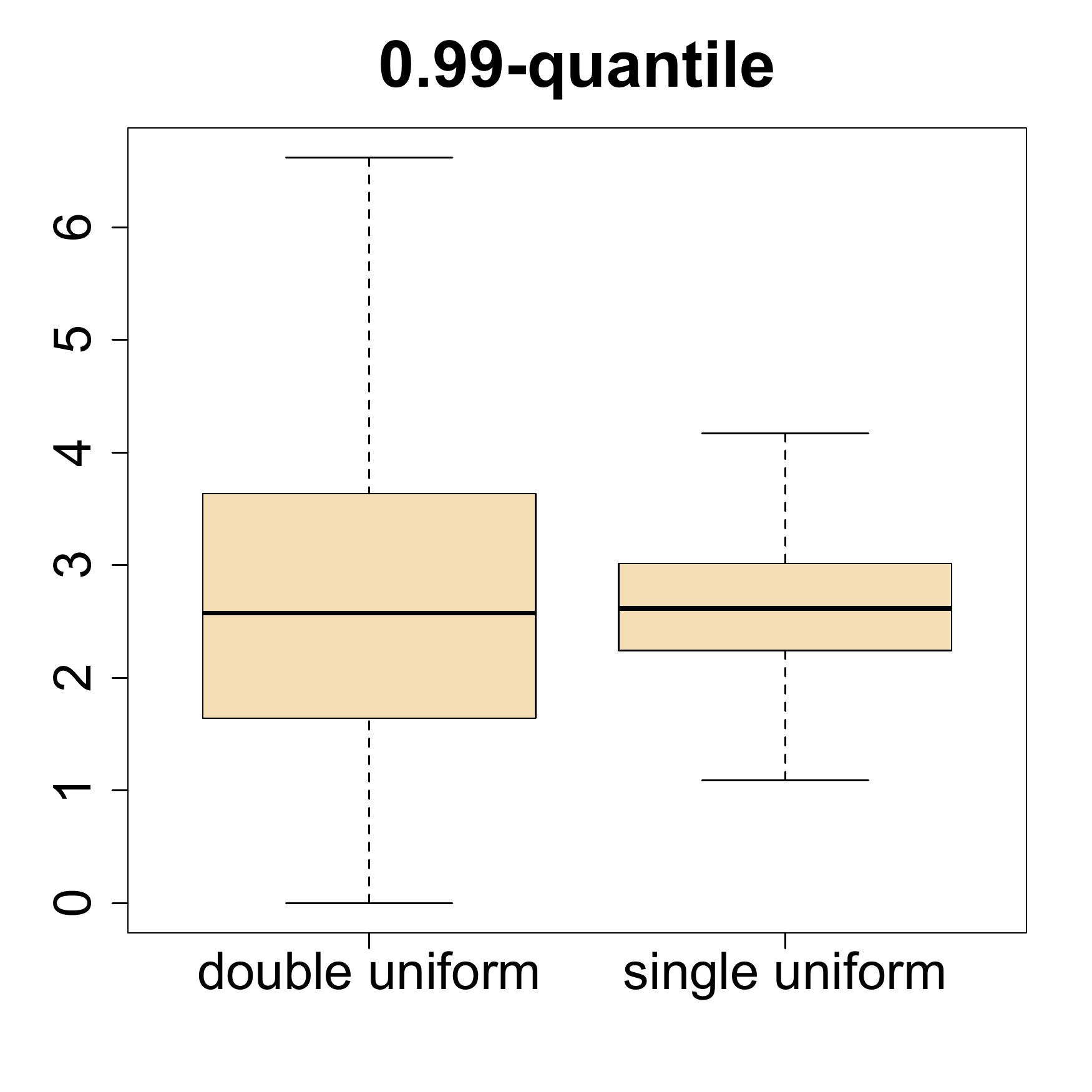}}
\end{picture} \caption{Boxplot of quantiles of mixture models with 20,000
parameter values from the {\it single} and {\it double uniform priors} when
$k=3$ {\em (top)} and $k=20$ {\em (bottom)}. The mean of the mixture is $0$ and its
variance is $1$.}\label{prior_quantile}
\end{figure}

To evaluate the difference between both modellings, using a uniform prior on
$\varphi^2$ and a Dirichlet distribution on the $p_i$'s, we generated $20,000$
samples from both priors. The resulting
component-wise parameter distributions are represented in Figure \ref{prior}.
As expected, under the {\it single uniform} prior, all $\eta_i$'s and
$\gamma_i$'s are uniformly distributed over the $k$-ball, are thus exchangeable
and, as a result, all density estimates are close. When using the {\it double uniform} priors, the components
are ordered through their spherical representation. As $k$ increases, the
ordering becomes stronger over a wider range, when compared with the first
uniform prior and the component-wise priors become more skewed away from the
global mean parameter value. This behaviour is reflected in the 
distribution of some quantiles of the mixture model, as seen in Figure
\ref{prior_quantile}.  When $k=3$, 
the supports of the mixture models are quite similar for both priors. When $k=20$,
the double uniform prior tends to put more mass around the global mean of
$0$, as shown by the median distribution, and it allows for mixtures with long
tails more readily than the single uniform prior. Therefore, the {\it double uniform
prior} is our natural choice for simulation studies in Section 4. 

\section{MCMC implementations}\label{sec:mcmc}

\subsection{A Metropolis-within-Gibbs sampler in the Gaussian case}\label{sec:debaZ}

Given the reparameterisations introduced in Section \ref{sec:miX}, and in
particular Section \ref{sec:tion} for the Gaussian mixture model, different
MCMC implementations are possible and we investigate in this section some of
these. To this effect, we distinguish between the single and double uniform priors.

Although the target density is not too dissimilar to the target explored by early Gibbs
samplers in \cite{diebolt:robert:1990a} and \cite{gelman:king:1990}, simulating
directly the new parameters implies managing constrained parameter spaces. The
hierarchical nature of the parameterisation also leads us to consider a block
Gibbs sampler that coincides with this hierarchy. Since the corresponding full
conditional posteriors are not in closed form, a Metropolis-within-Gibbs
sampler is implemented here with random walk proposals. In this approach, the
scales of the proposal distributions are automatically calibrated towards
optimal acceptance rates \citep{GORAGWG1997,GRSR2001,GRSR2009}.
Convergence of a simulated chain is assessed based on the rudimentary
convergence monitoring technique of \cite{GLMRUB1992}. The description of the
algorithm is provided by a pseudo-code representation in the Supplementary
Material (Figure \ref{fig:algoboX}).
Note that the Metropolis-within-Gibbs version does not rely on latent variables
and completed likelihood as in \cite{tanner:wong:1987} and
\cite{diebolt:robert:1990a}. Following the adaptive MCMC method in Section 3 of
\citet{GRSR2009}, we derive the optimal scales associated with proposal
densities, based on 10 batches with size 50. The scales $\epsilon$ are
identified by a subscript with the corresponding parameter (see, e.g., Table \ref{tab1}).

For the single reparameterisation, all steps in Figure \ref{fig:algoboX} are the same except that Steps 2.5 and 2.7 are ignored. When $k$ is not large, one potential proposal density for $((\varphi^2)^{(t)},(\eta_1^2)^{(t)},\ldots,(\eta_k^2)^{(t)})$ is a Dirichlet distribution, 
\[ 
((\varphi^2)',(\eta_1^2)',\ldots,(\eta_k^2)')\sim
\text{Dir}((\varphi^2)^{(t-1)}\epsilon,(\eta_1^2)^{(t-1)}\epsilon,\ldots,(\eta_k^2)^{(t-1)}\epsilon) \, .
\] 
Alternative proposal densities will be discussed along simulation studies in Section \ref{sim_study}.

\subsection{A Metropolis--Hastings algorithm for Poisson mixtures}\label{sec:mcmcois}

Since the full conditional posteriors corresponding to the Poisson mixture \eqref{eq:poimix} are not in closed form
under the new parameterisation,  these parameters can again be simulated by implementing a Metropolis-within-Gibbs
sampler. Following an adaptive MCMC approach, the scales of the proposal distributions are automatically calibrated
towards optimal acceptance rates \citep{gelman:gilks:roberts:1996}. The description of the algorithm is provided in
details by a pseudo-code in the Supplementary Material (Figure
\ref{fig:MCMCpois}). Note that the Metropolis-within-Gibbs version relies on completed likelihoods.

\subsection{Removing and detecting label switching}\label{sec:witch}

The standard parameterisation of mixture models contains weights
$\{p_i\}^k_{i=1}$ and component-wise parameters $\{\theta_i\}^k_{i=1}$ as shown
in \eqref{eq:mix}. The likelihood function is invariant under permutations of
the component indices. If an exchangeable prior is chosen on weights and
component-wise parameters, which is the case for some of our priors, the
posterior density is also invariant under permutations and the component-wise parameters are
not identifiable. This phenomenon, called {\it label switching}, is
well-studied in the literature
\citep{celeux:hurn:robert:2000,stephens:2000b,fruhwirth:2001,jasra:holmes:stephens:2005}.
The posterior distribution involves $k!$ symmetric global modes and
a Markov chain targetting this posterior is expected to explore all of
them. However, MCMC chains often fail to achieve this feature
\citep{celeux:hurn:robert:2000} and usually end up exploring one single
global mode of the target.

In our reparameterisations of mixture models of Sections \ref{sub:single} and
\ref{sub:duale}, each $\theta_i$ is a function of a novel
component-wise parameter from a simplex, conditional on the global parameter(s)
and the weights. The mapping between both parameterisations is a one-to-one map
conditional on the weights. In other words, there is a unique value for
$\theta_i$ given a particular set of values on this simplex and the weights.
Depending on the reparameterisation and the choice of the prior distribution,
the parameters on a simplex can be exchangeable (as, e.g., in a Poisson mixture) and with
the use of a uniform prior, label switching is expected to occur. When using
the double spherical representation in Section \ref{sub:duale}, the parameterisation
is not exchangeable, due to the choice of the orthogonal basis. However,
adopting an exchangeable prior on the weights (e.g., a Dirichlet distribution
with a common parameter) and uniform priors on all angular parameters leads to
an exchangeable posterior on the standard parameters of the mixture.  Therefore,
label switching should also occur with this prior modelling.

When an MCMC chain manages to jump between modes, the inference on each of the
mixture components becomes harder \citep{celeux:hurn:robert:2000,geweke:2007}.
To get component-specific inference and to give a meaning to each component,
various relabelling methods have been proposed in the literature (see, e.g.,
\citealp{fruhwirth:2006}). A first available alternative is to reorder labels
so that the mixture weights are in increasing order \citep{fruhwirth:2001}. A
second method proposed by, e.g., \citet{lee:mari:meng:robe:2008} is that labels
are reordered towards producing the shortest distance between the current
posterior sample and the (or a) maximum posterior probability (MAP) estimate. 

Note that the second method depends on the parameterisation of the model since
both MAP and distance vary with this parameterisation.
For instance, for the spherical representation of a Gaussian mixture model, the closeness of the $\gamma_i$'s
to the MAP cannot be determined via distance measures on $\varpi_i$'s, due to the
symmetric features of trigonometric functions. For such cases, we recommend to
transform the MCMC sample back to the standard parameterisation, then apply a
relabelling method on the standard parameters. (This step has no significant
impact on the overall computing time.)

Let us denote by $\mathfrak{S}_k$ the set of permutations on $\{1,\ldots,k\}$.
Then, given an MCMC sample for the new parameters, the second relabelling
technique is implemented as follows:

\begin{enumerate}
\item Reparameterise the MCMC sample to the standard parameterisation, $\{\btheta^{(t)},\bp^{(t)}\}_{t=1}^T$. 
\item Find the MAP estimate $(\btheta^*,\bp^*)$ by computing the posterior values of the sample.
\item For each $t$, reorder $(\btheta^{(t)},\bp^{(t)})$ as 
$(\widetilde{\btheta}^{(t)},\widetilde{\bp}^{(t)})= \delta^o(\btheta^{(t)},\bp^{(t)}) $  where $\delta^o=\argmin_{\delta\in\mathfrak{S}_k}  \| \delta(\btheta^{(t)},\bp^{(t)})-(\btheta^*,\bp^*) \| $.
\end{enumerate}

The resulting permutation at iteration $t$ is denoted by $r^{(t)}\in
\mathfrak{S}_k$. Label switching occurrences in an MCMC sequence can be
monitored via the changes in the sequence $r^{(1)},\ldots,r^{(T)}$. If the MCMC
chain fails to switch modes, the sequence is likely to remain at the same
permutation. On the opposite, if a MCMC chain moves between some of the $k!$
symmetric posterior modes, the $r^{(t)}$'s are expected to vary.

While the relabelling process forces one to label each posterior sample by its
distance from the MAP estimate, there exists an easier alternative to produce
estimates for component-wise parameters. This approach is achieved by $k$-mean
clustering on the population of all $\{\btheta^{(t)}_k,\bp^{(t)}\}_{t=1}^T$.
When using the Euclidean distance as in the MAP recentering,  which is the point process
representation adopted in \cite{stephens:2000}, clustering can be seen as a
natural solution without the cost of relabelling an MCMC sample. When posterior
modes are well separated, component-wise estimates from relabelling and
from k-mean clustering are expected to be similar. In the event of poor
switching, as exhibited for instance in some of our experiments,
a parallel tempering resolution can be easily added to the analysis, as
detailed in an earlier version of this work \citep{kamary:lee:robert:2016}.

\section{Simulation studies for Gaussian and Poisson mixtures}\label{sim_study}

In this section, we examine the performances of the above
Metropolis-within-Gibbs method, applied to both reparameterisations defined in
Section \ref{sec:tion}, for both simulated and real datasets. 

\subsection{The Gaussian case $k=2$}\label{sub:k=2}

In this specific case, there is no angle to consider. Two straightforward
proposals are compared over simulation experiments. One is based on Beta and
Dirichlet proposals:
$$
p^*\sim\mathcal{B}eta(p^{(t)}\epsilon_p, (1-p^{(t)})\epsilon_p)\,,\quad
({\varphi^2}^*,{\eta_1^2}^*,{\eta_2^2}^*)\sim
\mathcal{D}ir({\varphi^2}^{(t)}\epsilon,{\eta_1^2}^{(t)}\epsilon,{\eta_2^2}^{(t)}\epsilon)
$$
(this will be called Proposal 1) and another one is based on Gaussian random walks proposals (Proposal 2):
\begin{align*}
\log(p^*/(1-p^*))&\sim \mathcal{N}(\log(p^{(t)}/(1-p^{(t)})), \epsilon_p)\\
(\vartheta_1^*, \vartheta_2^*)^T&\sim \mathcal{N}(\chi_2^{(t)}, \epsilon_{\vartheta}I_2)\quad\text{with}\\
({\varphi^2}^*, {\eta_1^2}^*, {\eta_2^2}^*)&=({\exp(\vartheta_1^*)}/{\bar{\vartheta}^*},
{\exp(\vartheta_2^*)}/{\bar{\vartheta}^*}, {1}/{\bar{\vartheta}^*})\,,\\
  \chi_2^{(t)}&= (\log({\varphi^2}^{(t)}/{\eta_2^2}^{(t)}), \log({\eta_1^2}^{(t)}/{\eta_2^2}^{(t)}))\\
\text{and}\quad\bar{\vartheta}^*&=1+\exp(\vartheta_1^*)+\exp(\vartheta_2^*)\,.
 \end{align*}
The global parameters are proposed using Normal and Inverse-Gamma moves
$\mu^*\sim \mathcal{N}(\bar{x},\epsilon_{\mu})$ and ${\sigma^2}^*\sim \mathcal{IG}((n+1)/2,(n-1)\bar{\sigma}^2/2)$,
where $\bar{x}$ and $\bar{\sigma}^2$ are sample mean and variance respectively.
We present below some analyses and also explain how MCMC methods can be used to
fit mixture distributions. 
 
\begin{example}\label{ex:simulated_data}
In this experiment, a dataset of size $50$ is simulated from the mixture $0.65
\mathcal{N}(-8, 2)+0.35 \mathcal{N}(-0.5, 1)$, which implies that the true
value of $(\varphi^2, \eta_1, \eta_2)$ is $(0.813, 0.149, 0.406)$.  

First, ten chains were simulated with Proposal 1 and  different starting values. 
As can be seen in Figure \ref{fig1}, the estimated densities are almost
indistinguishable among the different chains and highest-posterior regions all include
the true values. The chains visited all posterior modes. The inference results about
parameters using Proposals 1 and 2 are compared in Figure \ref{fig2}. The true
values are identified by the empirical posterior distributions using both
proposals. We further note that the chain derived by Proposal 1 produces more
symmetric posteriors, in particular for $p, \varphi, \eta_1, \eta_2$. This
suggests that the chain achieves a better mixing behaviour.

The scales $\epsilon$ for Proposals 1 and 2 are determined by aiming at the
optimal acceptance rate of \cite{GORAGWG1997}, taken to be $0.44$ for small
dimensions of the parameter space. As shown in Table \ref{tab1}, an
adaptive Metropolis-within-Gibbs strategy manages to recover acceptance rates
close to optimal values. A second example in Section \ref{sec:OleFaz},
Supplementary Material, illustrates how this method using Proposal 1 behaves for
a dataset with a slightly larger sample size and unlike Figure \ref{fig2}
the chain fails to move between posterior modes.

\begin{figure}[!h]
\includegraphics[width=.33\textwidth,height=4cm]{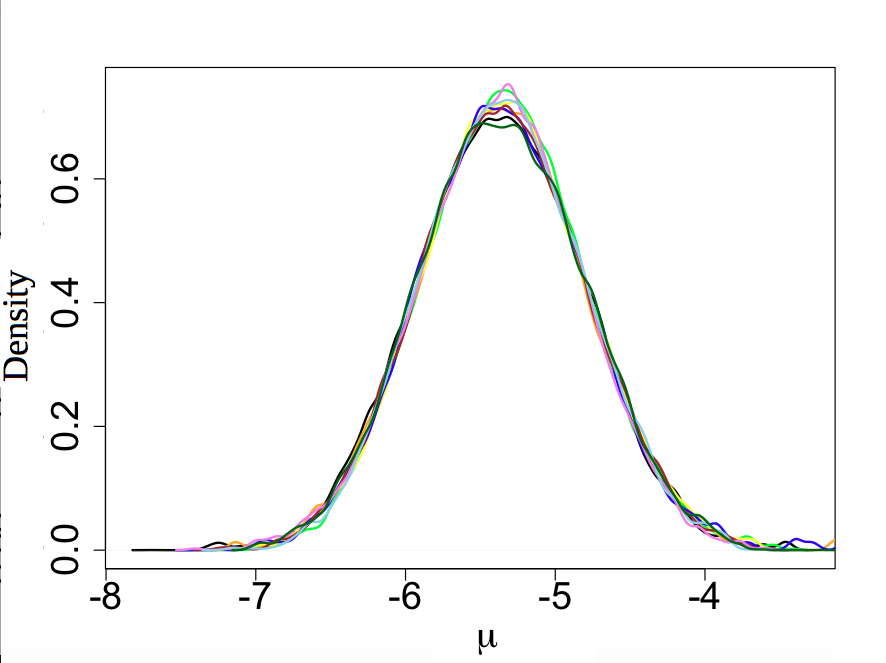}\includegraphics[width=.33\textwidth,height=4cm]{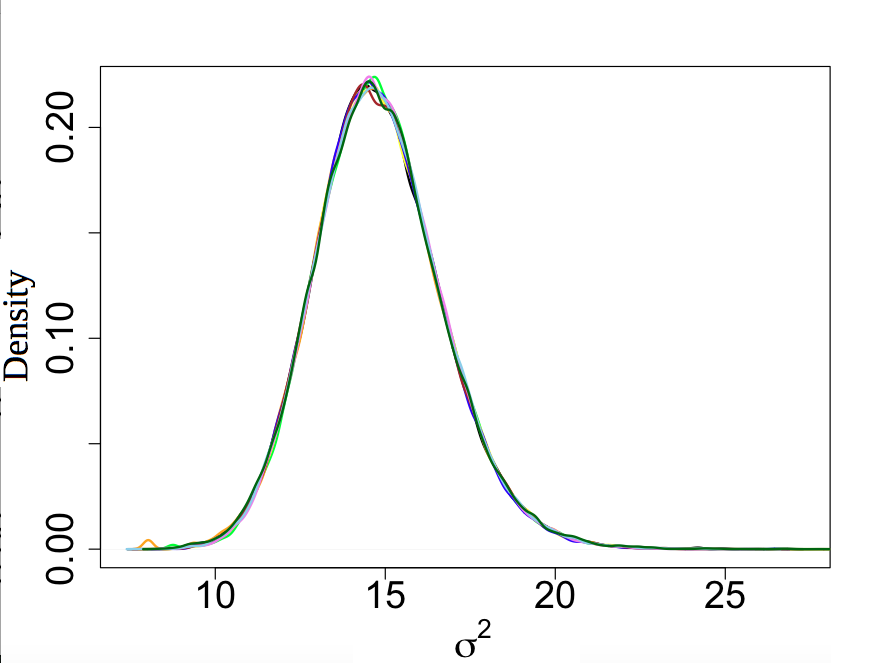}\includegraphics[width=.33\textwidth,height=4cm]{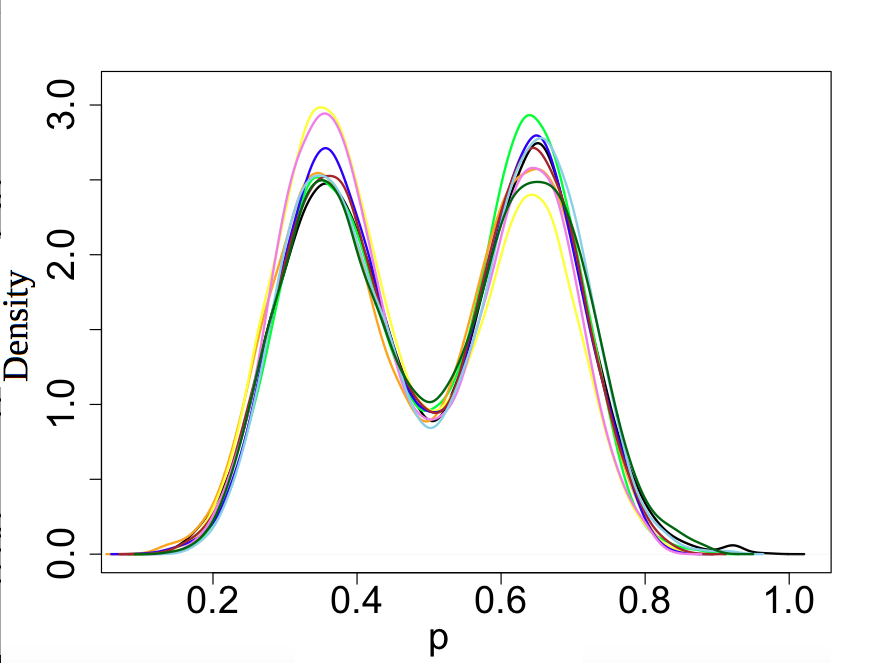}
\includegraphics[width=.33\textwidth,height=4cm]{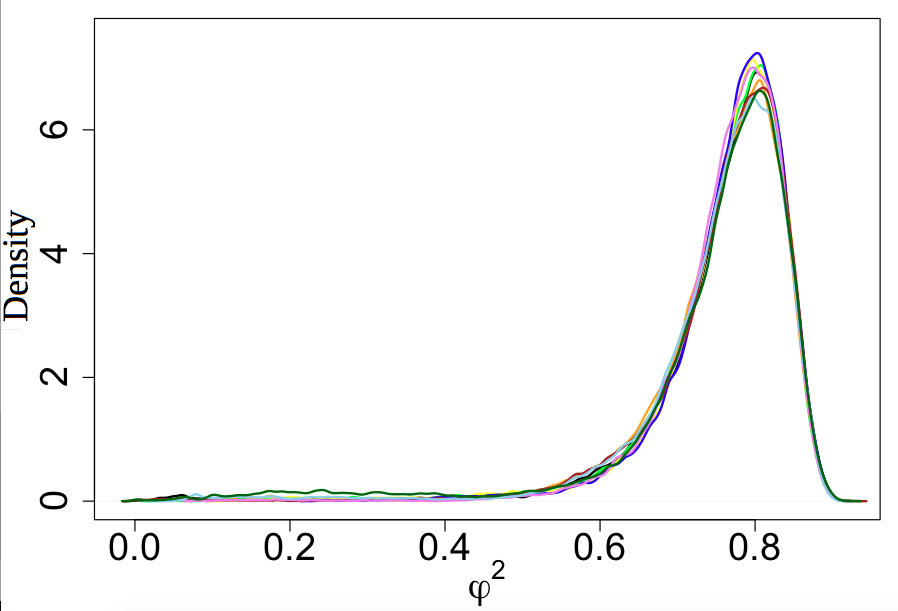}\includegraphics[width=.33\textwidth,height=4cm]{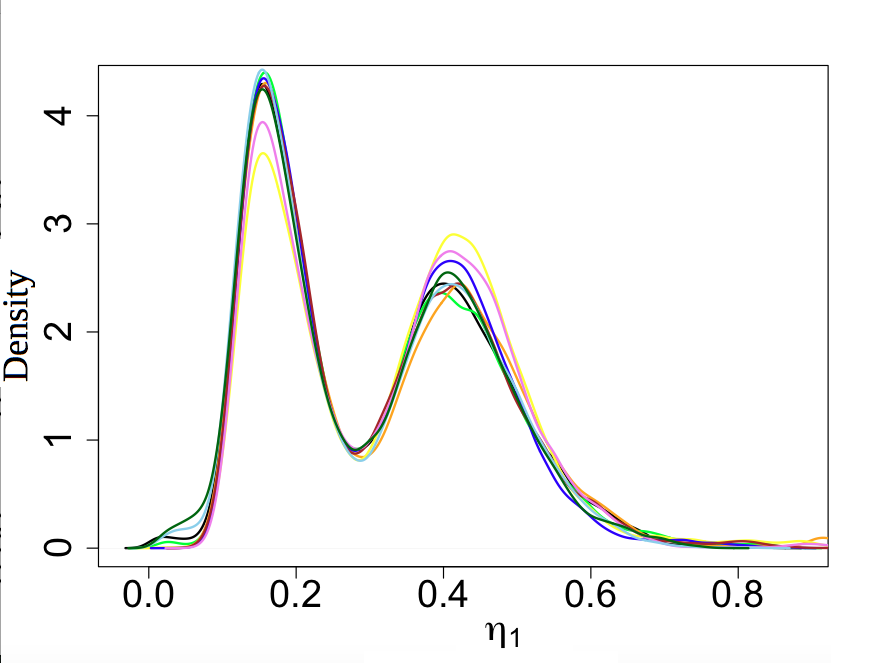}\includegraphics[width=.33\textwidth,height=4cm]{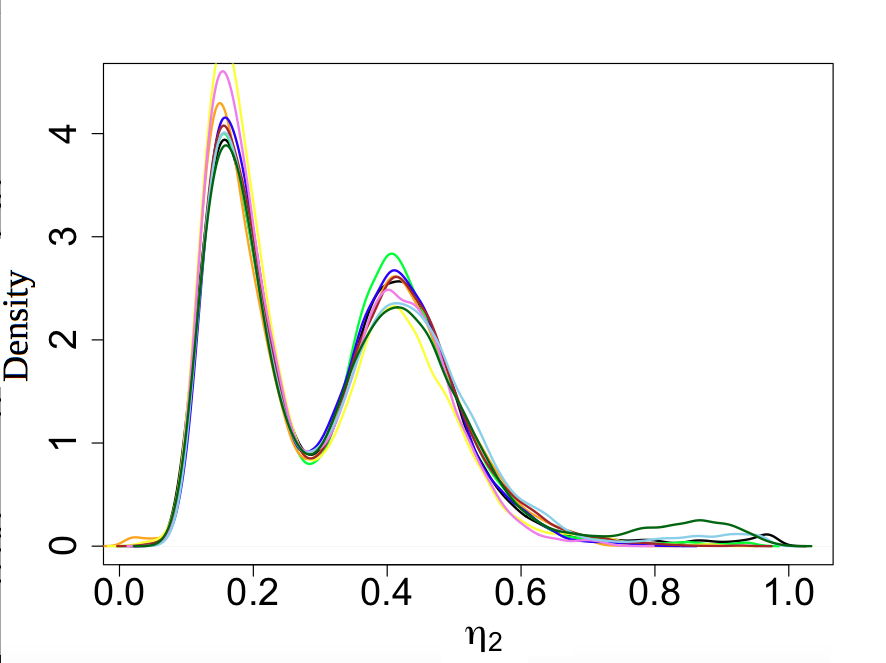}
\caption{{\bf Example \ref{ex:simulated_data}:} Kernel estimates of the
posterior densities of the parameters $\mu$, $\sigma$, $p$, $\varphi$,
$\eta_i$, based on $10$ parallel MCMC chains for Proposal 1 and $2\times10^5$
iterations, based on a single simulated sample of size $50$. The true value of
$(\mu,\sigma^2,p,\varphi^2, \eta_1, \eta_2)$ is $(-5.37,15.75,0.65,0.81, 0.15,
0.41)$.}
\label{fig1}
\end{figure}
 
\begin{figure}[!h]
\includegraphics[width=.33\textwidth,height=4cm]{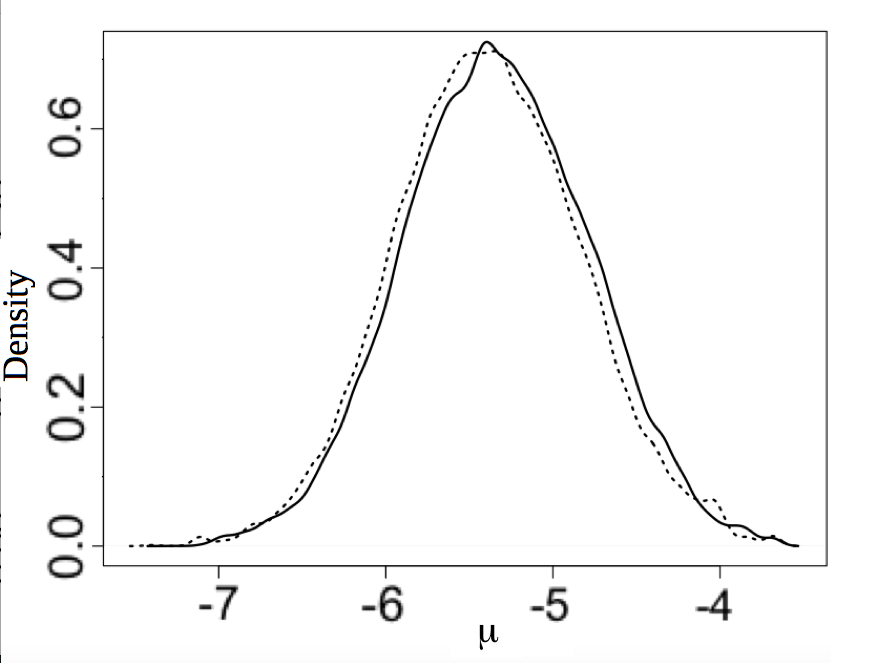}\includegraphics[width=.33\textwidth,height=4cm]{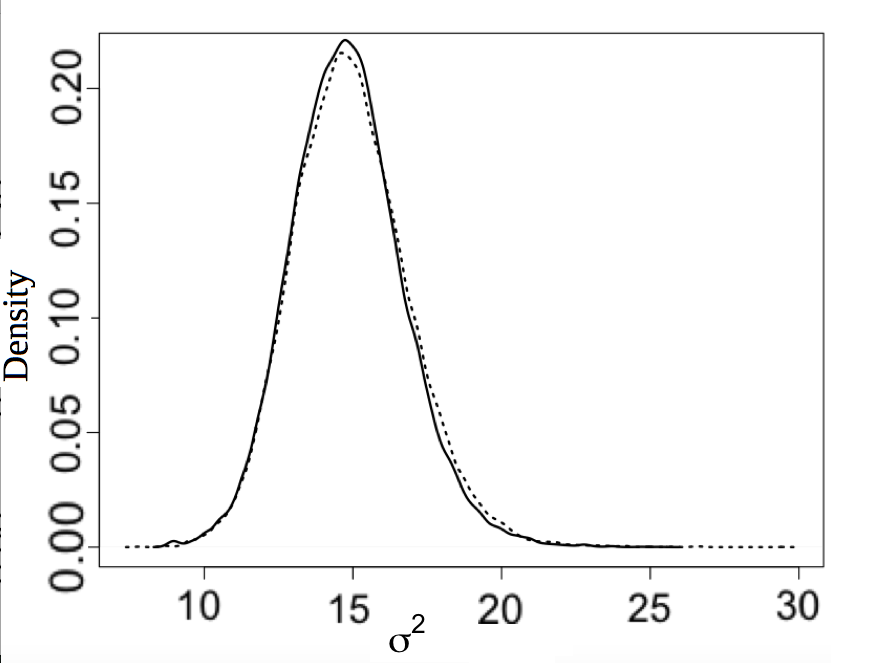}\includegraphics[width=.33\textwidth,height=4cm]{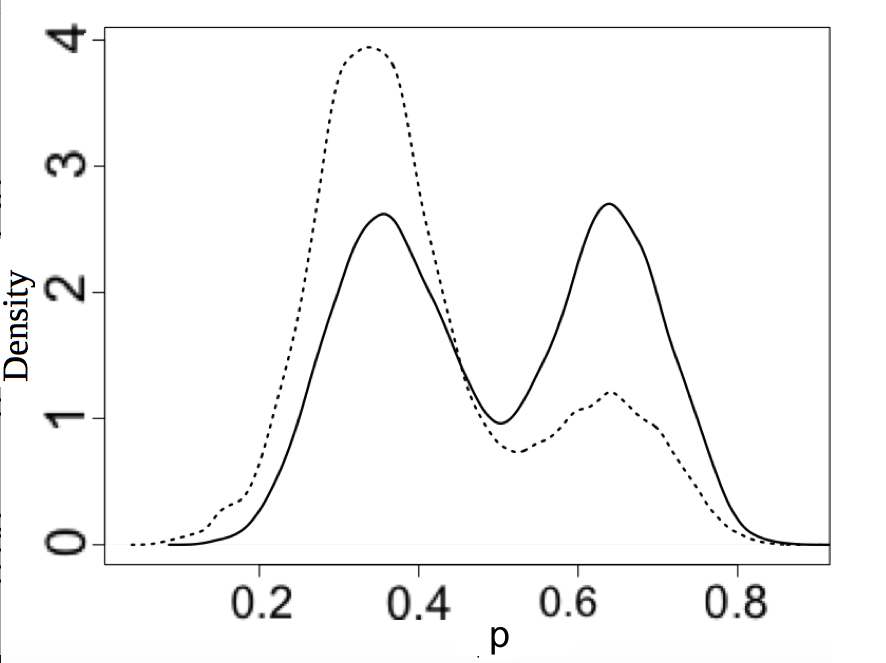}
\includegraphics[width=.33\textwidth,height=4cm]{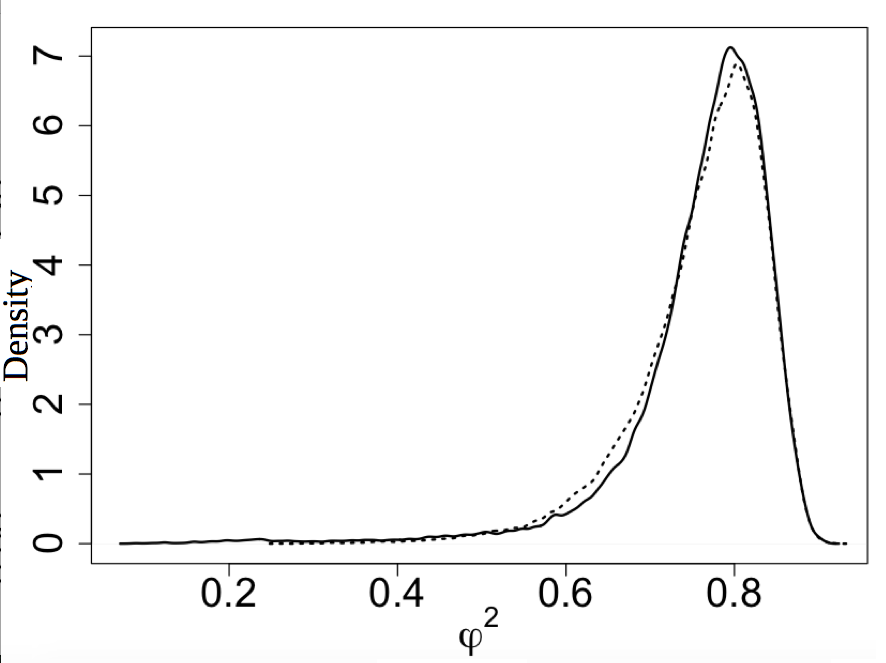}\includegraphics[width=.33\textwidth,height=4cm]{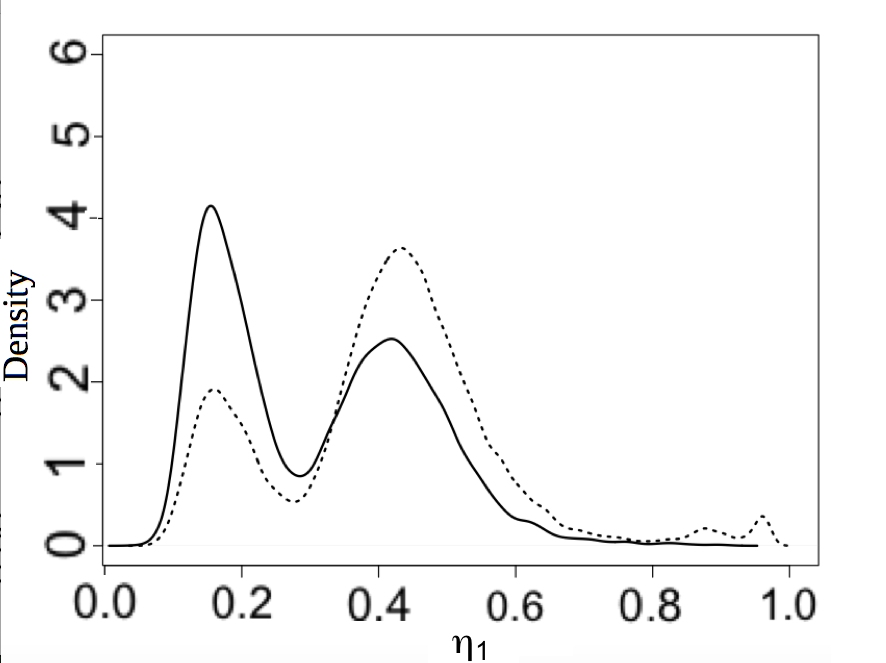}\includegraphics[width=.33\textwidth,height=4cm]{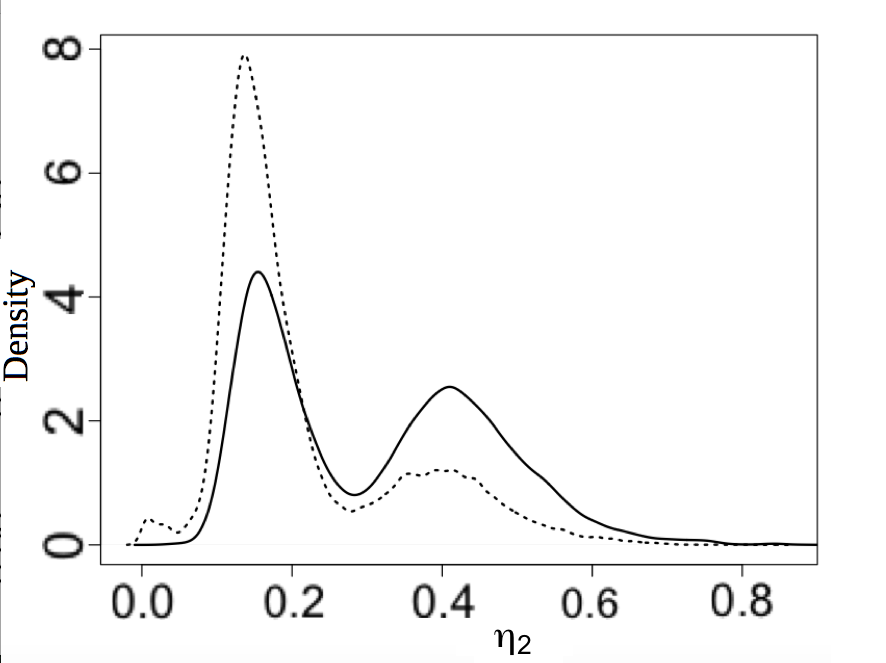}
\caption{{\bf Example \ref{ex:simulated_data}:} Comparison between MCMC samples from our 
algorithm using Proposal 1 ({\em solid line}) and Proposal 2 ({\em dashed line}), with $90,000$ iterations and the 
sample of Figure \ref{fig1}. The true value of $(\mu,\sigma^2,\varphi^2, \eta_1, \eta_2)$ is $(-5.375,15.747,0.813, 0.149, 0.406)$. }
\label{fig2}
\end{figure} 

\begin{table}
\centering 
\begin{tabular}{ c||  c c c c c c c} 
\hline  Proposal 1 & $ar_\mu$ & $ar_\sigma$ & $ar_p$ & $ar_{\varphi,\eta} $ & $\epsilon_{\mu}$ & $\epsilon_p$ & $\epsilon$ \\ 
& 0.40 & 0.47 & 0.45 & 0.24 & 0.56 & 77.06 & 99.94 \\ \hline 
Proposal 2  & $ar_\mu$ & $ar_\sigma$ & $ar_p$ & $ar_{\varphi,\eta}$ & $\epsilon_{\mu}$ & $\epsilon_{p}$ & $\epsilon_{\vartheta}$  \\ 
& 0.38 & 0.46 & 0.45 & 0.27 & 0.55 & 0.29 & 0.35\\ \hline 
\end{tabular}   
\caption{\small \label{tab1} {\bf Example \ref{ex:simulated_data}}: Acceptance rate ($ar$) and corresponding proposal scale ($\epsilon$) when the adaptive Metropolis-within-Gibbs sampler is used.}
\end{table}
\end{example}

\subsection{The general Gaussian mixture model}\label{sub:k}

We now consider the general case of estimating a mixture for
any $k$ when the variance vector $(\eta_1^2, \ldots, \eta_k^2)$ also has the
spherical coordinate system as represented in Section \ref{sub:duale}. All algorithms
used in this section are publicly available within our R package Ultimixt. The package Ultimixt contains
functions that implement adaptive determination of optimal scales and convergence monitoring based on \cite{GLMRUB1992} criterion. In addition, {Ultimixt} includes functions that summarise the simulations and compute point estimates of
each parameter, such as posterior mean and median. It also produces an estimated mixture density in numerical and
graphical formats. The output further provides graphical representations of the generated parameter samples.
  
\begin{example}\label{ex:simulated_data_K3} 
The sample is made of 50 simulation from the mixture 
$$
0.27 \mathcal{N}(-4.5, 1)+0.4 \mathcal{N}(10, 1)+0.33 \mathcal{N}(3, 1)\,.
$$ 
Since this is a Gaussian mixture with the common variance of 1, simulated
chains of component-wise mean parameters and weights are good indicators to
monitor whether the chain explores all posterior modes. Monitoring the chain
for an angle parameter $\varpi$ and $p_i$, we illustrate the motivation of
sampling $\eta$ and $\varpi$ through two steps in Figure 10, Supplementary
Material. 

From our simulation experience of the adaptive Metropolis-within-Gibbs
algorithm using only a random walk proposal (restrict to Step 2.8), the
simulated samples were quite close to the true values; however, the chain
visited only one of the posterior modes. This lack of label switching helps us
in producing point estimates directly from this MCMC output \citep{geweke:2007}
but this also shows an incomplete convergence of the MCMC sampler.
 
To help the chain visit all posterior modes, the proposals are restricted to
Step 2.4 of the Metropolis-within-Gibbs algorithm, namely using only a uniform
distribution $\mathcal{U}[0,2\pi]$. The MCMC samples on the $p_i$'s are both well-mixed
and exhibit strong exchangeability (see Figure \ref{figu3} in the
Supplementary Material). However, the corresponding acceptance rate is quite
low at $0.051$. To increase this rate, the random walk proposal of Step 2.8 on $\varpi$, namely
$\mathcal{U}(\varpi^{(t)}-\epsilon_{\varpi},\varpi^{(t)}+\epsilon_{\varpi} )$, is added
and this clearly improves performances, with acceptance rates all close to
$0.234$ and $0.44$. Almost perfect label switching occurs in this case (see
Figure \ref{figu4} in the Supplementary Material). Hence posterior samples for
$\eta$'s and $\varpi$'s are generated using an independent proposal plus a
random walk proposal in our adaptive Metropolis-within-Gibbs algorithm. 

The simulated chains are almost indistinguishable component-wise, due to
label switching. As described in Section \ref{sec:witch},
we relabelled the MCMC chains using both (a) a k-means clustering
algorithm and (b) a removal of label switching by permutations,
as presented in Section \ref{sec:witch}. Point estimates of the relabelled chain are shown in Table \ref{table1}
and the marginal posterior distributions of component-wise mean and standard
deviation are shown in Figure \ref{figu1}. Bayesian estimations computed by
both methods are almost identical and all parameters of the mixture
distribution are accurately estimated.

\begin{figure}[!h]
\centering\setlength{\unitlength}{1cm}
\begin{picture}(16,8)
\put(0,4){\includegraphics[width=5cm,height=4cm]{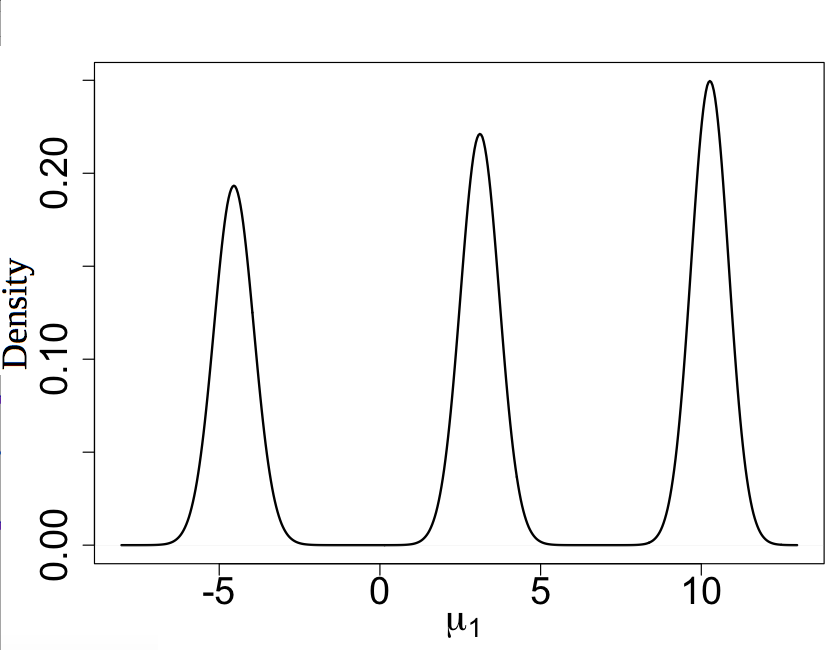}}
\put(5,4){\includegraphics[width=5cm,height=4cm]{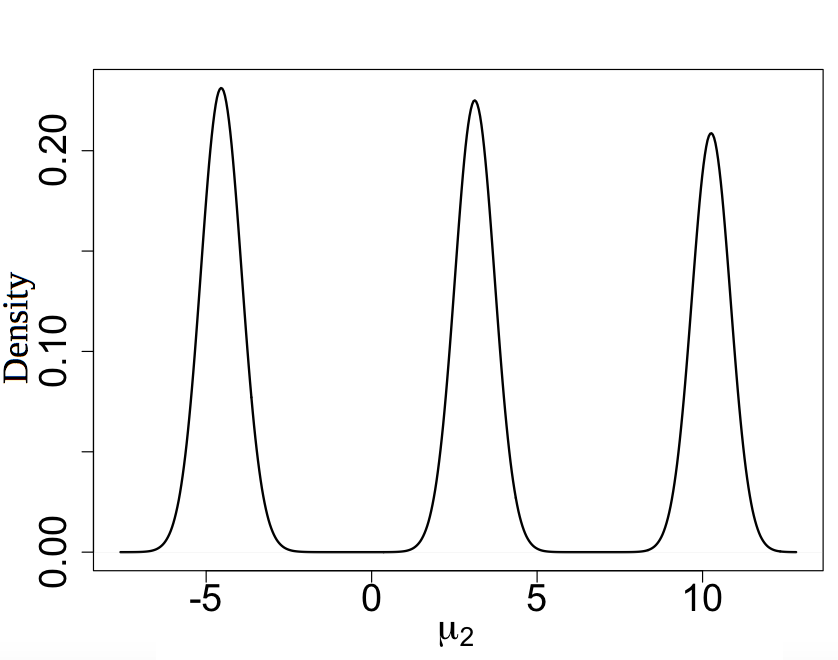}}
\put(10,4){\includegraphics[width=5cm,height=4cm]{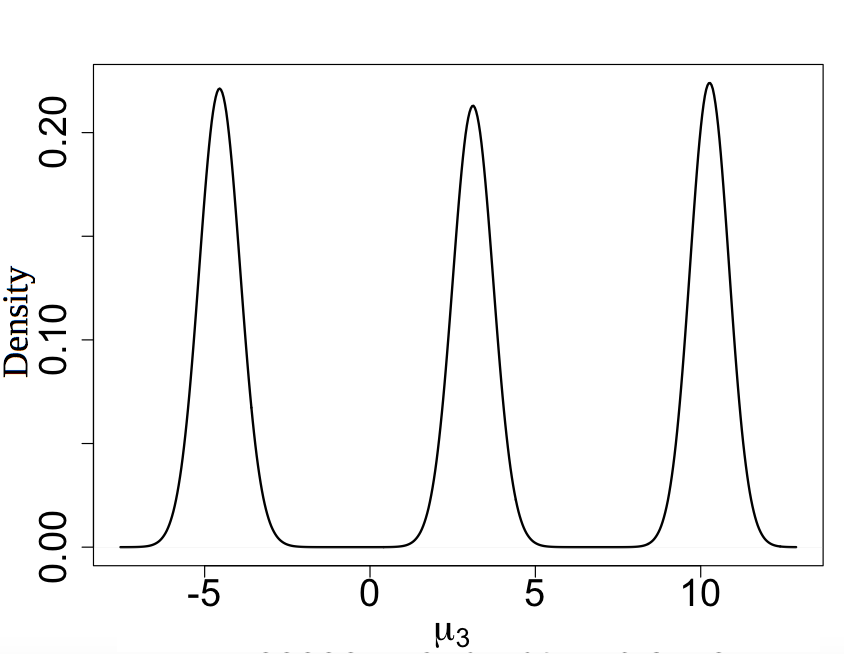}}
\put(0,0){\includegraphics[width=5cm,height=4cm]{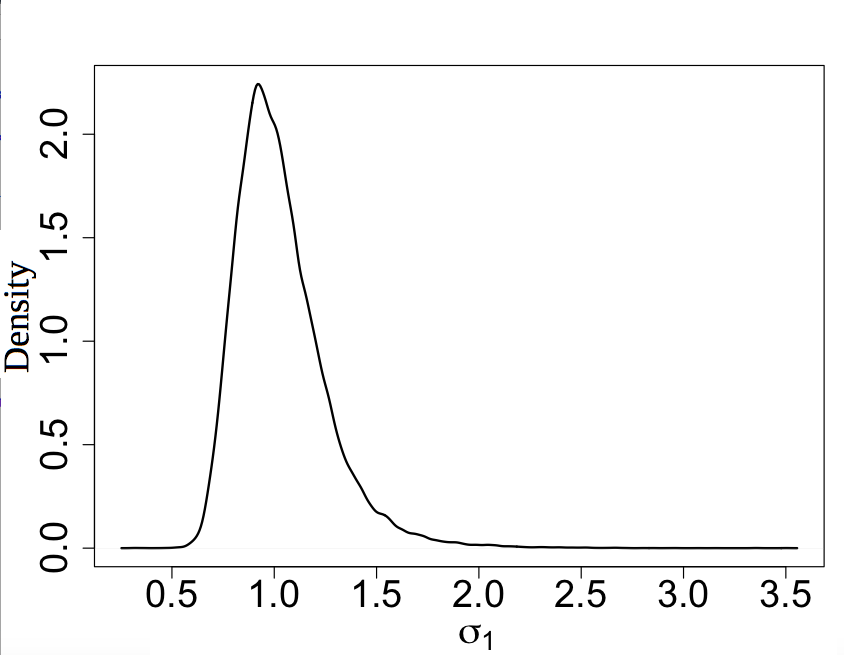}}
\put(5,0){\includegraphics[width=5cm,height=4cm]{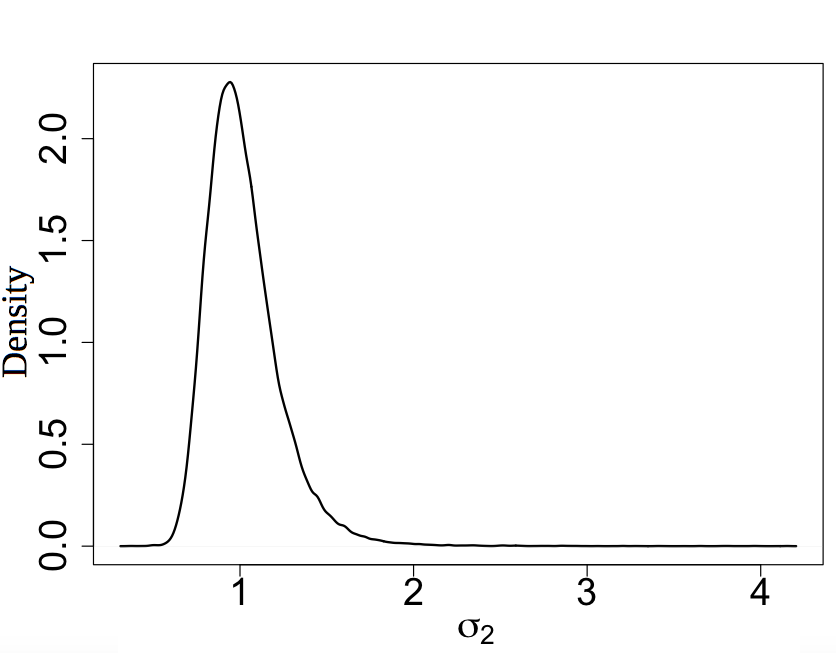}}
\put(10,0){\includegraphics[width=5cm,height=4cm]{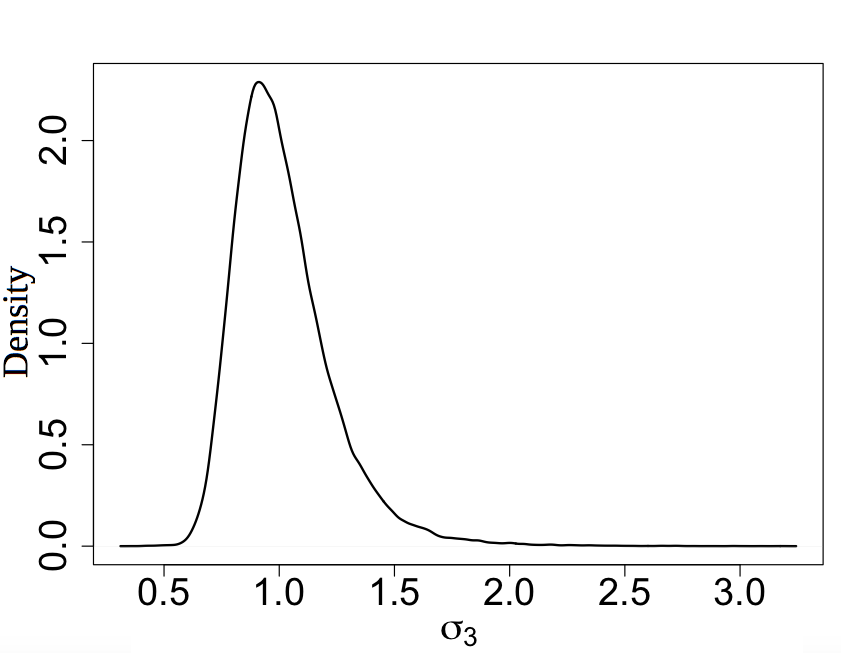}}
\end{picture}
\caption{{\bf Example \ref{ex:simulated_data_K3}:} Estimated marginal posterior densities of component means and
standard deviations, based on $10^5$ MCMC iterations.} 
\label{figu1} 
\end{figure}
 
\begin{table} 
\scriptsize 
\begin{tabular}{c c}
\begin{tabular}{|c||c c c || c c c|} 
\hline 
& \multicolumn{3}{c||}{\bf k-means clustering} & \multicolumn{3}{c|}{\bf Relabelled using MAP}\\
\cline{2-7} & $\varpi$ & $\xi_1$ & $\xi_2$ & $\varpi$ & $\xi_1$ & $\xi_2$\\  Median & 3.54 & 0.97 & 0.73 &
3.32 & 0.94 & 0.83 \\ Mean & 3.53 & 0.98 & 0.72 &  3.45  & 0.94 & 0.82 \\\cline{2-7} &$p_1$ & $p_2$ & $p_3$ & $p_1$ &
$p_2$ & $p_3$  \\ Median & 0.40 & 0.27 & 0.33 & 0.41 & 0.27 & 0.33\\ Mean& 0.41 & 0.27  & 0.33 & 0.41 & 0.27
& 0.33\\ \cline{2-7} &$\mu_1$ & $\mu_2$ & $\mu_3$ & $\mu_1$ & $\mu_2$ & $\mu_3$ \\  Median & 10.27 &  -4.55
& 3.11  & 10.27 & -4.55 & 3.11 \\ Mean& 10.27 & -4.54  & 3.12  & 10.26 & -4.45 & 3.11 \\\cline{2-7}
&$\sigma_1$ & $\sigma_2$ & $\sigma_3$ & $\sigma_1$ & $\sigma_2$ & $\sigma_3$\\ Median & 0.93 &  1.04 & 1.01
& 0.93 & 1.04 & 1.03 \\ Mean& 0.95 & 1.08  & 1.05  & 0.95 & 1.07 & 1.05 \\ \hline\hline
\end{tabular} 
& \begin{tabular}{c}

\begin{tabular}{|c|| c c c|}
\cline{1-4} & \multicolumn{3}{c|}{\bf Global parameters} \\ \cline{1-4}
& $\mu$ & $\sigma$ & $\varphi$  \\ Median & 3.98 &  6.03 & 0.98 \\ Mean& 3.98 & 6.02 &
0.99 \\ \hline \end{tabular} 
\\ \\ 
\begin{tabular}{|c c c c c c |} 
\hline \multicolumn{6}{|l|}{\bf
Proposal scales} \\ \hline
$\epsilon_\mu$ & $\epsilon_\sigma$ & $\epsilon_p$ & $\epsilon_\varphi$ & $\epsilon_\varpi$ &
$\epsilon_\xi$ \\ 0.33 &  0.06 & 190 & 160 & 0.09 & 0.39 \\ \hline\hline
\multicolumn{6}{|l|}{\bf Acceptance rates} \\ \hline
$ar_\mu$ & $ar_\sigma$ & $ar_p$ & $ar_\varphi$ & $ar_\varpi$ & $ar_\xi$ \\
0.22 &  0.34 & 0.23 & 0.43 & 0.42 & 0.22\\ \hline
\end{tabular} \end{tabular}
\end{tabular}
\caption{\small {\bf Example }\ref{ex:simulated_data_K3}: Point estimators of the parameters
of a mixture of 3 components, proposal scales and corresponding acceptance rates.} 
 \label{table1}
\end{table} 
\end{example}
 
 \begin{example}\label{ex:pork}  
Computer aid tomography (CT) scanning is frequently used in animal science to
study the tissue composition of an animal. Figure \ref{fig:pork} (a) shows the
CT scan image of the cross-section of pork carcass in 256 grey-scale units.
Different tissue types produce different intensity-level observations on the CT
scan. Pixels attributed to fat tend to have grey scale readings 0-100, muscle
101-220, and bone 221-256. \citet{thompson:kinghorn:1992}, \citet{lee:2009} and
\citet{mcgrory:2013} modelled the composition of the three tissues of a pig
carcass using Gaussian mixture models and a model with six components
was favoured. In this paper, a random subset of 2000 observations from the original data,
made of 36326 observations, is used and estimation of the mixture model is
compared to estimates based on the Gibbs sampler of
{\sf bayesm} by \citet{PTRSI2010} and on the EM algorithms of {\sf mixtools} by
\citet{BCHY2009}. The data-dependent priors of {\sf bayesm} on the standard parameters are 
$$
\mu_i \sim N(\bar{\mu},10\sigma_i)\,, \quad \sigma_i^2 \sim
\text{IG}(\nu,3) \quad \text{and} \quad (p_1, \ldots,p_k)\sim
\mathcal{D}ir(\alpha_0,\ldots,\alpha_0) 
$$
where $\text{IG}(\nu,3)$ is the Inverse-Gamma distribution with scale
parameter 3 and degrees of freedom $\nu$. The hyperparameters
$\bar{\mu}$ and $\nu$ are derived from the data. Marginal
prior distributions of standard parameters using either our double uniform prior (Section
\ref{sub:duale}) or priors obtained by {\sf bayesm} are compared graphically in Figure
\ref{fig_prior}. While the priors for $\mu_i$
and $\sigma_i$ yielded by {\sf bayesm} do not vary with $k$, our
marginal posteriors get more skewed toward $0$ with $k$ but has a longer tail to provide
flexible supports for component-wise location and scale parameters. We stress that we fixed
the global mean and variance to 0 and 1 here, implying that the outcome will be more
variable when the Jeffrey prior is called. 

\begin{figure}[!h] 
\centering\setlength{\unitlength}{1cm}
\begin{picture}(16,9.5)
\put(0,5){\includegraphics[width=6cm,height=4.5cm]{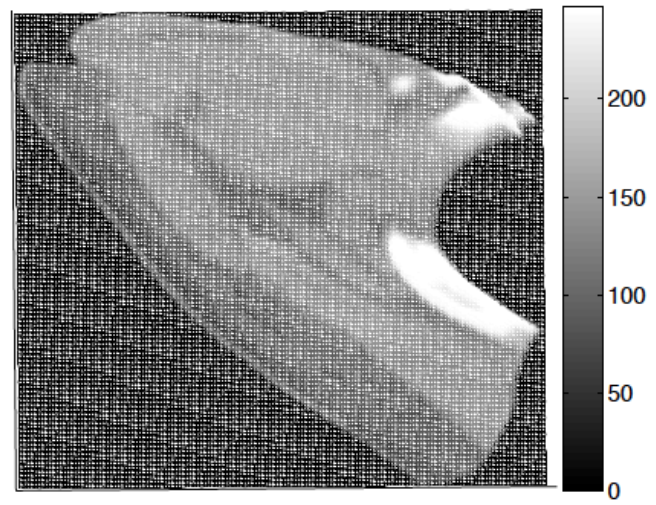}}
\put(8,5){\includegraphics[width=8cm,height=4.5cm]{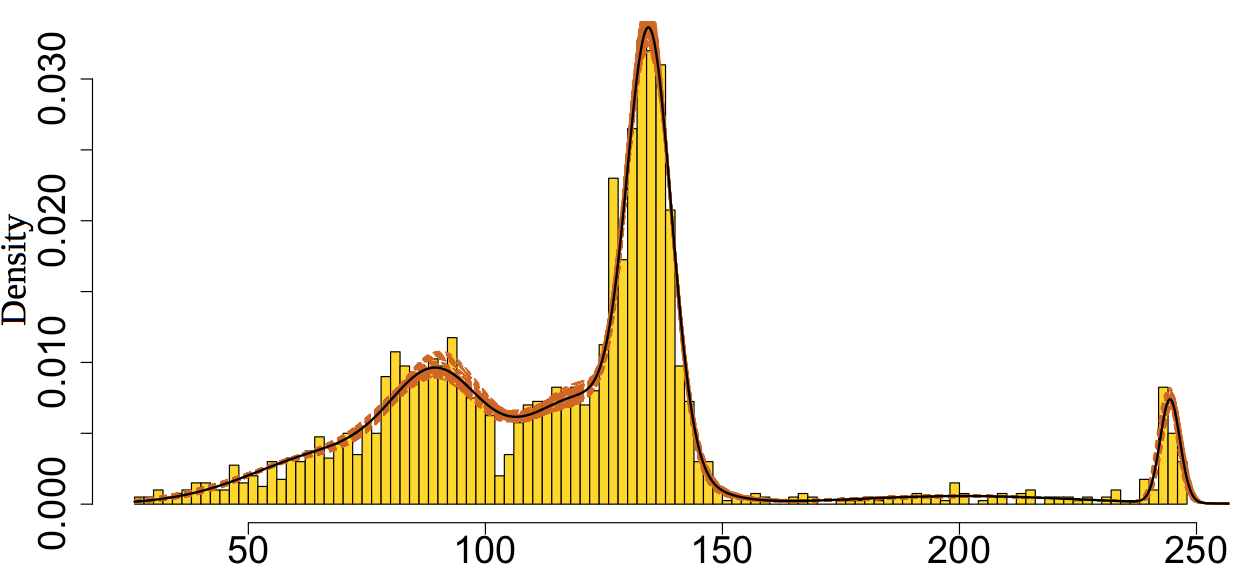}}
\put(0,0){\includegraphics[width=8cm,height=4.5cm]{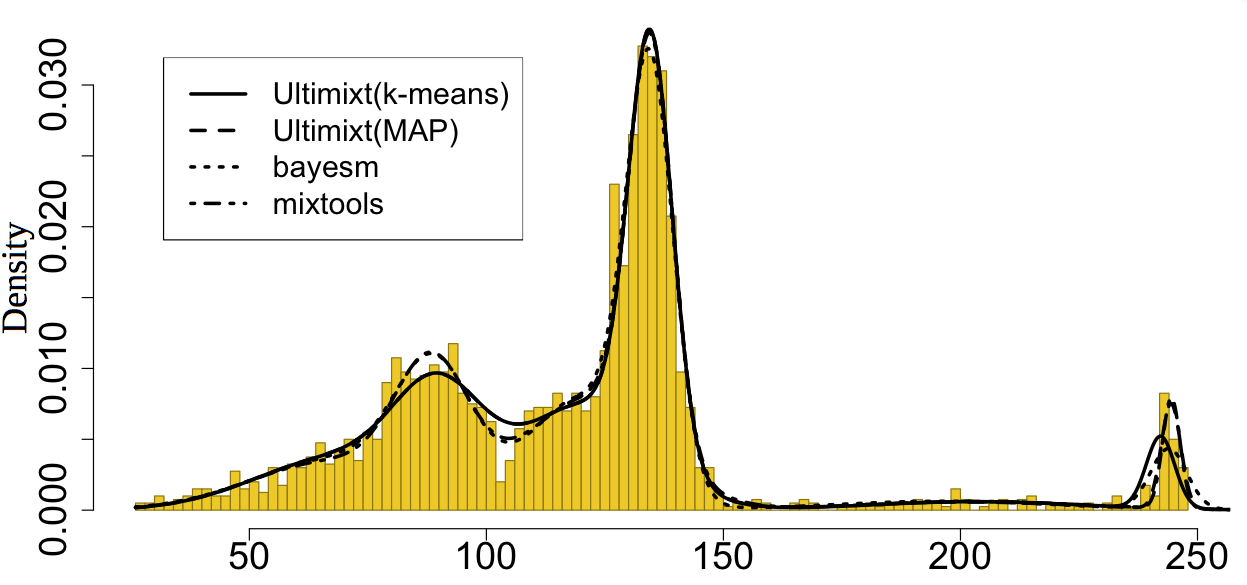}}
\put(8,0){\includegraphics[width=8cm,height=4.5cm]{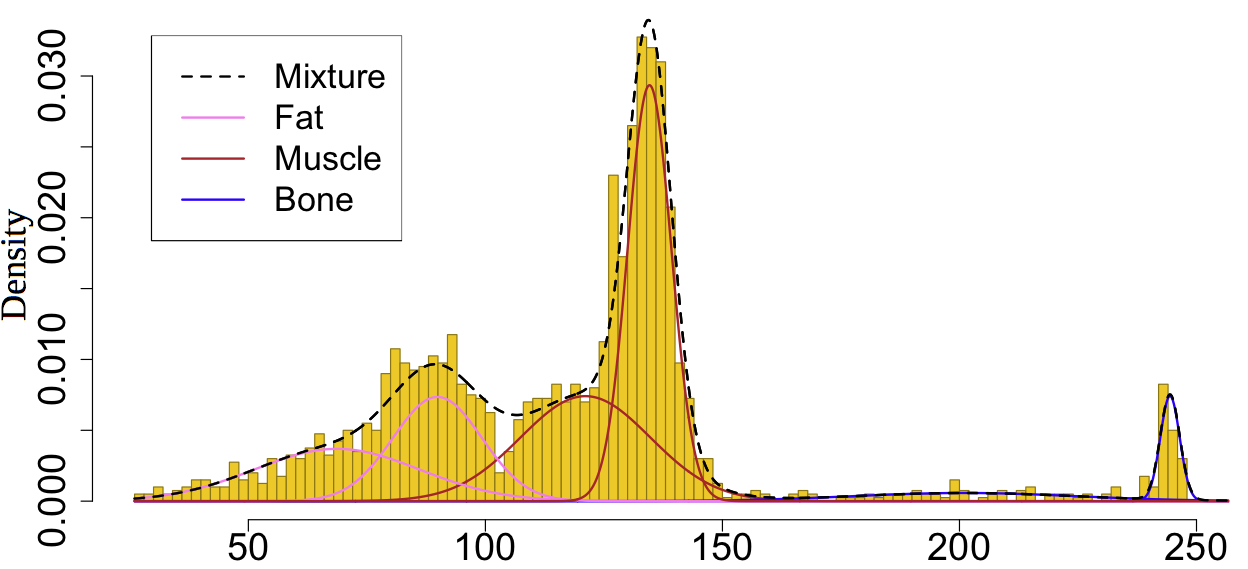}}
\put(7,9){(a} \put(15.2,9){(b} \put(7,4.3){(c} \put(15.2,4.3){(d}
\end{picture} \caption{{\bf CT image data and the analysis result:} (a) The CT
image of a cross-section from a pork carcass in grey-scale units. The right bar
describes the grey-scale, 0-256. (b) Representation of last 500 MCMC iterations
as mixture densities with the overlaid average curve for $k=6$ components
({\em dark line}) (c) Comparison between the mixture density estimates obtained
by {\sf Ultimixt}, {\sf mixtools} and {\sf bayesm} (d) Mixture model
overlapping with distributions of each components: Two violet, brown and blue
lines are distributions representing fat tissue, muscle and bone,
respectively.} \label{fig:pork}
\end{figure}
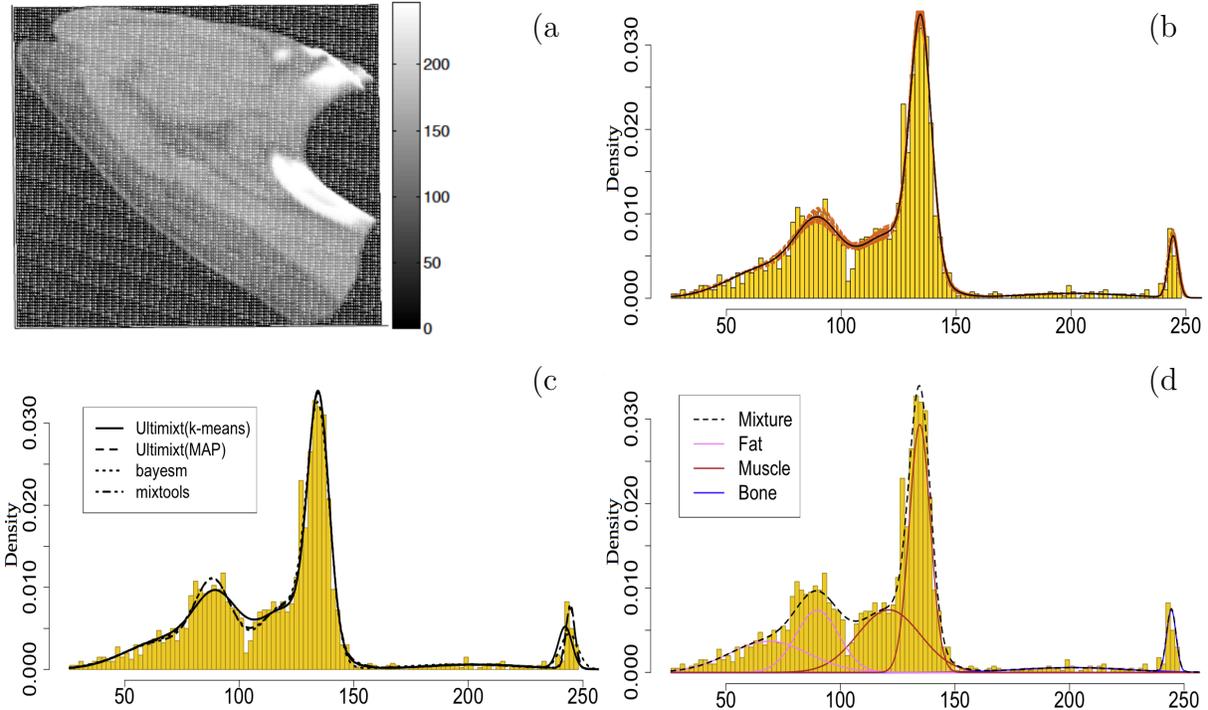 

\begin{figure}[!h] 
\centering\setlength{\unitlength}{1cm}
\begin{picture}(18,6)
\put(0,3){\includegraphics[width=5.5cm,height=3cm]{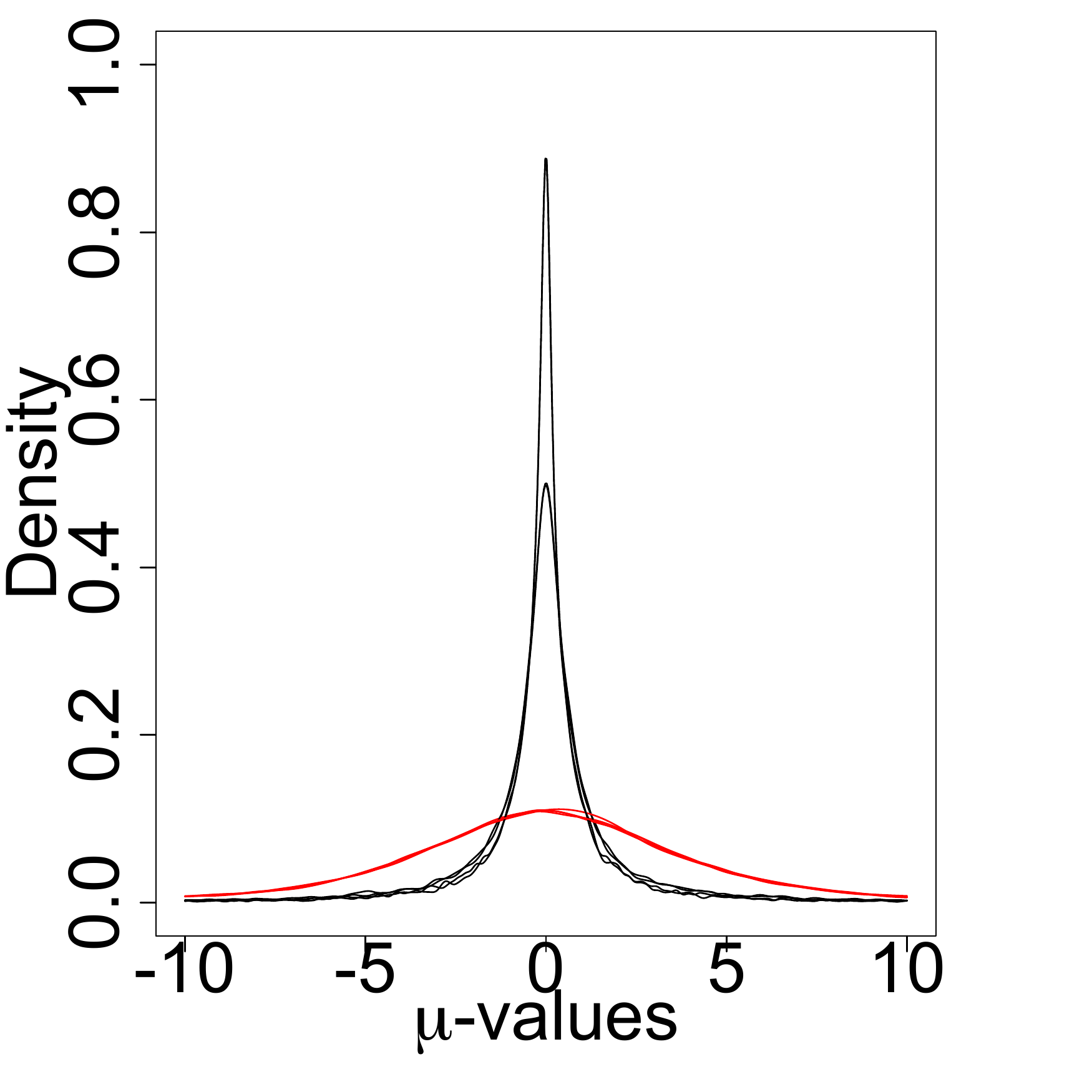}}
\put(5.5,3){\includegraphics[width=5.5cm,height=3cm]{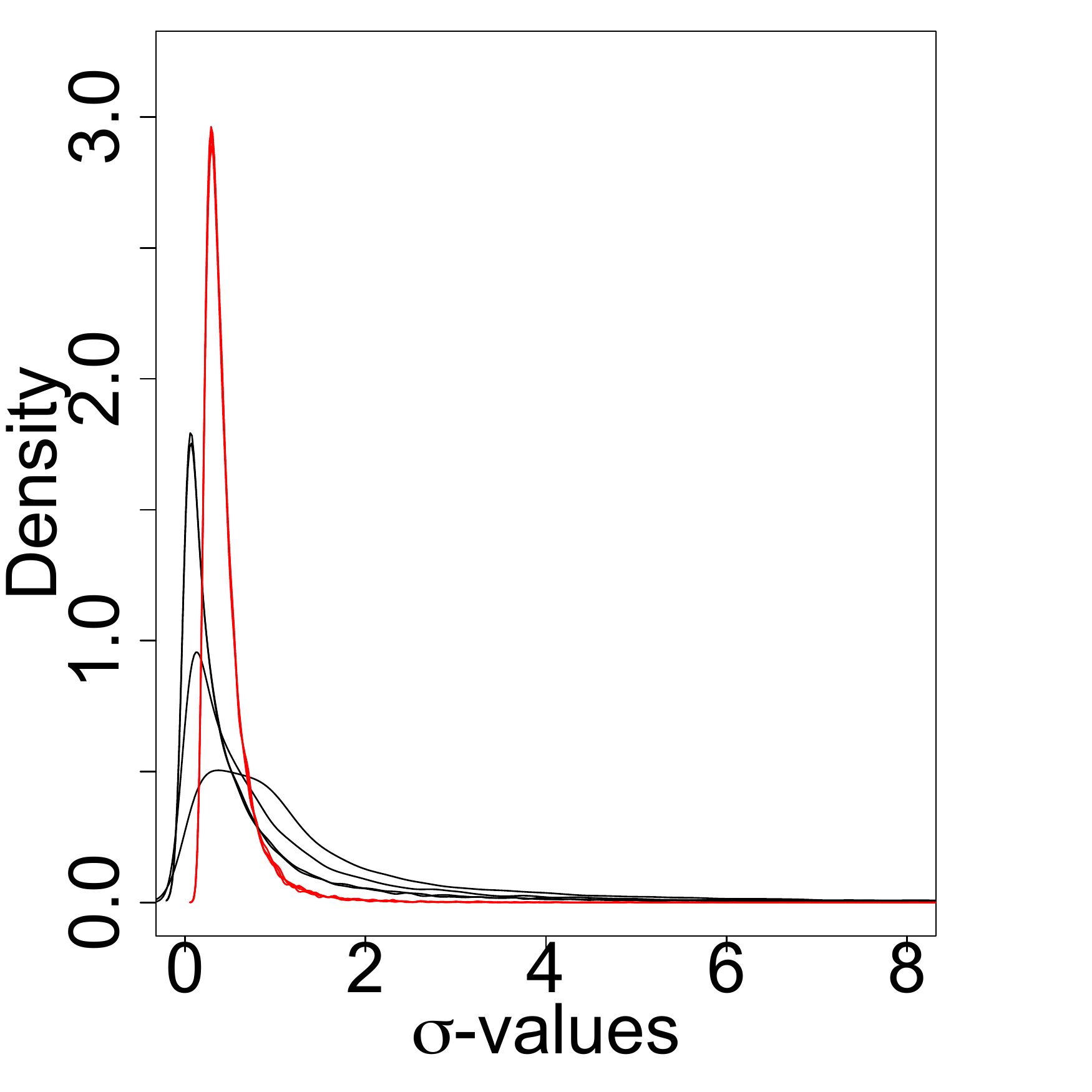}}
\put(11,3){\includegraphics[width=5.5cm,height=3cm]{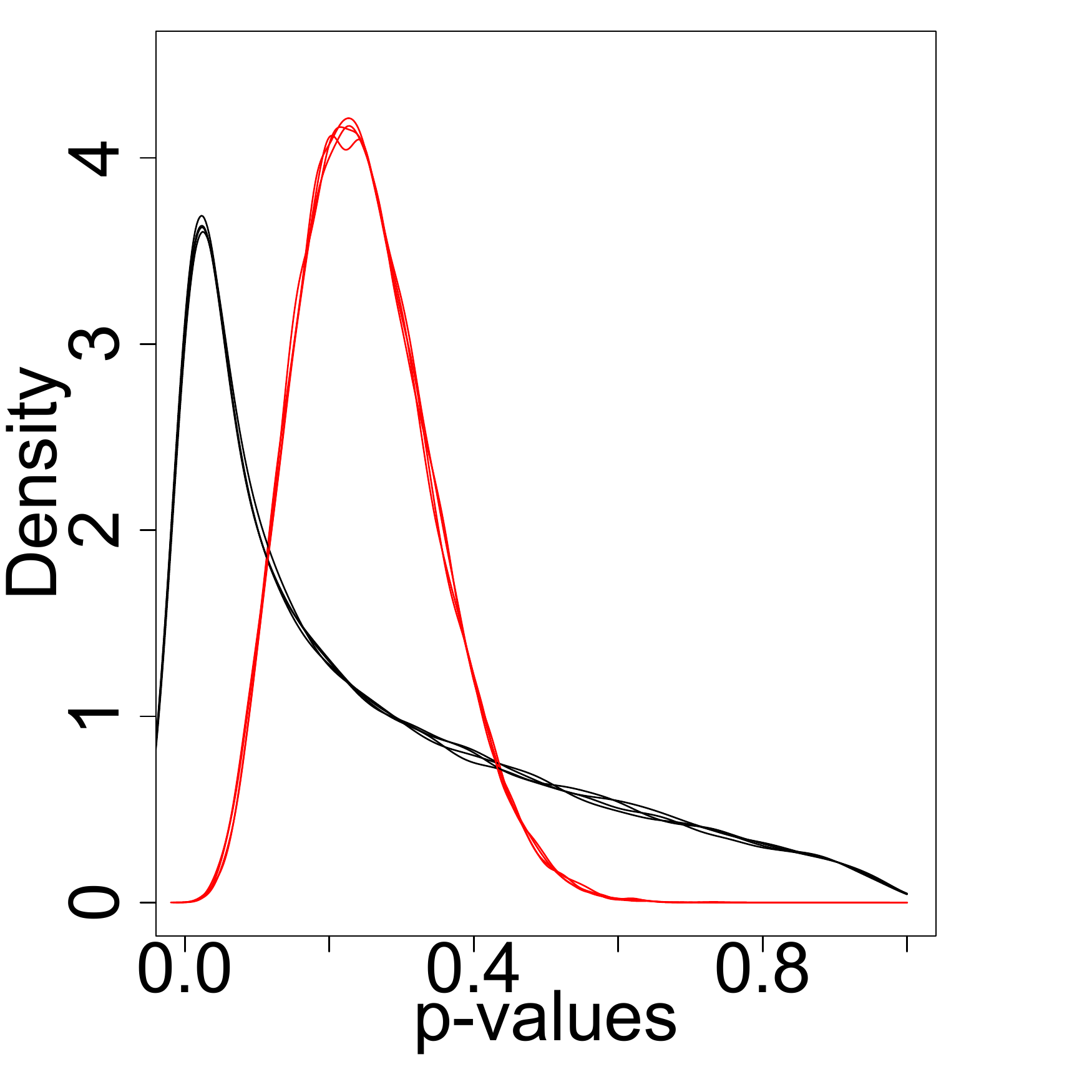}} 
\put(0,0){\includegraphics[width=5.5cm,height=3cm]{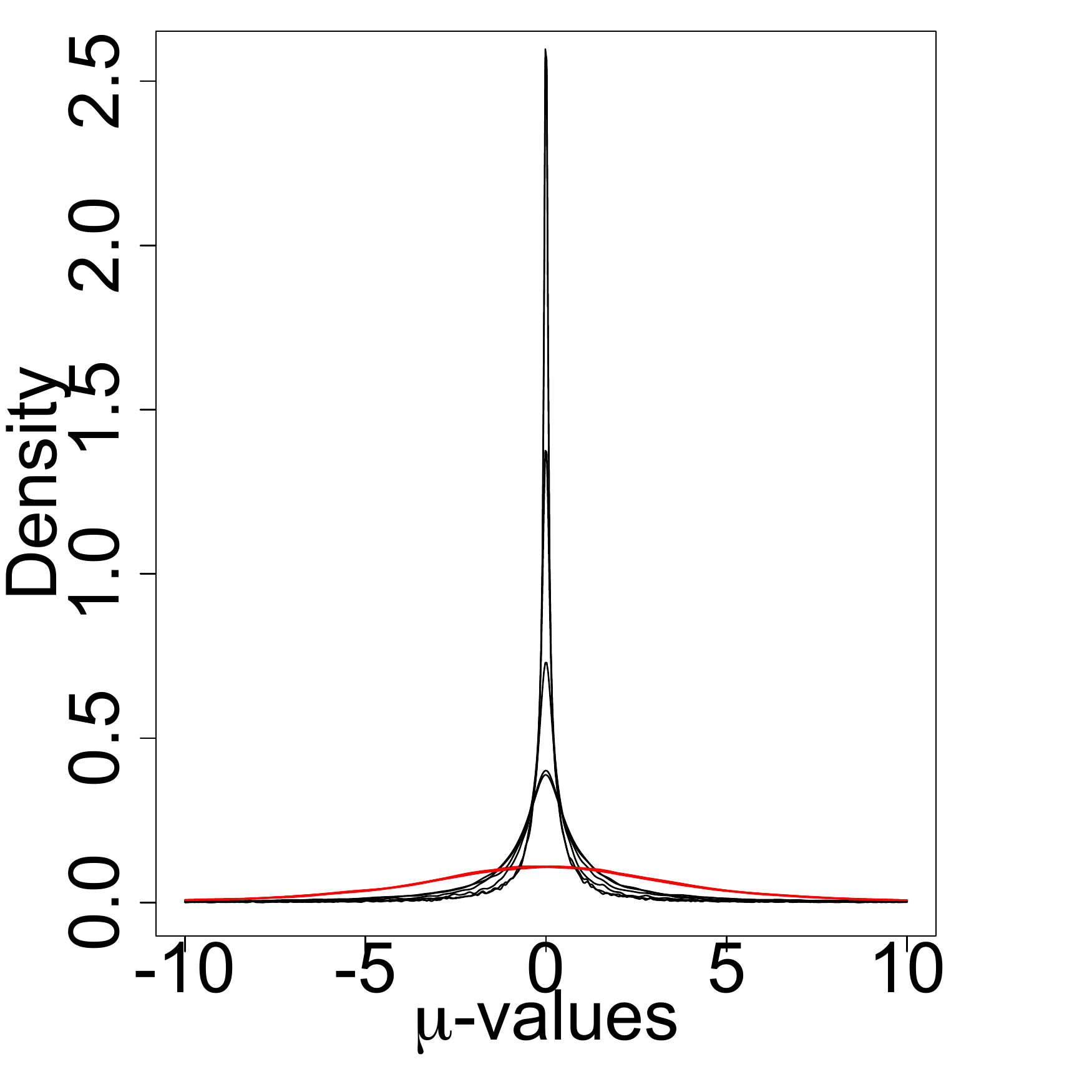}} 
\put(5.5,0){\includegraphics[width=5.5cm,height=3cm]{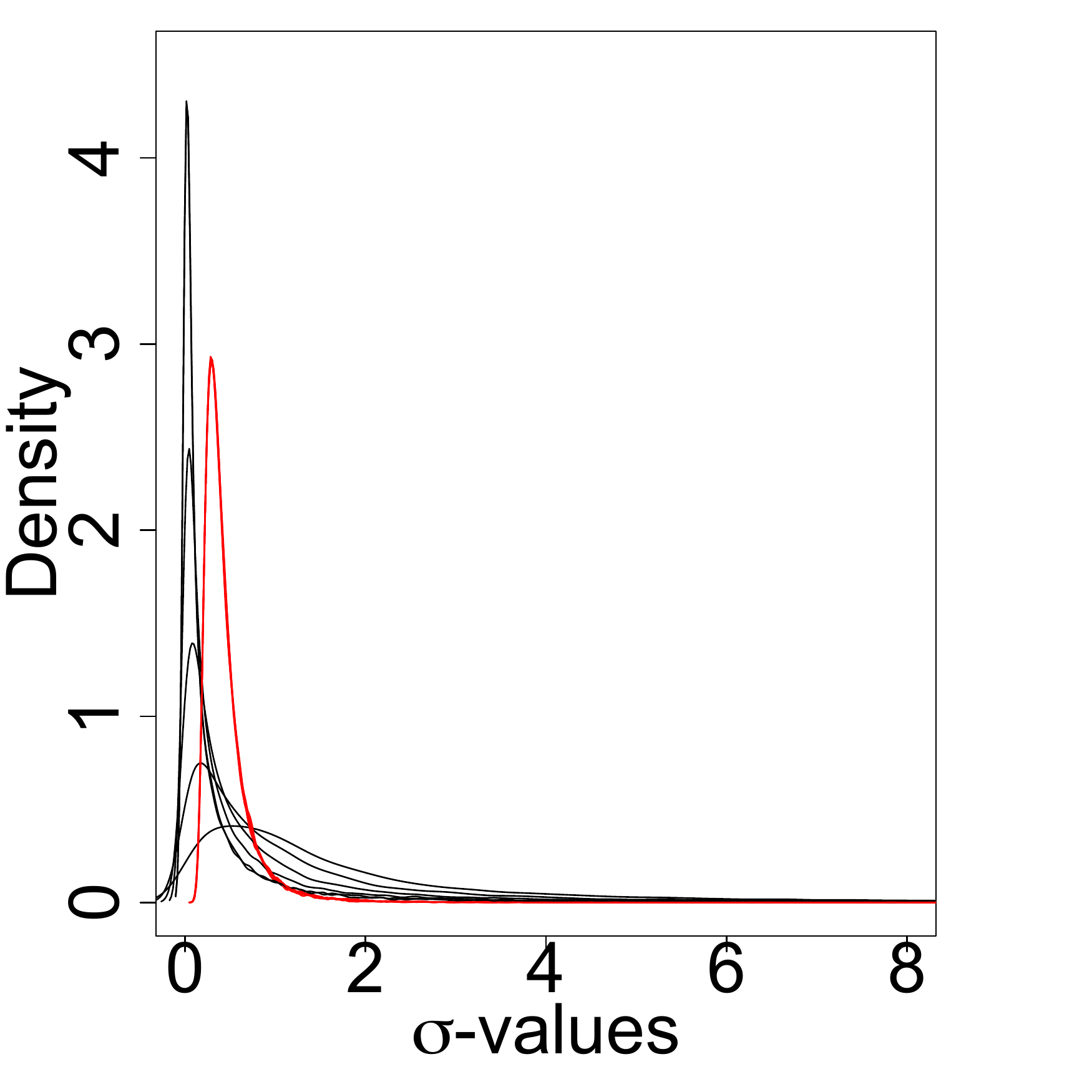}}
\put(11,0){\includegraphics[width=5.5cm,height=3cm]{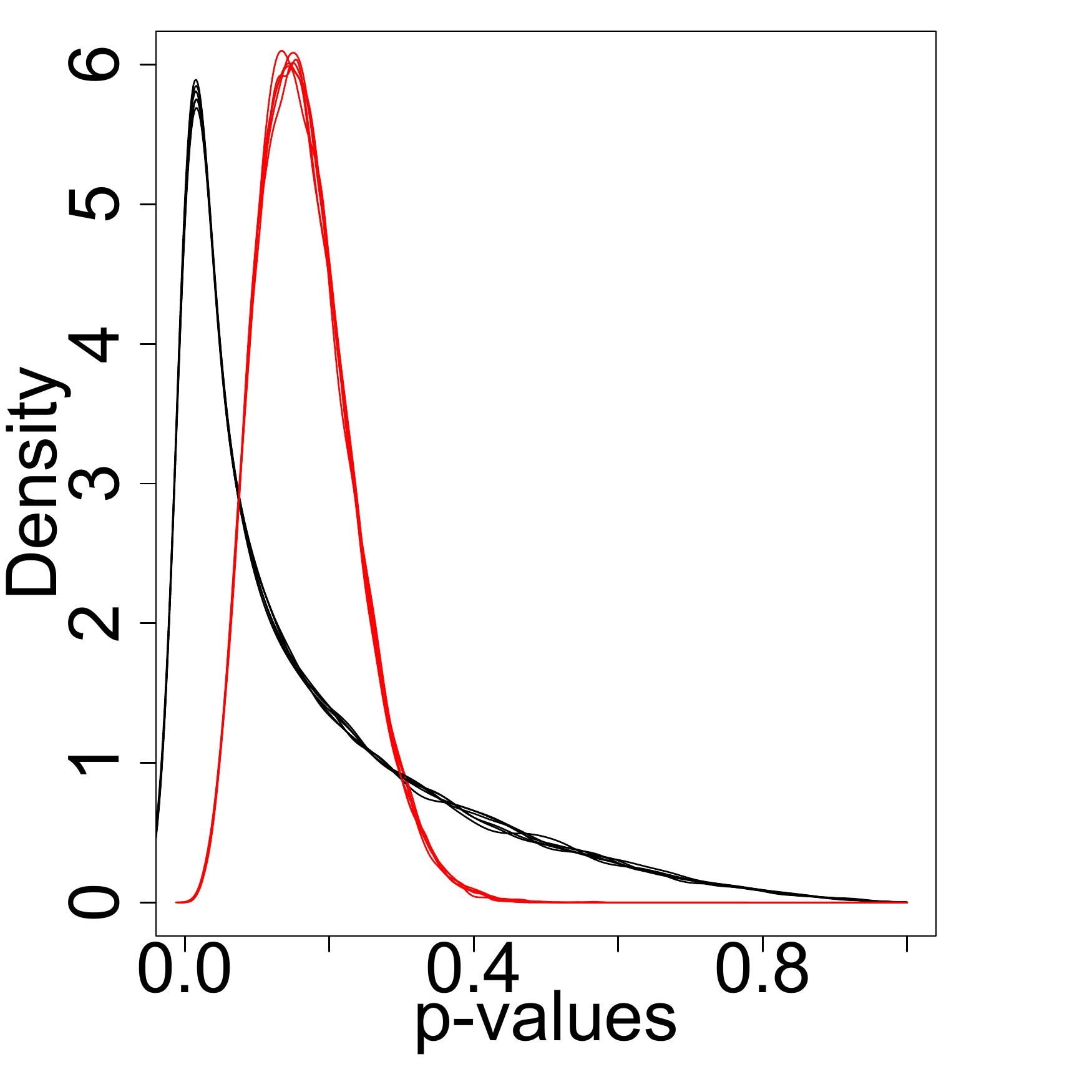}}
\end{picture} 
\caption{{\bf CT image data :} Density estimate of 20,000 draws
of $\mu_i,\sigma_i$ and $p_i$ ($i=1,\ldots,k$) from the prior by {\sf bayesm}
{\em (red lines)} and our double uniform prior {\em (black lines)} assuming a global mean
of 0 and variance of 1 when $k=4$ {\em (first row)} and $k=6$ {\em (second
row)}. For the prior by {\sf bayesm} hyperparameters $\alpha_0=5$,
$\bar{\mu}=0$ and $\nu=3$ are obtained using {\sf bayesm}.} \label{fig_prior}
\end{figure} 

Following the analysis of this data by \citet{mcgrory:2013}, a mixture model of
six Gaussian components is considered. The resulting means, medians and $95\%$
credible intervals of the parameters of the mixture components are displayed in
Table \ref{table:pork}, Supplementary Material, along with estimates based on
the Gibbs sampler of {\sf bayesm} \citep{PTRSI2010} and on the EM algorithms of
{\sf mixtools} \citep{BCHY2009}, with our approach being produced by {\sf
Ultimixt} \citep{kamary:lee:2015}. The MCMC sample from {\sf Ultimixt} is again
summarised by both $k$-means clustering and post-MCMC relabelling using the MAP
estimates. As can be seen from Figure \ref{fig:pork} (with exact values in
Table \ref{table:pork} from the Supplementary Material), the estimates from the
three packages {\sf Ultimixt}, {\sf mixtools} and {\sf bayesm} are relatively
similar and tissue composition similar to the findings of \citet{mcgrory:2013}
is observed. Figure \ref{fig:pork} (d) shows how the composition of tissues is
modelled by six Gaussian components which can be interpreted as follows: two
components correspond to fat (33\%), two to muscle (59\%), one to bone (4\%)
and the remaining component models the mixed tissue of muscle and bone (4\%).
Among the six components, the biggest component has the weight of $34\%$ and
corresponds to muscle. In the intended application, this is the quantity of
interest: the higher this percentage, the higher the meat quality of the animal.

\end{example}

\subsection{Poisson mixtures}\label{poi_study}

The following example demonstrates how a weakly informative
prior for a Poisson mixture is associated with a MCMC algorithm. 
Under the constraint, $\sum^k_{i=1} \gamma_i=1$, the Dirichlet
prior with the common parameter is used on local parameters $\gamma_i$.
Any other vague proper prior on this compact space is also suitable.

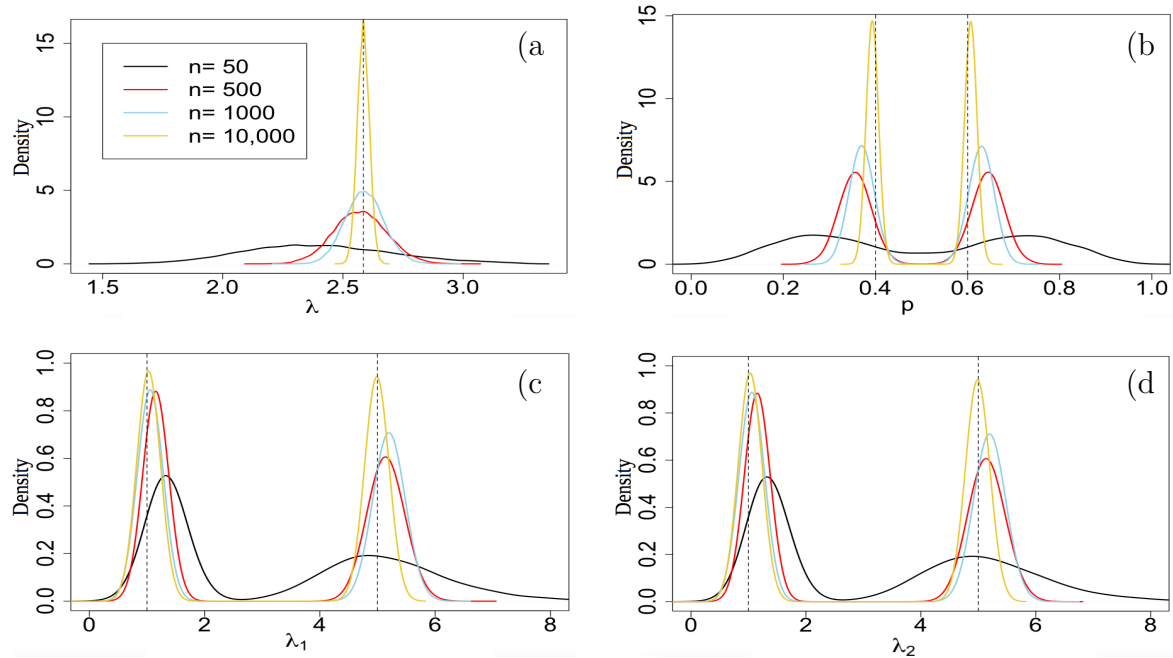
\begin{figure}[h!] 
\centering\setlength{\unitlength}{1cm}
\begin{picture}(13,9)
\put(-1,4.5){\includegraphics[width=7.5cm,height=4.5cm]{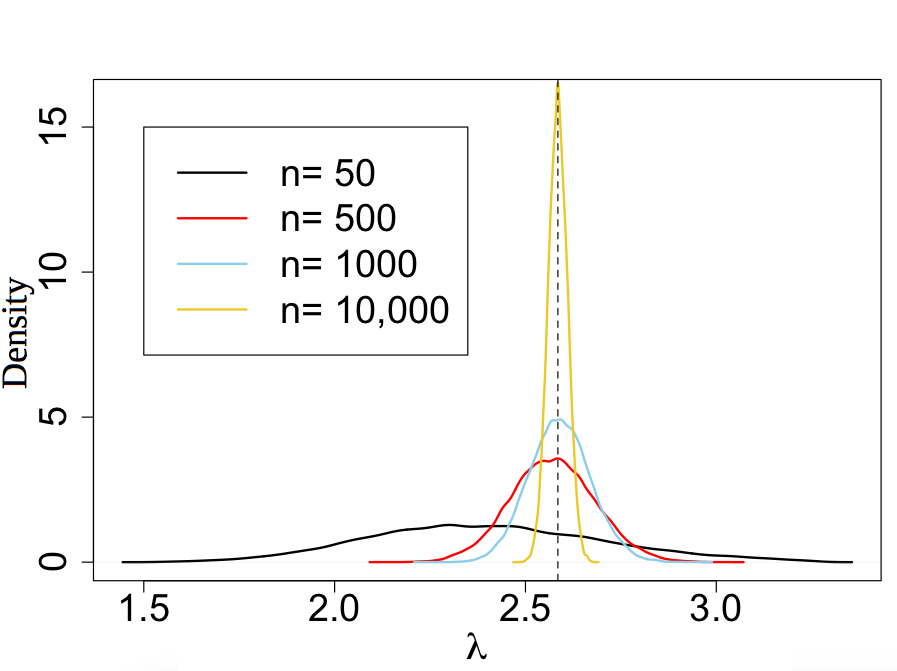}}
\put(7,4.5){\includegraphics[width=7.5cm,height=4.5cm]{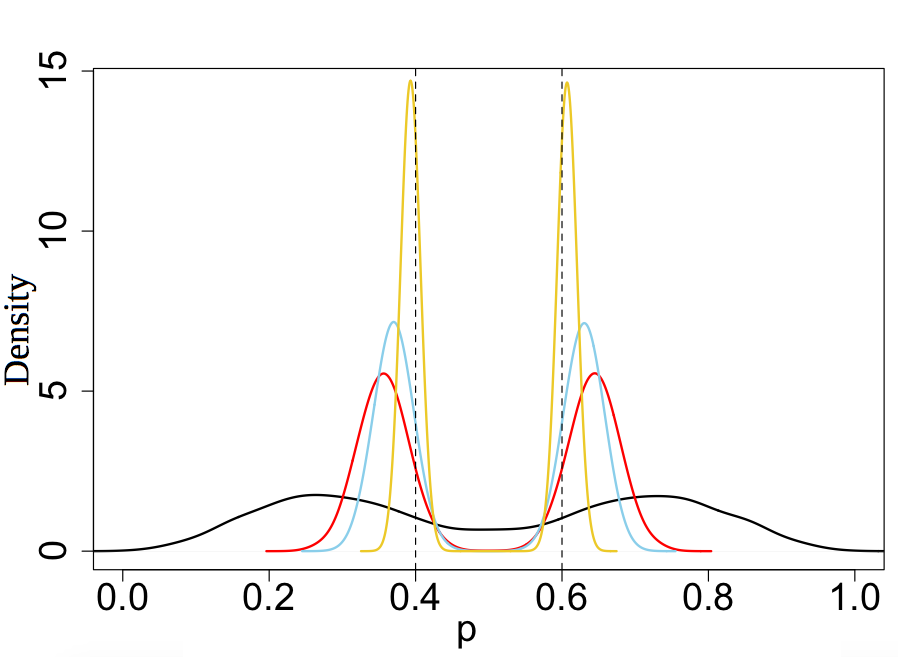}}
\put(-1,0){\includegraphics[width=7.5cm,height=4.5cm]{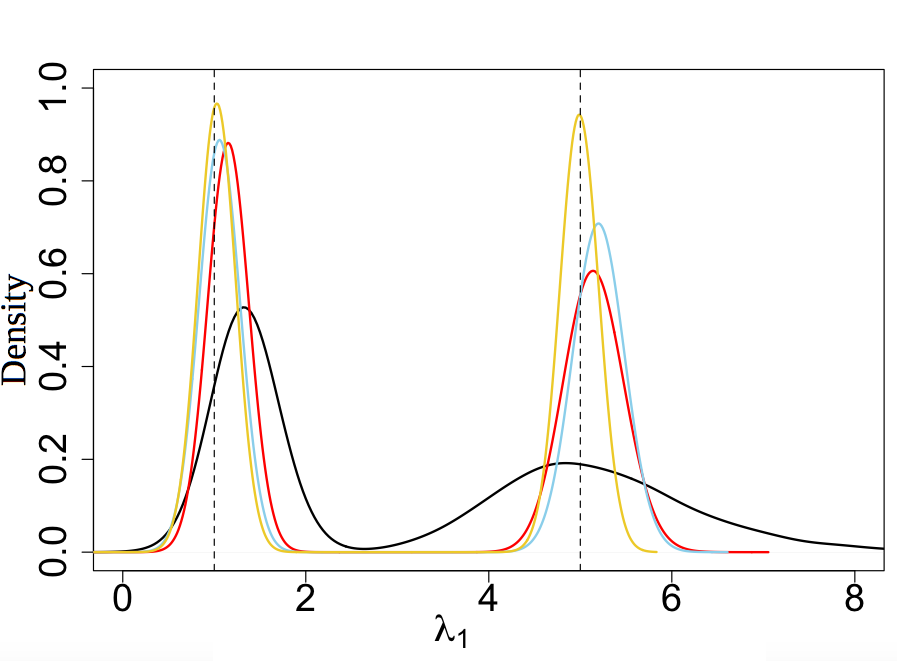}}
\put(7,0){\includegraphics[width=7.5cm,height=4.5cm]{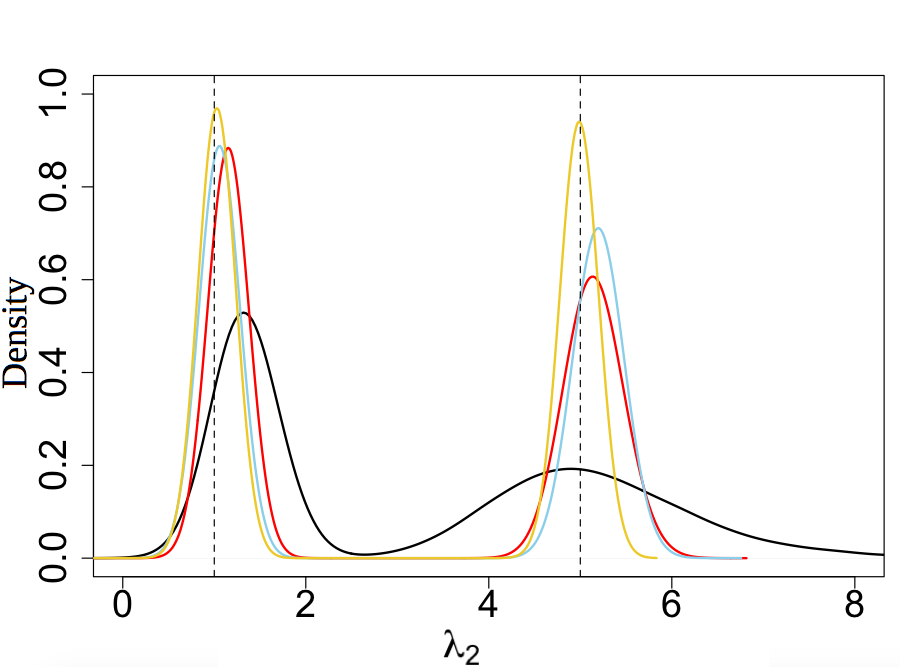}}
\put(5.7,8){(a} \put(13.8,8){(b} \put(5.7,3.5){(c} \put(13.8,3.5){(d}
\end{picture} \caption{{\bf Mixture of two Poisson distributions \ref{ex1}:} Comparison between the empirical densities of $5\,10^4$ MCMC simulations of (a) the global mean, (b) the weight and (c)-(d) component means. True values are indicated by dashed lines. The different colors in all graphs correspond to the different sample sizes indicated in (a). 
} 
\label{fig:Poi1}
\end{figure}

\begin{figure}[h!]
\centering\setlength{\unitlength}{1cm}
\begin{picture}(13,9)
\put(-1,4.5){\includegraphics[width=7.5cm,height=4.5cm]{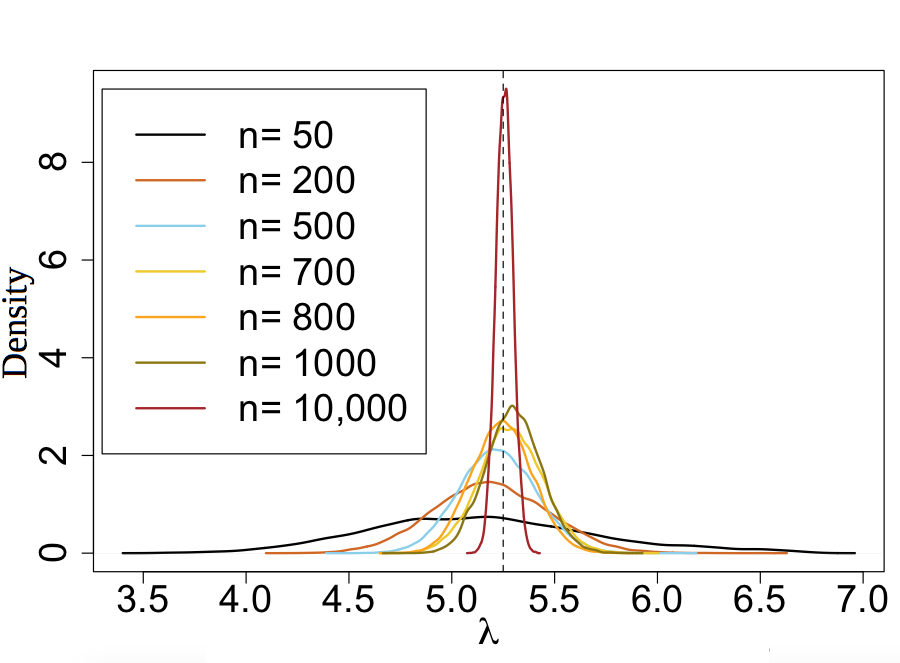}}
\put(7,4.5){\includegraphics[width=7.5cm,height=4.5cm]{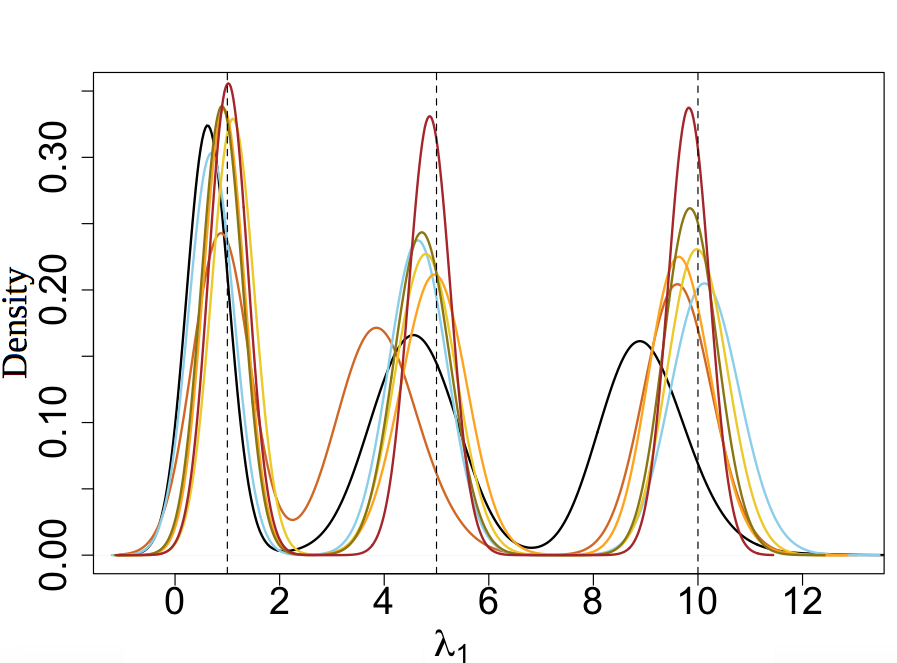}}
\put(-1,0){\includegraphics[width=7.5cm,height=4.5cm]{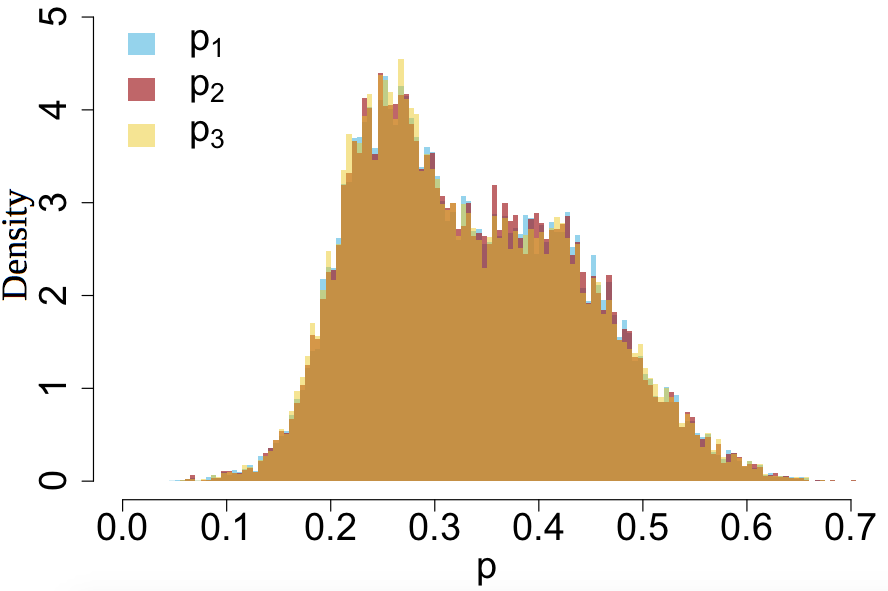}}
\put(7,0){\includegraphics[width=7.5cm,height=4.5cm]{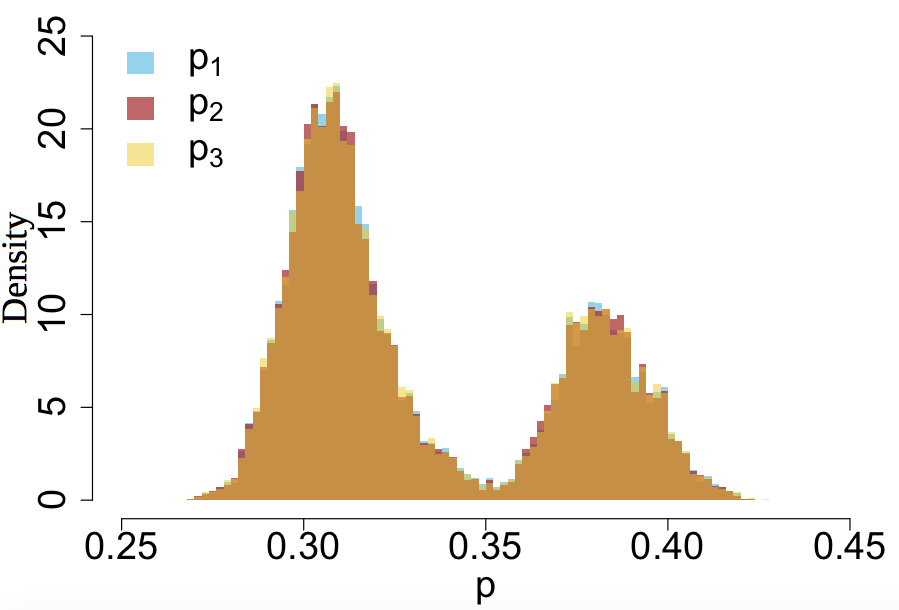}}
\put(5.7,8){(a} \put(13.8,8){(b} \put(5.7,3.5){(c} \put(13.8,3.5){(d}
\end{picture} 
\caption{{\bf Mixture of three Poisson distributions:} Comparison between the
empirical densities of $5\,10^4$ MCMC simulations of (a) the global mean, (b) the
component mean $\lambda_1$ and weights for two samples of size (c) $n=200$ and
(d) $n=10^4$. True values are indicated by dashed lines. The different colors
in all graphs correspond to the different sample sizes indicated in (a). 
total number of MCMC iterations is $5\,10^4$ with a burn-in of $10^3$ iterations.
}
\label{fig:Poi3}
\end{figure}

\begin{example}\label{ex1}
The following two Poisson mixture models are considered for various sample sizes, from $50$ to $10^4$. 

\begin{itemize}
\item[ ] {\bf Model 1}: $0.6 \mathcal{P}(1)+0.4 \mathcal{P}(5)$
\item[ ] {\bf Model 2}: $0.3 \mathcal{P}(1)+0.4 \mathcal{P}(5)+0.3 \mathcal{P}(10)$
\end{itemize}

Figures \ref{fig:Poi1} and \ref{fig:Poi3} display the performances of the Metropolis-within-Gibbs sampler (see
also Figure \ref{fig:MCMCpois} in Supplementary Material). The convergence of the resulting
sequence of estimates to the true values is illustrated by the figures as the number of data points
increases. While label switching occurs with our prior modelling, as shown in both Figures \ref{fig:Poi1} and \ref{fig:Poi3}, the point estimate of each parameter subjected to label switching (component weights and means) can be computed by relabelling the MCMC draws. We then derive point estimates by clustering over the parameter space, using k-mean clustering, resulting in close agreement with the true values.

\end{example}
  
\section{Conclusion}\label{sec:con}

We have introduced a novel parametrisation for location-scale mixture models. By
expressing the parameters in terms of the global mean and global variance of
the mixture of interest, it has been shown that the remaining parameters vary
within a compact set. This reparameterisation makes the use of a
well-defined uniform prior possible for these parameters (as well as any proper prior)
and we established that an improper prior reproducing the Jeffreys prior on location-scale parameters
induces a proper posterior distribution for a minimal sample size. We illustrated the
implications of this new prior modelling and of the resulting MCMC algorithms
on some standard distributions, namely mixtures of Gaussian, Poisson and
exponential distributions and their compound extensions. While the notion of a
{\em non-informative} or {\em objective} prior is mostly open to
interpretations and somehow controversial, we argue we have defined in this
paper what can be considered as the first reference prior for mixture models.
We have shown further that relatively standard simulation algorithms are able to
handle these new parametrisations, as exhibited by our {Ultimixt} R package,
and that they can manage the computing issues connected with label switching. 

While the extension to non-Gaussian cases with location-scale features
is shown here to be conceptually straightforward, considering this reparameterisation
in higher dimensions is delicate when made in terms of the covariance matrices. Indeed,
even though we can easily set the variance matrix of the mixture model as a reference parameter,
reparameterising the component variance matrices against this reference matrix
and devising manageable priors remains an open problem that we are currently exploring.


\footnotesize
\hyphenation{Post-Script Sprin-ger}

\normalsize
\appendix
\section*{Supplementary material}

\section{Proof of Lemma \ref{lem:mings}}

The population mean is given by
$$\BE_{\btheta,\bp}[X]=  \sum_{i=1}^k p_i \BE_{f(\cdot|\theta_i) }[X] = \sum_{i=1}^k p_i\mu_i$$
where $\BE_{f(\cdot|\theta_i) }[X]$ is the expected value component $i$. Similarly, the population variance is given by
$$\varm_{\btheta,\bp}(X)=\sum^k_{i=1} p_i \BE_{f(\cdot|\theta_i)}[X^2]- \BE_{\btheta,\bp}[X]^2 = 
\sum_{i=1}^k p_i (\sigma_i^2 + \mu_i^2)-\BE_{\btheta,\bp}[X]^2\,,$$
which concludes the proof.

\section{Proof of Lemma \ref{lem:const}}

The result is a trivial consequence of Lemma \ref{lem:mings}. The population mean is
$$\BE_{\btheta,\bp}[X]=\sum_{i=1}^k p_i\mu_i=\sum_{i=1}^k p_i (\mu+\sigma\alpha_i)=\mu+\sigma\sum_{i=1}^k p_i\alpha_i $$
and the first constraint follows. The population variance is
\[ \begin{array}{c c l}
\varm_{\btheta,\bp}(X)&=&\displaystyle\sum_{i=1}^k p_i\sigma_i^2 + \sum_{i=1}^k p_i(\mu_i^2-\BE_{\btheta,\bp}[X]^2) \\
&=&\displaystyle\sum_{i=1}^k p_i\sigma^2\tau_i^2 +\sum_{i=1}^k 
p_i (\mu^2+2\sigma\mu\alpha_i+\sigma^2\alpha_i^2-\mu^2 ) \\
&=&\displaystyle\sum_{i=1}^k p_i\sigma^2\tau_i^2 +\sum_{i=1}^k p_i\sigma^2\alpha_i^2
\end{array} \]
The last equation simplifies to the second constraint above.

\section{Proof of Theorem \ref{th:proper}}

When $n=1$, it is easy to show that the posterior is not proper. The marginal likelihood is then 
\[ \begin{array}{c c l}
M_k(x_1)&=&\displaystyle\sum^k_{i=1} \int p_i f(x_1|\mu+\sigma\alpha_i,\sigma^2\tau_i^2) \pi(\mu,\sigma,\bp,\balpha,\btau) \, \text{d}(\mu,\sigma,\bp,\balpha,\btau) \\ 
&=& \displaystyle\sum^k_{i=1} \int \left\{\int \dfrac{p_i}{\sqrt{2\pi}\sigma^2\tau_i } \exp \left(
\dfrac{-(x_{1}-\mu-\sigma\alpha_i)^2}{2\tau_i^2\sigma^2} \right) \, \text{d}(\mu,\sigma) \right\} 
\pi(\bp,\balpha,\btau) \, \text{d}(\bp,\balpha,\btau) \\
&=& \displaystyle\sum^k_{i=1} \int \left\{\int_0^{\infty} \dfrac{p_i}{\sigma}  \, \text{d}\sigma \right\} 
\pi(\bp,\balpha,\btau) \, \text{d}(\bp,\balpha,\btau) \\
\end{array} \]
The integral against $\sigma$ is then not defined.

For two data-points, $x_1, x_2 \sim \sum^k_{i=1} p_i f(\mu+\sigma\alpha_i, \sigma^2\tau_i^2)$, the associated marginal
likelihood is 
\[ \begin{array}{c c l}
M_k(x_1,x_2) &=& \displaystyle\int \prod^2_{j=1} \left\{ \sum^k_{i=1} p_i f(x_j| \mu+\sigma\alpha_i, \sigma^2\tau_i^2 )
\right\} \pi(\mu,\sigma,\bp,\balpha,\btau) \, \text{d}(\mu,\sigma,\bp,\balpha,\btau)  \\
&=& \displaystyle\sum^k_{i=1}\displaystyle\sum^k_{j=1} \displaystyle\int p_i p_j f(x_1| \mu+\sigma\alpha_i, \sigma^2\tau_i^2 )f(x_2| \mu+\sigma\alpha_j, \sigma^2\tau_j^2 )  \pi(\mu,\sigma,\bp,\balpha,\btau) \, \text{d}(\mu,\sigma,\bp,\balpha,\btau)  ~.
\end{array} \]
If all those $k^2$ integrals are proper, the posterior distribution is proper. An arbitrary integral
$(1\le i,j\le k)$ in this sum leads to
\[ \begin{array} {l l}
\displaystyle\int &p_i p_j f(x_1| \mu+\sigma\alpha_i, \sigma^2\tau_i^2 )f(x_2| \mu+\sigma\alpha_j, \sigma^2\tau_j^2 )  \pi(\mu,\sigma,\bp,\balpha,\btau) \, \text{d}(\mu,\sigma,\bp,\balpha,\btau)  \\
=& \displaystyle\int \Bigg\{\int \dfrac{p_i p_j}{2\pi \sigma^3 \tau_i \tau_j} \exp \Bigg[
\dfrac{-(x_{1}-\mu-\sigma\alpha_i)^2}{2\tau_i^2\sigma^2} + \dfrac{-(x_{2}-\mu-\sigma\alpha_j)^2}{2\tau_j^2\sigma^2}
\Bigg] \text{d}(\mu,\sigma) \Bigg\} \pi(\bp,\balpha,\btau) \, \text{d}(\bp,\balpha,\btau) \\
=&  \displaystyle\int \Bigg\{\int_0^{\infty} \dfrac{p_i p_j}{\sqrt{2\pi} \sigma^2 \sqrt{\tau^2_i +\tau^2_j}} \exp \Bigg[
\dfrac{-1}{2(\tau_i^2+\tau_j^2)} \Bigg ( \dfrac{1}{\sigma^2}(x_{1}-x_{2})^2 + \dfrac{2}{\sigma}(x_{1}-x_{2})(\alpha_i-\alpha_j) \\
&\qquad +(\alpha_i-\alpha_j)^2 \Bigg) \Bigg]  \, \text{d}\sigma \Bigg \} \pi(\bp,\balpha,\btau) \,
\text{d}(\sigma,\bp,\balpha,\btau)\,.
\end{array} \]
Substituting $\sigma=1/z$, the above is integrated with respect to $z$, leading to
\[ \begin{array}{l l}
 \displaystyle\int &\Bigg\{ \displaystyle\int_0^{\infty} \dfrac{p_i p_j}{\sqrt{2\pi} \sqrt{\tau^2_i +\tau^2_j}} \exp \Bigg (
\dfrac{-1}{2(\tau_i^2+\tau_j^2)} \Bigg ( z^2(x_{1}-x_{2})^2 + 2z(x_{1}-x_{2})(\alpha_i-\alpha_j) \\
&\qquad +(\alpha_i-\alpha_j)^2 \Bigg) \Bigg) \, \text{d}z \Bigg\} \pi(\bp,\balpha,\btau) \, \text{d}(\bp,\balpha,\btau) \\
= & \displaystyle\int \Bigg\{ \displaystyle\int_0^{\infty} \dfrac{p_i p_j}{\sqrt{2\pi} \sqrt{\tau^2_i +\tau^2_j}} \exp \Bigg (
\dfrac{-(x_{1}-x_{2})^2}{2(\tau_i^2+\tau_j^2)} \Big ( z+\dfrac{\alpha_i-\alpha_j}{x_{1}-x_{2}} \Big ) ^2  \Bigg) 
 \, \text{d} z \Bigg\} \pi(\bp,\balpha,\btau)  \,\text{d}(\bp,\balpha,\btau) \\
=& \displaystyle\int \frac{p_i p_j}{ |x_1-x_2|}  \Phi\left(
-\dfrac{\alpha_i-\alpha_j}{x_1-x_2}\dfrac{|x_1-x_2|}{\sqrt{\tau_i^2+\tau_j^2}}\right)  \pi(\bp,\balpha,\btau) ~
\text{d}(\bp,\balpha,\btau)\,,
\end{array}\]
where $\Phi$ is the cumulative distribution function of the standardised Normal distribution. 
Given that the prior is proper on all remaining parameters of the mixture, it follows that the integrand is 
bounded by $1/|x_{1}-x_{2}|$, and so it integrates against the remaining components of $\btheta$.  
Now, consider the case $n\ge 3$.
Since the posterior $\pi(\btheta|x_1,x_2)$ is proper, it constitutes a proper prior when considering only the
observations $x_3,\ldots,x_n$. Therefore, the posterior is almost everywhere proper.

\section{Proof of Theorem \ref{thm:poimix}}

Considering one single positive observation $x_1$, the marginal likelihood is
\begin{align*}
M_k(x_1)&=\sum_{i=1}^k\int p_if(x_1|\lambda \gamma_i/p_i)\pi(\lambda, \pmb\gamma, \pmb{p})d(\lambda, \pmb\gamma, \pmb{p})\\
&=\sum_{i=1}^k\int \left(\int_0^\infty p_i \exp(-\lambda \gamma_i/p_i)(\lambda \gamma_i/p_i)^{x_1}/x_1!\pi(\lambda)d\lambda\right)\pi(\pmb\gamma, \pmb{p})d(\pmb\gamma, \pmb{p})\\
&=\sum_{i=1}^k\int \left(p_i (\gamma_i/p_i)^{x_1}/{x_1}!\int_0^\infty \exp(-\lambda \gamma_i/p_i)\lambda^{x_1-1} d\lambda\right)\pi(\pmb\gamma, \pmb{p})d(\pmb\gamma, \pmb{p})\\
&=\sum_{i=1}^k\int \left(p_i/x_1\right)\pi(\pmb\gamma, \pmb{p})d(\pmb\gamma, \pmb{p})
\end{align*}
Since the prior is proper on all remaining parameters of the mixture, the
integrals in the above sum are all finite. The posterior $\pi(\lambda,
\pmb\gamma, \pmb{p}|x_1)$ is therefore proper and it constitutes a proper prior
when considering further observations $x_2, \ldots , x_n$. Therefore, the resulting posterior for the sample
$(x_1,\ldots,x_n)$ is proper.

\section{Proof of Theorem \ref{thm:expmix}}

Considering one single observation $x_1$, the marginal likelihood is
\begin{align*}
M_k(x_1)&=\sum_{i=1}^k\int p_if(x_1|p_i/\lambda \gamma_i)\pi(\lambda, \pmb\gamma, \pmb{p})d(\lambda, \pmb\gamma,
\pmb{p})\\
&=\sum_{i=1}^k\int \left(\int_0^\infty p_i \exp(-\nicefrac{p_ix}{\lambda \gamma_i})
\nicefrac{p_i}{\lambda \gamma_i} \pi(\lambda)d\lambda\right)\pi(\pmb\gamma, \pmb{p})\text{d}(\pmb\gamma, \pmb{p})\\
&=\sum_{i=1}^k\int p_i \int_0^\infty \nicefrac{p_i}{\lambda\gamma_i}\exp(-\lambda \gamma_i x_1/p_i)
\text{d}\lambda\pi(\pmb\gamma, \pmb{p})\text{d}(\pmb\gamma, \pmb{p})\\
&=\sum_{i=1}^k\int \left(p_i/x_1\right)\pi(\pmb\gamma, \pmb{p})\text{d}(\pmb\gamma, \pmb{p})\,.
\end{align*}
Since the prior is proper on all remaining parameters of the mixture, the
integrals in the above sum are all finite. The posterior $\pi(\lambda,
\pmb\gamma, \pmb{p}|x_1)$ is therefore proper and it constitutes a proper prior
when considering further observations $x_2, \ldots , x_n$. Therefore, the resulting posterior for the sample
$(x_1,\ldots,x_n)$ is proper.

\section{Pseudo-code representations of the MCMC algorithms for the Normal and Poisson cases}
\label{sec:psudo}

\begin{figure}[H]
\fbox{
  \parbox{1\textwidth}{
    {\bf Metropolis-within-Gibbs algorithm for a reparameterised Gaussian mixture} 
\begin{description}
\item [1] Generate initial values $(\mu^{(0)},\sigma^{(0)},\bp^{(0)},\varphi^{(0)},\xi_1^{(0)},\ldots,\xi^{(0)}_{k-1},\varpi^{(0)}_1,\ldots,\varpi^{(0)}_{k-2} )$.
\item[2] For $t=1,\ldots,T $, the update of
$(\mu^{(t)},\sigma^{(t)},\bp^{(t)},\varphi^{(t)},\xi_1^{(t)},\ldots,\xi^{(t)}_{k-1},\varpi^{(t)}_1,\ldots,\varpi^{(t)}_{k-2}
)$ \\is as follows;

\begin{description}
\item[2.1] Generate a proposal $\mu'\sim \mathcal{N}(\mu^{(t-1)},\epsilon_{\mu})$ and update $\mu^{(t)}$ against $\pi(\cdot|\bx,\sigma^{(t-1)},\bp^{(t-1)},\varphi^{(t-1)},\boldsymbol\xi^{(t-1)},\boldsymbol\varpi^{(t-1)})$.

\item[2.2] Generate a proposal $\log(\sigma)'\sim \mathcal{N}(\log(\sigma^{(t-1)}),\epsilon_{\sigma})$ and update
$\sigma^{(t)}$ against $\pi(\cdot|\bx,\mu^{(t)},\bp^{(t-1)},\varphi^{(t-1)},\boldsymbol\xi^{(t-1)},\boldsymbol\varpi^{(t-1)})$.

\item[2.3] Generate proposals $\xi'_i\sim \mathcal{U}[0,\pi/2 ]$, $i=1,\cdots,k-1$, and update $(\xi^{(t)}_1,\ldots,\xi^{(t)}_{k-1})$ against $\pi(\cdot|\bx,\mu^{(t)},\sigma^{(t)}, \bp^{(t-1)},\varphi^{(t-1)}, \boldsymbol\varpi^{(t-1)})$. 

\item[2.4] Generate proposals $\varpi'_i\sim \mathcal{U}[0,\pi ]$, $i=1,\cdots,k-3$, and $\varpi'_{k-2}\sim \mathcal{U}[0,2\pi ]$. Update $(\varpi^{(t)}_1,\ldots,\varpi^{(t)}_{k-2})$ against $\pi(\cdot|\bx,\mu^{(t)},\sigma^{(t)}, \bp^{(t-1)}, \varphi^{(t-1)},\boldsymbol\xi^{(t)})$. 

\item[2.5] Generate a proposal $(\varphi^2)'\sim \mathcal{B}eta((\varphi^2)^{(t)}\epsilon_{\varphi}+1,
(1-(\varphi^2)^{(t)})\epsilon_{\varphi}+1)$ and update $ \varphi^{(t)}$ against $\pi(\cdot|\bx,\mu^{(t)},\sigma^{(t)}, \bp^{(t-1)}, \boldsymbol\xi^{(t)},\boldsymbol\varpi^{(t)})$. 

\item[2.6] Generate a proposal $\bp' \sim \text{Dir}(p_1^{(t-1)}\epsilon_p+1, \ldots, p_k^{(t-1)}\epsilon_p+1)$, and
update $\bp^{(t)}$ against $\pi(\cdot|\bx,\mu^{(t)},\sigma^{(t)},\varphi^{(t)},\boldsymbol\xi^{(t)},\boldsymbol\varpi^{(t)})$.

\item[2.7] Generate proposals $\xi'_i\sim U[\xi^{(t)}_i-\epsilon_{\xi},\xi^{(t)}_i+\epsilon_{\xi} ]$, $i=1,\cdots,k-1$,
and update $(\xi^{(t)}_1,\ldots,\xi^{(t)}_{k-1})$ against $\pi(\cdot|\bx,\mu^{(t)},\sigma^{(t)},\bp^{(t)},\varphi^{(t)}, \boldsymbol\varpi^{(t)})$. 

\item[2.8] Generate proposals $\varpi'_i\sim U[\varpi^{(t)}_i-\epsilon_{\varpi},\varpi^{(t)}_i+\epsilon_{\varpi} ]$,
$i=1,\cdots,k-2$, and update $(\varpi^{(t)}_1,\ldots,\varpi^{(t)}_{k-2})$ against $\pi(\cdot|\bx,\mu^{(t)},\sigma^{(t)},\bp^{(t)},\varphi^{(t)}, \boldsymbol\xi^{(t)})$.

\end{description} \end{description}
}}
\caption{\label{fig:algoboX}
Pseudo-code representation of the Metropolis-within-Gibbs algorithm used in this paper for reparameterisation (ii) of
the Gaussian mixture model,
based on two sets of spherical coordinates. For simplicity's sake, we denote $\bp^{(t)}=(p_1^{(t)},\ldots,p_k^{(t)})$,
$\bx=(x_1,\ldots,x_n)$, $\boldsymbol\xi^{(t)}=(\xi_1^{(t)},\ldots,\xi^{(t)}_{k-1})$ and $\boldsymbol\varpi^{(t)}=(\varpi^{(t)}_{1},\ldots,\varpi^{(t)}_{k-2})$.
}
\end{figure}

\begin{figure}[H]
\fbox{
  \parbox{1\textwidth}{
    {\bf Metropolis-within-Gibbs algorithm for a reparameterised Poisson mixture}
\begin{description}
\item [1] Generate initial values $(\lambda^{(0)},\pmb{\gamma}^{(0)},\bp^{(0)})$.
\item[2] For $t=1,\ldots,T $, the update of $(\lambda^{(t)},\pmb{\gamma}^{(t)},\bp^{(t)})$ \\is as follows;

\begin{description}
\item[2.1] Generate a proposal $\lambda'\sim \mathcal{N}(\log(\overline{X}), \epsilon_{\lambda} )$ and update $\lambda^{(t)}$ against $\pi(\cdot|\bx,\pmb{\gamma}^{(t-1)},\bp^{(t-1)})$.

\item[2.2] Generate a proposal $\pmb\gamma' \sim \text{Dir}(\gamma_1^{(t-1)}\epsilon_\gamma+1, \ldots, \gamma_k^{(t-1)}\epsilon_\gamma+1)$, and
update $\pmb\gamma^{(t)}$ against $\pi(\cdot|\bx,\lambda^{(t)},\bp^{(t-1)})$.

\item[2.3] Generate a proposal $\bp' \sim \text{Dir}(\bp_1^{(t-1)}\epsilon_p+1, \ldots, \bp_k^{(t-1)}\epsilon_p+1)$, and
update $\bp^{(t)}$ against $\pi(\cdot|\bx,\lambda^{(t)},\pmb\gamma^{(t)})$.
\end{description} \end{description}
}}
\caption{\label{fig:MCMCpois}
Pseudo-code representation of the Metropolis-within-Gibbs algorithm used to approximate the posterior distribution of
the reparameterisation of the Poisson mixture. For
$i=1, \ldots, k$, we denote $\pmb{\gamma}=(\gamma_1, \ldots, \gamma_k)$,
$\bp^{(t)}=(p_1^{(t)},\ldots,p_k^{(t)})$ and
$\bx=(x_1,\ldots,x_n)$. $\overline{X}$ is the empirical mean of the observations.
}
\end{figure}

\newpage

\section{Convergence graphs for Example \ref{ex:simulated_data_K3}}

\begin{figure}[!h]
\centering\setlength{\unitlength}{1cm}
\begin{picture}(16,5)
\put(0,0){\includegraphics[width=7.7cm,height=4cm]{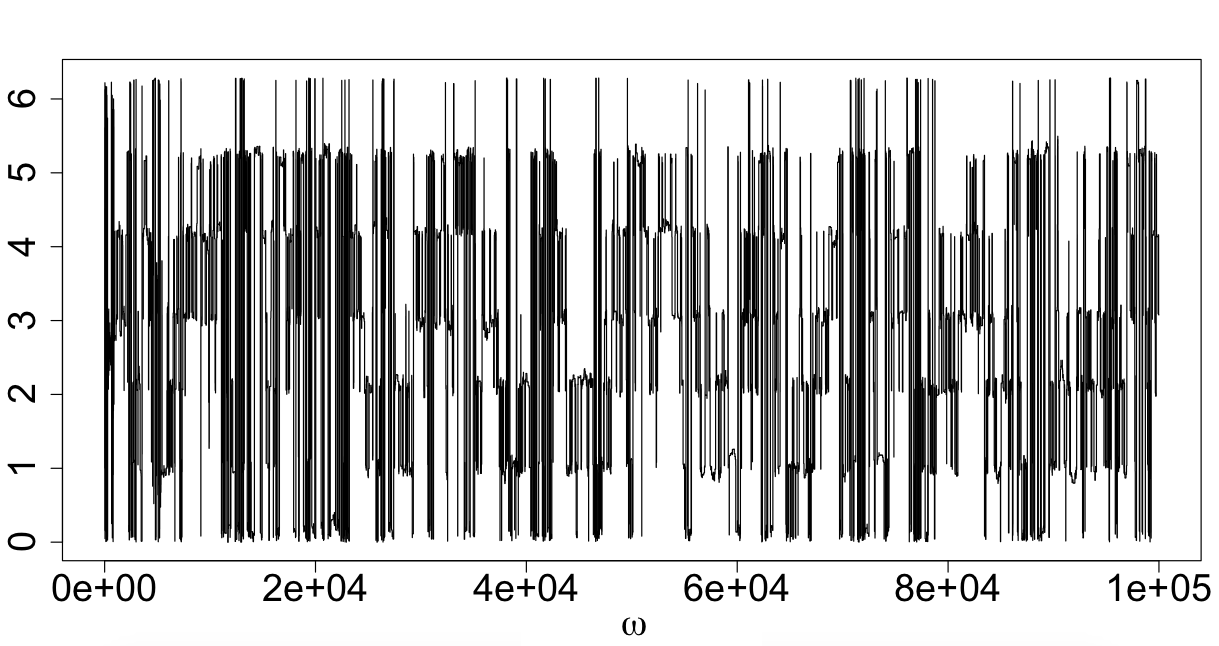}}
\put(8,0){\includegraphics[width=8cm,height=4cm]{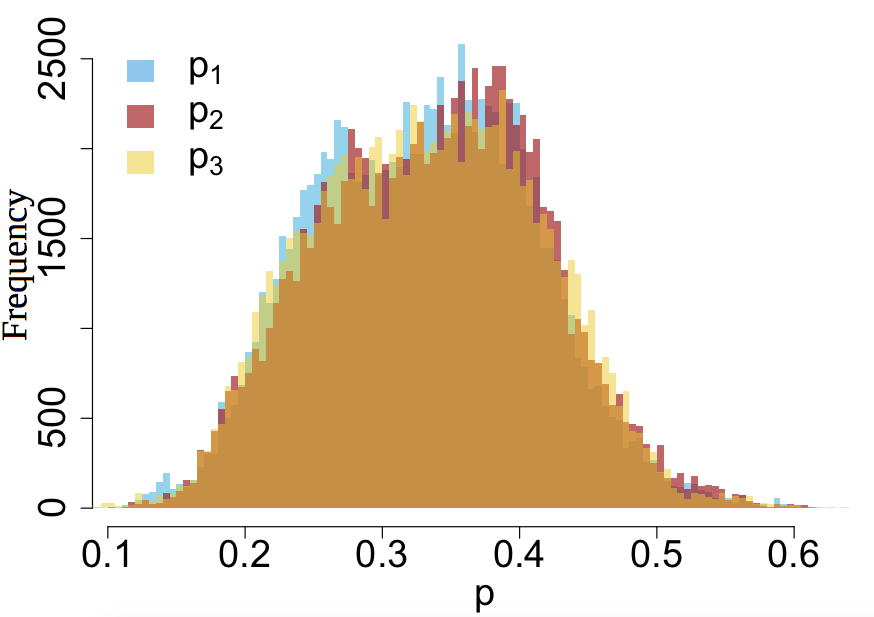}}
\end{picture}
\caption{{\bf Example \ref{ex:simulated_data_K3}:} {\em (Left)} Evolution of the sequence $(\varpi^{(t)})$ and 
{\em (Right)} histograms of the simulated weights based on $10^5$ iterations of an
adaptive Metropolis-within-Gibbs algorithm with independent proposal on $\varpi$. }
\label{figu3} 
\end{figure}
 
\begin{figure}[!h]
\includegraphics[width=5.3cm,height=4.5cm]{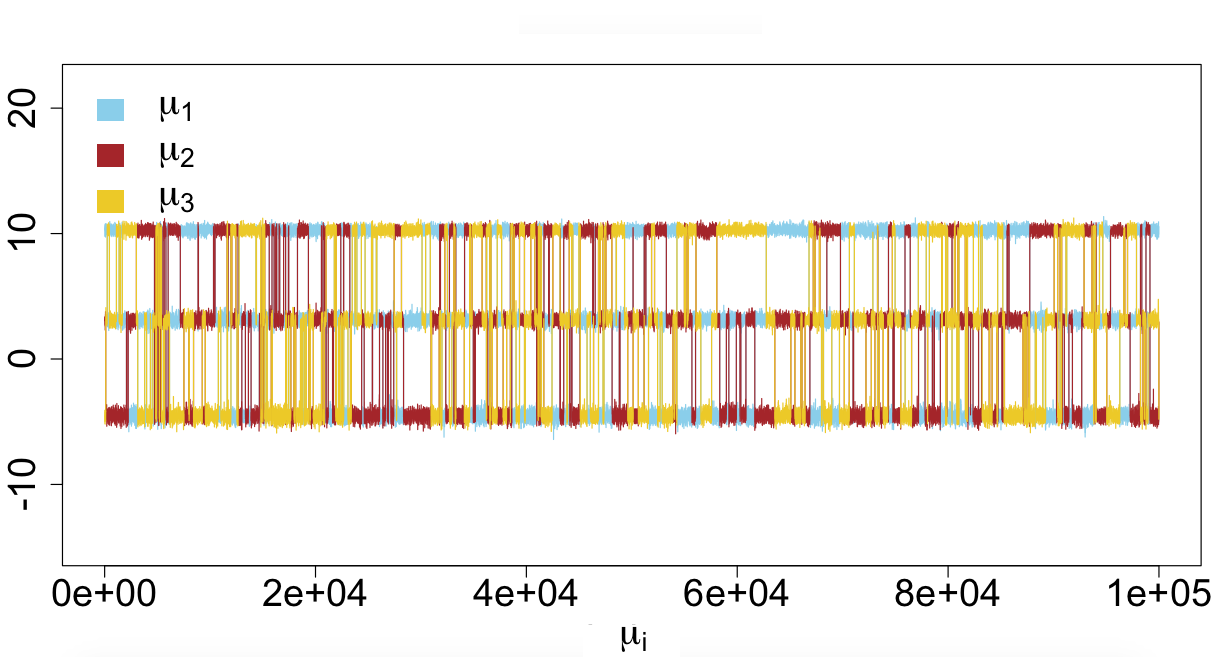}
\includegraphics[width=5.3cm,height=4.5cm]{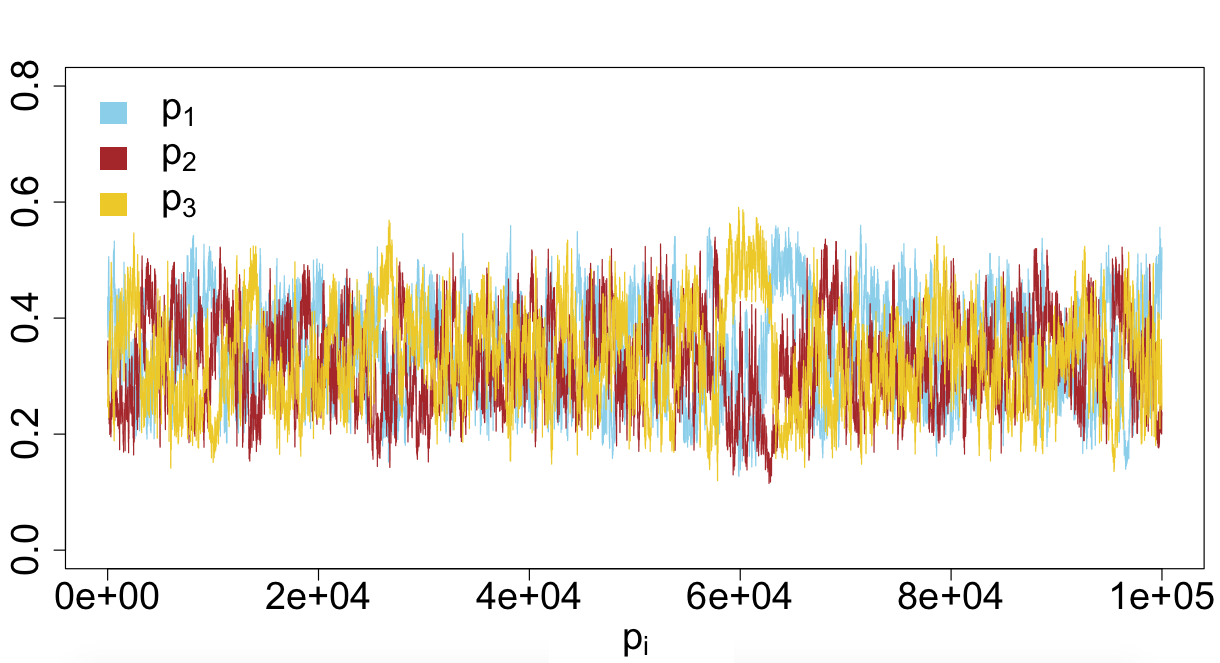}
\includegraphics[width=5.3cm,height=4.5cm]{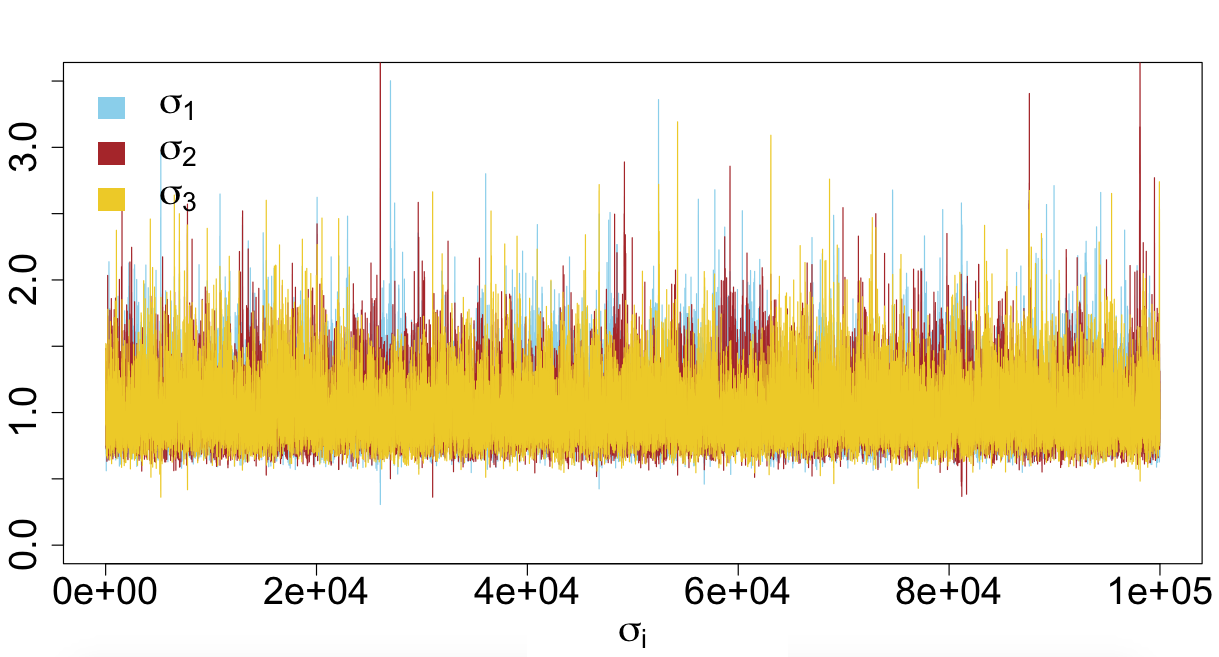}
\caption{{\bf Example \ref{ex:simulated_data_K3}:} Traces of $10^5$ simulations from the posterior
distributions of the component means, standard deviations and weights, involving an additional random walk proposal on
$\varpi$. } 
\label{figu4} 
\end{figure}

\section{An illustration of the proposal impact on Old Faithful}
\label{sec:OleFaz}

We analysed the R benchmark Old Faithful dataset, using the 272 observations of 
eruption times and a mixture model with two components. The empirical mean and
variance of the observations are $(3.49, 1.30)$.

\begin{figure}[H]
\includegraphics[width=.33\textwidth,height=3.5cm]{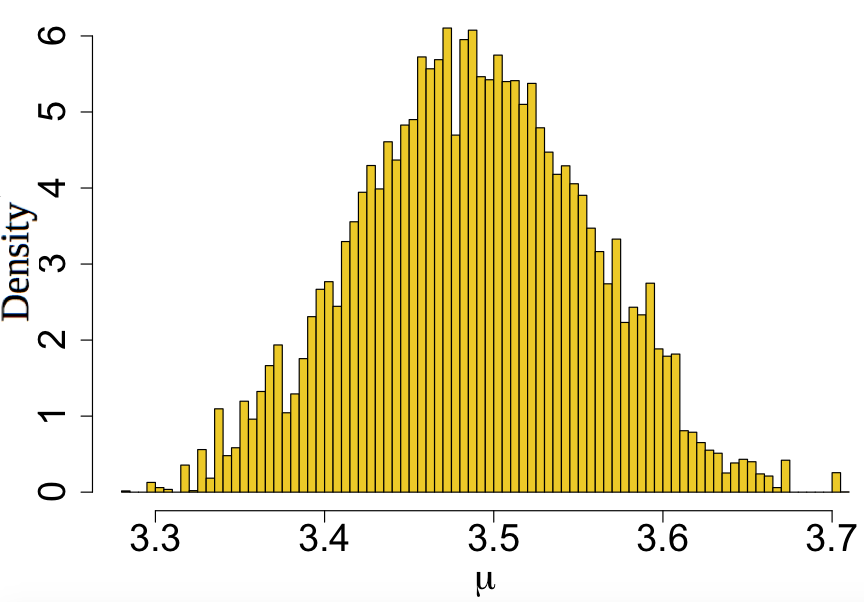}\includegraphics[width=.33\textwidth,height=3.5cm]{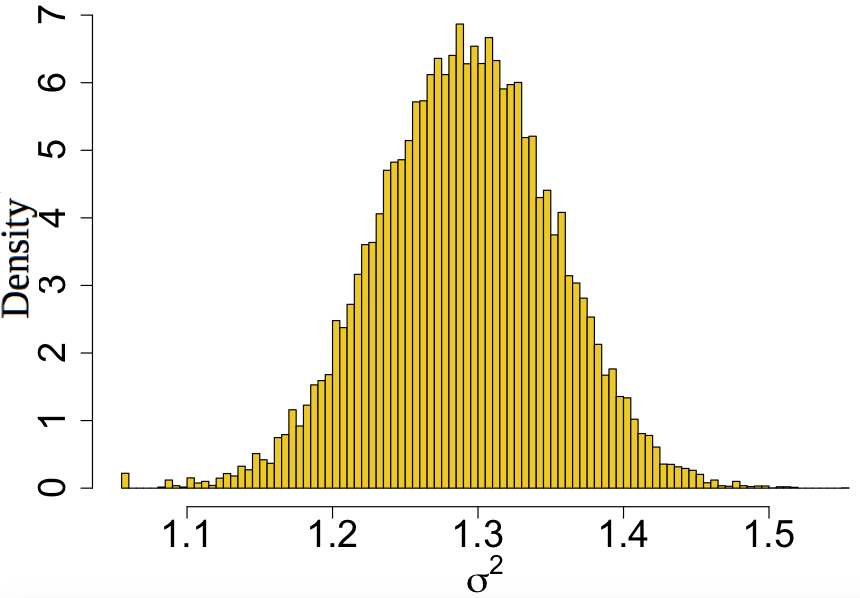}\includegraphics[width=.33\textwidth,height=3.5cm]{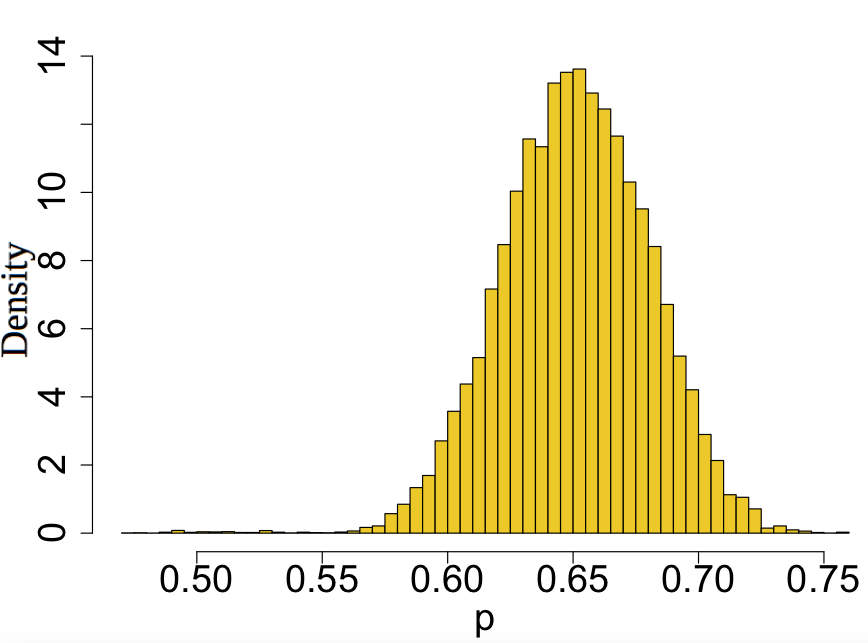}
\includegraphics[width=.33\textwidth,height=3.5cm]{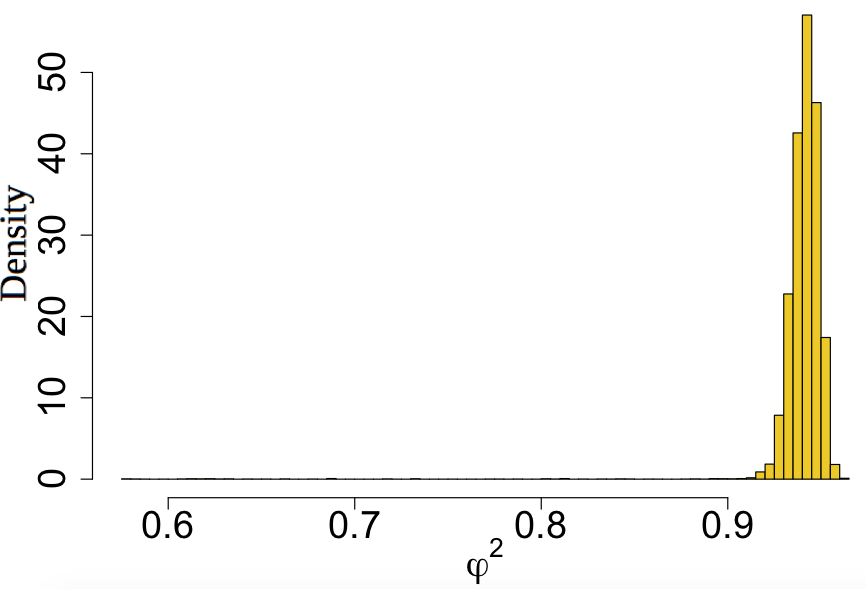}\includegraphics[width=.33\textwidth,height=3.5cm]{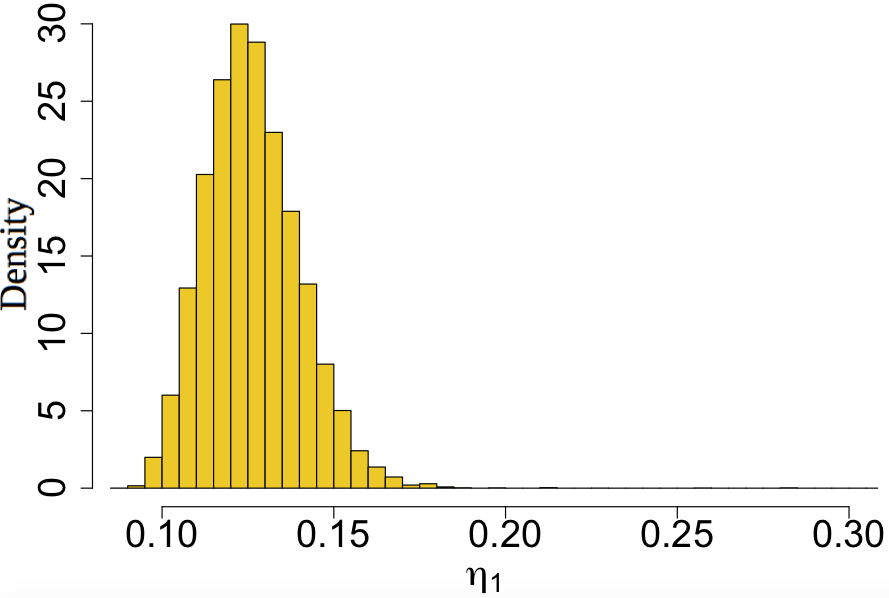}\includegraphics[width=.33\textwidth,height=3.5cm]{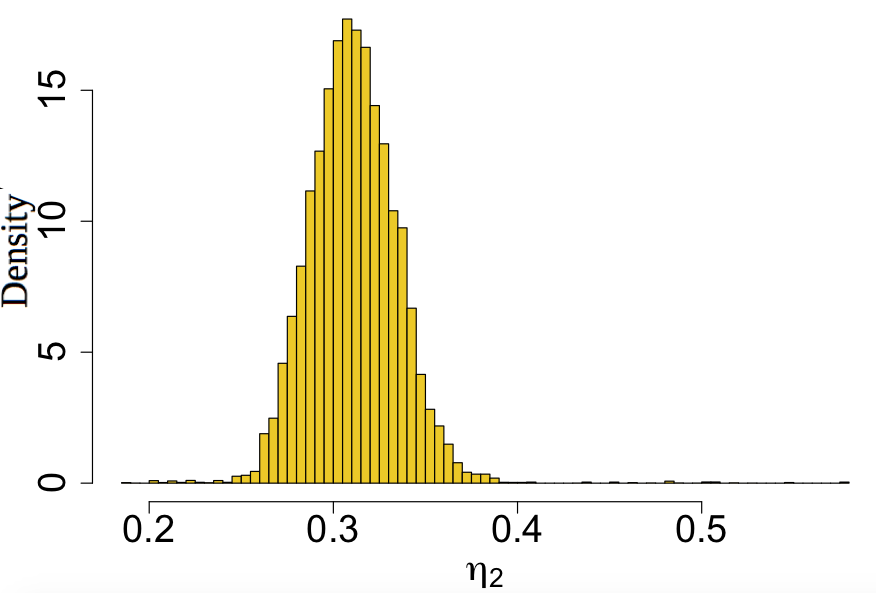}
\caption{{\bf Old Faithful dataset:} 
Posterior distributions of the parameters of a two-component mixture distribution based on $50,000$ MCMC iterations.}
\label{fig3}
\end{figure}

When using Proposal 1, the optimal scales $\epsilon_\mu, \epsilon_p, \epsilon$ after $50,000$ iterations are
$0.07,\allowbreak 501.1,\allowbreak 802.19$, respectively. The posterior distributions of the generated samples shown in Figure \ref{fig3} demonstrate a strong concentration of $(\mu,\sigma^2)$ near the empirical mean and variance. There is a strong indication that the chain gets trapped into a single mode of the posterior density. 

\section{Values of estimates of the mixture parameters behind the CT dataset}

\begin{table}[H]
\scriptsize
 \begin{tabular}{l l}
\begin{tabular}{|c || c c c c c c|} 
\hline 
&  \multicolumn{5}{c}{\bf k-means clustering ({\sf Ultimixt})} & \\  \cline{2-7}
 & $p_1$& $p_2$ & $p_3$ & $p_4$ & $p_5$ & $p_6$ \\ 
Median &   0.16 & 0.17 & 0.25 &   0.34 & 0.04&  0.04\\ 
Mean &   0.16 & 0.17 & 0.25 & 0.34 & 0.03 &  0.04 \\ 
\cline{2-7} &$\mu_1$ & $\mu_2$ & $\mu_3$ & $\mu_4$ & $\mu_5$ & $\mu_6$\\
 Median & 68.75 & 89.88 & 121.0 & 134.6 & 201.3 & 244.4  \\ 
Mean & 68.68 & 89.88 & 121.1 & 134.6 & 203.3  &242.2 \\
\cline{2-7} &$\sigma_1$ & $\sigma_2$ & $\sigma_3$ & $\sigma_4$ & $\sigma_5$ & $\sigma_6$ \\
Median & 17.37 &9.380 &13.66& 4.613  &23.62& 2.055 \\
Mean &17.38 & 9.388 & 13.69& 4.615 & 22.72& 2.995 \\
\cline{1-7}
&  \multicolumn{5}{c}{\bf Relabelled using MAP ({\sf Ultimixt})} & \\ \cline{2-7}
&$p_1$ & $p_2$ & $p_3$ & $p_4$ & $p_5$ & $p_6$  \\
  Median & 0.16 &0.17 & 0.25  & 0.34 & 0.03 & 0.04    \\ 
Mean &  0.16 & 0.17 & 0.25 & 0.34 &0.03 &  0.04    \\ 
2.5\% & $0.15$ & $0.16$ & $0.24$ & $0.33$ & $0.03$ & $0.04$     \\
 97.5\% & 0.17 & 0.18 & 0.27 & 0.36 & 0.04 & 0.05  \\
 \cline{2-7} &$\mu_1$ & $\mu_2$ & $\mu_3$ & $\mu_4$ & $\mu_5$ & $\mu_6$ \\
 Median & 68.71 & 89.91 & 121.0 & 134.6 & 201.0 & 244.4   \\ 
  Mean & 68.67 & 89.92  &121.1 & 134.6 & 201.1 & 244.3   \\ 
  2.5\% & 65.89 & 88.85  & 120.4 & 134.2  & 196.3 & 243.9  \\  
  97.5\%  & 71.10 & 90.82 & 122.5 & 135.1 & 206.3 &  244.9 \\ 
  \cline{2-7} &$\sigma_1$ & $\sigma_2$ &$\sigma_3$ & $\sigma_4$ & $\sigma_5$ & $\sigma_6$\\
  Median & 17.37 &  9.386 & 13.62 &  4.615 & 23.62  & 2.054  \\ 
  Mean & 17.38  & 9.391 & 13.71 & 4.621 &  23.66 & 2.047  \\ 
  2.5\% & 16.43 & 8.673   & 12.83 & 4.370 & 21.89  &  1.822  \\ 
  97.5\% &  18.38  & 10.12  & 14.80 &  4.871 & 25.67 & 2.220  \\ \hline
  \end{tabular}
&  \begin{tabular}{|c || c c c c c c|} 
\hline 
& \multicolumn{5}{c}{\bf Gibbs sampler ({\sf bayesm})}& \\ \cline{2-7}
 & $p_1$ & $p_2$ & $p_3$ & $p_4$ & $p_5$ & $p_6$ \\ 
Mean & 0.17 & 0.18 & 0.21 & 0.36 & 0.04 & 0.04  \\
\cline{2-7} & $\mu_1$ & $\mu_2$ & $\mu_3$ & $\mu_4$ & $\mu_5$ & $\mu_6$  \\ 
Mean &  69.74 & 88.99 & 119.94 & 134.48 & 196.69 & 243.94 \\
\cline{2-7} &$\sigma_1$ &$\sigma_2$ & $\sigma_3$ & $\sigma_4$ & $\sigma_5$ & $\sigma_6$ \\
Mean & 18.44 & 8.048 & 10.85 & 4.956 & 24.08 &  3.808  \\ \hline
 
& \multicolumn{5}{c}{\bf EM estimate ({\sf mixtools})} & \\ \cline{2-7}
 &$p_1$ & $p_2$ & $p_3$ & $p_4$ & $p_5$ & $p_6$ \\
 & 0.15 & 0.18 & 0.27 & 0.33 & 0.04 & 0.04 \\
\cline{2-7} &$\mu_1$ & $\mu_2$ & $\mu_3$ & $\mu_4$ & $\mu_5$ &$\mu_6$ \\ 
  &   67.62 &  88.57 & 121.9 & 134.6 & 203.2 & 244.5 \\
\cline{2-7} &$\sigma_1$ & $\sigma_2$ & $\sigma_3$ & $\sigma_4$ & $\sigma_5$ & $\sigma_6$  \\
 & 17.42 & 7.818 &13.84 &  4.579 & 23.27 & 1.841 \\ \hline
\end{tabular} \end{tabular}
\caption{\small \label{table:pork} {\bf CT image dataset}: Estimates of the parameters of a mixture of 6 components.}
\end{table} 

\section{R codes}

File ``allcodes.tar.gz" contains all R codes used to produce the examples
processed in the paper (this is a GNU zipped tar file). In addition, the R
package Ultimixt is deposited on CRAN and contains R functions to estimate
univariate mixture models from one of the three families studied in this paper.
\end{document}